\documentclass[11pt]{article}
\usepackage[utf8]{inputenc}
\usepackage[margin=0.8in]{geometry}
\usepackage{amsmath,amsfonts,bm}
\usepackage{color}
\usepackage{xargs,xstring}
\usepackage{setspace}
\usepackage{natbib}
\usepackage{graphicx}
\usepackage{float}
\usepackage{accents}
\usepackage[subrefformat=parens,labelformat=parens]{subfig}
\usepackage[inline]{enumitem}

\usepackage{longtable}
\usepackage{array}
\newcolumntype{L}[1]{>{\raggedright\let\newline\\\arraybackslash\hspace{0pt}}m{#1}}
\newcolumntype{C}[1]{>{\centering\let\newline\\\arraybackslash\hspace{0pt}}m{#1}}
\newcolumntype{R}[1]{>{\raggedleft\let\newline\\\arraybackslash\hspace{0pt}}m{#1}}

\usepackage{titlesec}
\titleformat*{\section}{\large\bfseries}
\titleformat*{\subsection}{\normalsize\bfseries}
\titleformat*{\subsubsection}{\normalsize\bfseries}
\titleformat*{\paragraph}{\normalsize\bfseries}
\titleformat*{\subparagraph}{\normalsize\bfseries}

\title{Soft Tensor Regression}
\author{}
\author{Georgia Papadogeorgou$^1$, Zhengwu Zhang$^2$, David B. Dunson$^3$}
\date{}

\newcommand\numberthis{\addtocounter{equation}{1}\tag{\theequation}}

\newcommand{\softer}{{Softer}}
\newcommand{\parafac}{PARAFAC}
\newcommandx{\subs}{{j_1j_2\dots j_K}}
\newcommandx{\subss}{{\underaccent{\tilde}{j}}}
\newcommandx{\tensor}{\bm X}
\newcommandx{\B}{\bm B}
\newcommandx{\C}{\bm C}
\newcommandx{\Rp}[1][1=k]{\mathbb{R}^{p_{#1}}}
\newcommandx{\betakj}[2][1=T,2=T]{\beta_{\IfEqCase{#1}{{T}{k} {F}{k'}},
                                         \IfEqCase{#2}{{T}{\subss} {F}{\subss'}}}}
\newcommandx{\gammakj}[2][1=T,2=T]{\gamma_{\IfEqCase{#1}{{T}{k} {F}{k'}},
                                   j_{\IfEqCase{#1}{{T}{k} {F}{k'}}}\IfEqCase{#2}{{F}{'}}}}
\newcommandx{\wkj}{w_{k,j_k}}
\newcommandx{\tod}[1][1=T]{^{(\IfEqCase{#1}{{T}{d} {F}{d'}})}}
\newcommandx{\lambdafrac}[1][1=F]{\frac{2b_\lambda^2}{\IfEqCase{#1}{{T} {b_\tau}}(a_\lambda - 1)(a_\lambda - 2)}}

\newcommandx{\vect}[1][1=x]{\mathrm{vec}(#1)}
\newcommand{\hadamard}{\circ}
\newcommand{\Var}{\mathrm{Var}}
\newcommand{\Cov}{\mathrm{Cov}}

\newcommand{\E}{\mathbb{E}}

\newcommand{\independent}{\perp\!\!\!\perp}
\newcommand{\addvar}{\frac{a_\sigma}{b_\sigma }}
\newcommandx{\neigh}[1][1={}]{\mathcal{B}_{\epsilon^{#1}}^{\infty}\big(\B^0\big)}

\newcommand{\numtraits}{15}

\usepackage{amsthm}

\newtheoremstyle{myremark}
    {\topsep}                    
    {\topsep}                    
    {}                           
    {}                           
    {\bfseries}                   
    {.}                          
    {.5em}                       
    {}  

\theoremstyle{plain}

\newtheorem{prop}{Proposition}

\theoremstyle{myremark}
\newtheorem{remark}{Remark}

\usepackage[nokeyprefix]{refstyle}
\usepackage{varioref}
\usepackage{xr-hyper}
\usepackage[colorlinks=TRUE,allcolors=blue]{hyperref}
\usepackage[capitalise,noabbrev,compress]{cleveref}

\crefformat{equation}{#2\textup{(#1)}#3}
\crefformat{section}{\S#2#1#3}
\crefformat{subsection}{\S#2#1#3}
\crefformat{subsubsection}{\S#2#1#3}
\crefformat{prop}{Proposition #1}
\crefname{appsec}{Appendix}{Appendices}

\allowdisplaybreaks
\begin{document}

\maketitle
\vspace{-20pt}
\begin{center}
\footnotesize
$^1$Department of Statistics, University of Florida \\
$^2$Department of Statistics and Operations Research, University of North Carolina at Chapel Hill \\
$^3$Department of Statistical Science, Duke University
\end{center}

\begin{abstract}
Statistical methods relating tensor predictors to scalar outcomes in a regression model generally vectorize the tensor predictor and estimate the coefficients of its entries employing some form of regularization, use summaries of the tensor covariate, or use a low dimensional approximation of the coefficient tensor. However, low rank approximations of the coefficient tensor can suffer if the true rank is not small.
We propose a tensor regression framework which assumes a \textit{soft} version of the parallel factors (PARAFAC) approximation. In contrast to classic \parafac{}, where each entry of the coefficient tensor is the sum of products of row-specific contributions across the tensor modes, the soft tensor regression (\softer) framework allows the row-specific contributions to vary around an overall mean. We follow a Bayesian approach to inference, and show that softening the \parafac{} increases model flexibility, leads to improved estimation of coefficient tensors, more accurate identification of important predictor entries, and more precise predictions, even for a low approximation rank. From a theoretical perspective, we show that employing \softer{} leads to a weakly consistent posterior distribution of the coefficient tensor, {\it irrespective of the true or approximation tensor rank}, a result that is not true when employing the classic \parafac{} for tensor regression.
In the context of our motivating application, we adapt \softer{} to symmetric and semi-symmetric tensor predictors and analyze the relationship between brain network characteristics and human traits.
\end{abstract}

\textit{keywords:} adjacency matrix, Bayesian, brain connectomics, graph data, low rank, network data, parafac, tensor regression


\section{Introduction}

In many applications, data naturally have an array or tensor structure. When the tensor includes the same variable across two of its modes, it is often referred to as a network. Graph or network dependence is often summarized via an adjacency matrix or tensor. For example, data might correspond to an $R \times R \times p$ array containing $p$ features measuring the strength of connections between an individual's $R$ brain regions.
In tensor data analysis, interest often lies in characterizing the relationship between a tensor predictor and a scalar outcome within a regression framework. Estimation of such regression models most often requires some type of parameter regularization or dimensionality reduction since the number of entries of the tensor predictor is larger than the sample size.

In this paper, we propose a soft tensor regression (\softer{}) framework for estimating a high-dimensional linear regression model with a tensor predictor and scalar outcome. \softer{} directly accommodates the predictor's structure by basing the coefficient tensor estimation on the parallel factors approximation, similarly to other approaches in the literature. However, in contrast to previously developed methodology, \softer{} adaptively expands away from its low-rank mean to adequately capture and flexibly estimate more complex coefficient tensors. \softer's deviations from the underlying low-rank, tensor-based structure are interpretable as variability in the tensor's row-specific contributions.

\subsection{Tensor regression in the literature}

Generally, statistical approaches to tensor regression fall in the following categories: they estimate the coefficients corresponding to each tensor entry with entry-specific penalization, regress the scalar outcome on low-dimensional summaries of the tensor predictor, or estimate a coefficient tensor assuming a low-rank approximation.

A simple approach to tensor regression vectorizes the tensor predictor and fits a regression model of the outcome on the tensor's entries while performing some form of variable selection or regularization. Examples include \cite{Cox2003functional} and \cite{Craddock2009disease} who employed support vector classifiers to predict categorical outcomes based on the participants' brain activation or connectivity patterns. Other examples in neuroscience include \cite{Mitchell2004learning, Haynes2005predicting, OToole2005partially, Polyn2005category} and \cite{Richiardi2011decoding} (see \cite{Norman2006beyond} for a review). However, this regression approach to handle tensor predictors is, at the least, unattractive, since it fails to account for the intrinsic array structure of the predictor, effectively flattening it prior to the analysis.

Alternatively, dimensionality reduction can be performed directly on the tensor predictor reducing it to low dimensional summaries. In such approaches, the expectation is that these summaries capture all essential information while decreasing the number of parameters to be estimated.
For example, \cite{Zhang2019tensor} and \cite{Zhai2019predicting} use principal component analysis to extract information on the participants' structural and functional brain connectivity, and use these principal components to study the relationship between brain network connections and outcomes within a classic regression framework. However, this approach could suffer due to its unsupervised nature which selects principal components without examining their relationship to the outcome. Moreover, the performance of the low-dimensional summaries is highly dependent on the number and choice of those summaries, and the interpretation of the estimated coefficients might not be straightforward.

\cite{Ginestet2017hypothesis} and \cite{Durante2018bayesian} developed hypothesis tests for differences in the brain connectivity distribution among subgroups of individuals, employed in understanding the relationship between categorical outcomes and binary network measurements. 
Even though related, such approaches do not address our interest in building regression models with tensor predictors.

An attractive approach to tensor regression performs dimension reduction on the coefficient tensor.
Generally, these approaches exploit a tensor's Tucker decomposition \citep{Tucker1966} and its restriction known as the parallel factors (\parafac{}) or canonical decomposition.
According to the \parafac{}, a tensor is the sum of $D$ rank-1 tensors, and each entry can be written as the sum of $D$ products of row-specific elements. The minimum value of $D$ for which that holds is referred to as the tensor's rank. Note that the word ``row'' along a tensor mode is used here to represent rows in the classic matrix sense (slice of the tensor along the first mode), columns (slice of the tensor along the second mode), or slices along higher modes.

Within the frequentist paradigm, \cite{Hung2013matrix} suggested a bi-linear logistic regression model in the presence of a matrix predictor. For a tensor predictor, \cite{Zhou2013tensor} and \cite{Li2018tucker} exploited the \parafac{} and Tucker decompositions respectively, and proposed low rank approximations to the coefficient tensor. \cite{Guhaniyogi2017bayesian} proposed a related Bayesian tensor regression approach for estimating the coefficient tensor.
Even though these approaches perform well for prediction in these high-dimensional tensor settings, they are bound by the approximation rank in the sense that they cannot capture any true coefficient tensor. Moreover, these approaches are not directly applicable for identifying important connections.
In this direction, \cite{Wang2014clinical}  imposed sparsity in the coefficient tensor of a multi-linear logistic regression model by penalizing the \parafac{} contributions. \cite{Wang2018symmetric} proposed a related approach to identify small brain subgraphs that are predictive of an individual's cognitive abilities. Further, \cite{Guha2018bayesian} assumed a \parafac{} decomposition of the mean of the coefficient tensor and used a spike-and-slab prior distribution to identify brain regions whose connections are predictive of an individual's creativity index.

Low-rank approximations to the coefficient tensor of a linear tensor regression model provide a supervised approach to estimating the relationship between a tensor predictor and a scalar outcome. However, such approximations can lead to a poorly estimated coefficient tensor, misidentification of important connections, and inaccurate predictions, if the true rank of the coefficient tensor is not small. As we will illustrate in \cref{sec:parafac}, this performance issue arises due to the inflexibility of the \parafac{} approximation which specifies that each row has a \textit{fixed} contribution to all coefficient entries that involve it, leading to an overly rectangular or block structure of the estimated coefficient tensor. If the true coefficient tensor does not exhibit such a block structure, a large number of components $D$ might be necessary in order to adequately approximate it. Due to this rigid structure, we refer to the \parafac{} approximation to estimating the coefficient tensor as the \textit{hard} \parafac.

Recently, there have been efforts to relax the hard \parafac{} structure employing non-parametric methodology using kernels, Gaussian-processes, or neural networks \citep{Signoretto2013learning, Suzuki2016minimax, Kanagawa2016gaussian, Imaizumi2016doubly, Maruhashi2018learning}. Even though these methods are very promising, we focus on regression models which include the tensor predictor linearly, and focus on flexibly estimating the linear functional. The ideas of \parafac{} {\it softening} presented here could be extended to non-linear settings and situations outside the scope of tensor-regression.

\subsection{Our contribution}

With our main focus being improved estimation and inference over model coefficients, we aim to address the inflexibility of the hard \parafac{} in the linear tensor regression setting. Towards this goal, we propose a hierarchical modeling approach to estimate the coefficient tensor. Similarly to the hard \parafac{}, each entry of the coefficient tensor is the sum of products of row-specific contributions. However, our model specification allows {\it a row's contribution} to the coefficients that involve it to be {\it entry-specific} and to vary around a row-specific mean. This row-specific mean resembles the row-specific contribution in the hard \parafac{} approximation, and the entry-specific row contributions can be conceived as random effects. Conceptually, the row-specific mean can be thought of as a row's overall importance, and the entry-specific deviations represent small variations in the row's importance when interacting with the rows of the other tensor modes.
Allowing for the row contributions to vary by entry leads to the softening of the hard structure in the \parafac{} approximation, and for this reason, we refer to it as the \textit{soft} \parafac{}. 
We refer to the tensor regression model that uses the soft \parafac{} for estimation of the coefficient tensor as \textit{Soft Tensor Regression} (\softer).

We follow a fully Bayesian approach to inference which allows for straightforward uncertainty quantification in the coefficient estimates and predictions. By studying the induced prior distribution on the coefficient tensor, we choose sensible values for the hyperparameters. Importantly, the flexible structure of the soft \parafac{} allows for \softer{} to capture {\it any true coefficient tensor} without increasing the base rank, in contrast to existing models that use strictly low-rank approximations.
We explicitly show this by proving that the imposed prior structure on the coefficient tensor has full support on a large class including true tensors of {\it any rank}, and its posterior distribution is consistent for {\it any} base, approximation rank.
We also illustrate it in simulations, where we show that the performance of \softer{} is quite robust to the choice of the approximation rank.
Due to its increased flexibility, \softer{} performs better than strictly low-rank models in identifying important entries of the tensor predictor. We extend \softer{} to generalized linear models and symmetric tensors, widening its applicability.


We use the soft tensor regression framework in a study of the relationship between brain structural connectomics and human traits for participants in the Human Connectome Project (HCP; \cite{VanEssen2013}). In the last few years, HCP has played a very important role in expanding our understanding of the human brain by providing a database of anatomical and functional connections and individual demographics and traits on over a thousand healthy subjects. Data availability and increased sample sizes have allowed researchers across various fields to develop and implement new tools in order to analyze these complex and rich data (see \cite{Cole2014article, McDonough2014network, Smith2015positive, Riccelli2017surface, Croxson2018structural} among many others).
Using data from the HCP, exploiting state-of-the-art connectomics processing pipelines \citep{Zhang2018mapping}, and within an adaptation of the supervised \softer{} framework for symmetric tensor predictors, we investigate the relationship between structural brain connection characteristics and a collection of continuous and binary human traits.

\section{Tensor regression} 
\label{sec:tensor_regression}

In this section, we introduce some useful notation and regression of scalar outcomes on a tensor predictor.

\subsection{Some useful notation}

Let $a \in \Rp[1]$ and $b \in \Rp[2]$. Then $a \otimes b \in \mathbb{R}^{p_1 \times p_2}$ is used to represent the outer product of $a$ and $b$ with dimension $p_1 \times p_2$ and entries $[a\otimes b]_{ij} = a_i b_j$. Similarly, for vectors $a_k \in \Rp$, $k = 1,2, \dots, K$, the outer product $a_1 \otimes a_2 \otimes \dots \otimes a_K$ is a $K$-mode tensor $\bm A$ of dimensions $p_1, p_2, \dots, p_K$ and entries $\bm A_{\subs} = \prod_{k = 1}^K a_{k, j_k}$.
For two tensors $\bm A_1, \bm A_2$ of the same dimensions, we use $\bm A_1 \hadamard \bm A_2$ to represent the Hadamard product, defined as the element-wise product of the two tensors. Further, we use $\langle \bm A_1, \bm A_2 \rangle_F$ to represent the Frobenius inner product, which is the sum of the elements of $\bm A_1 \hadamard \bm A_2$. When the tensors are vectors (1-mode), the Frobenius inner product is the classic dot product.

For a $K$-mode tensor $\bm A$ of dimensions $p_1, p_2, \dots, p_K$, the phrase ``$j^{th}$ slice of $\bm A$ along mode $k$'' is used to refer to the $(K-1)$-mode tensor $\bm G$ with dimensions $p_1, \dots, p_{k-1}, p_{k + 1}, \dots, p_K$ and entries $\bm G_{j_1 \dots j_{k-1} j_{k +1} \dots j_K} = \bm A_{j_1 \dots j_{k-1} j j_{k +1} \dots j_K}$. For example, the $j^{th}$ slice of a $p_1 \times p_2$ matrix along mode 1 is the matrix's $j^{th}$ row. As a result, we refer to ``slice-specific'' quantities as ``row-specific''even when that slice is not along mode 1. For example, for a $p_1 \times p_2$ matrix, the mean entry of the $j^{th}$ row along mode 2 is the mean of the $j^{th}$ column. Remembering that we use ``row'' to refer to slices (and not necessarily to rows in the classic matrix sense) will be useful when discussing the hard \parafac{} in \cref{sec:parafac} and when introducing the soft \parafac{} in \cref{sec:soft_tensor_regression}.

\subsection{Regression of scalar outcome on tensor predictor}

Let $Y_i$ be a continuous outcome, $\C_i = (C_{i1}, C_{i2}, \dots, C_{ip})^T$ scalar covariates, and $\tensor_i$ a $K$-mode tensor of dimensions $p_1, p_2, \dots, p_K$ with entries $[\tensor_i]_\subs = X_{i,\subs}$, for unit $i = 1, 2, \dots N$. Even though our development is presented here for continuous outcomes, the relationship between tensor predictors and binary or categorical outcomes can be similarly evaluated by considering an appropriate link function as we do in \cref{sec:application}. We study the relationship between the outcome and the scalar and tensor predictors by assuming a linear model
\begin{equation}
    Y_i = \mu + \C_i^T \bm \delta + \sum_{j_1 = 1}^{p_1} \sum_{j_2 = 1}^{p_2} \dots \sum_{j_K = 1}^{p_K} X_{i,\subs} \beta_{\subs} + \epsilon_i, \ \epsilon_i \sim N(0, \tau^2),
    \label{eq:tensor_regression}
\end{equation}
where $\bm \delta \in \Rp[]$ and $\beta_{\subs} \in \mathbb{R}$.
Alternatively, organizing all coefficients $\beta_{\subs}$ in a tensor $\B$ of equal dimensions to $\tensor$ and $\subs$ entry equal to $\beta_{\subs}$, the same model can be written as
\begin{equation}
    Y_i = \mu + \C_i^T\bm \delta + \langle \tensor_i, \B \rangle_F + \epsilon_i.
    \label{eq:tensor_regression2}
\end{equation}

Since the coefficient tensor $\B$ includes $\prod_{k = 1}^K p_k$ coefficients, it is infeasible to estimate it  without some form of regularization or additional structure. Penalization or variable selection approaches based on the vectorization of the tensor predictor are implemented directly on model \cref{eq:tensor_regression}, ignoring the predictor's tensor structure.
Alternatively, one approach to account for the predictor's inherent structure is to assume a low-rank approximation to $\B$ based on the \parafac{} decomposition, discussed in \cref{sec:parafac}.

\section{Tensor regression using the hard \parafac{} approximation}
\label{sec:parafac}

Under the \parafac{} decomposition, a tensor $\B \in \mathbb{R}^{p_1 \times p_2 \times \dots p_K}$ can be written as
\begin{equation}
    \B = \sum_{d = 1}^D \beta_1^{(d)} \otimes \beta_2^{(d)} \otimes \dots \otimes \beta_{K}^{(d)}
    \label{eq:parafac}
\end{equation}
for some integer $D$ and $\beta_k^{(d)} \in \Rp$. The minimum value of $D$ for which $\B$ equals its representation \cref{eq:parafac} is referred to as its rank. For matrices ($2-$mode tensors), this decomposition is equivalent to the singular value decomposition, and $D$ is the matrix rank. We refer to the $d^{th}$ term in the sum as the \parafac{}'s $d^{th}$ {\it component}.

The tensor \parafac{} decomposition leads to a natural approximation of the coefficient tensor in \cref{eq:tensor_regression2} by assuming that the coefficient tensor is in the form \cref{eq:parafac} for some \textit{small} value of $D$, potentially much smaller than its true rank. Then, the $\prod_{k = 1}^K p_k$ coefficients in $\B$ are approximated using $D\sum_{k = 1}^K p_k$ parameters leading to a large reduction in the number of quantities to be estimated.


However, this reduction in the number of parameters might come at a substantial price if the rank $D$ used in the approximation is smaller than the tensor's true rank. 
According to \cref{eq:parafac}, the $(\subs)$ entry of $\B$ is equal to
\begin{equation}
    \B_{\subs} = \sum_{d = 1}^D \beta_{1,j_1}^{(d)}\beta_{2,j_2}^{(d)}\dots \beta_{K,j_K}^{(d)}.
    \label{eq:hardB_entry}
\end{equation}
In turn, \cref{eq:hardB_entry} implies that row $j_k$ along mode $k$ has \textit{fixed importance}, expressed as fixed row contributions $\beta_{k,j_k}\tod$, to all coefficient entries $\B_{\subs}$ that include $j_k$, irrespective of the remaining indices. We refer to $\beta_{k,j_k}\tod$ as the $d^{th}$ $j_k$-row contribution along mode $k$.
This is best illustrated by considering a rank-1 2-mode tensor (matrix) $\B = \beta_1 \otimes \beta_2$ for vectors $\beta_1 \in \Rp[1]$ and $\beta_2 \in \Rp[2]$. Then, $\B_{j_1j_2} = \beta_{1,j_1} \beta_{2,j_2}$, and the same entry $\beta_{1,j_1}$ is used in $\B_{j_1j_2}$ irrespective of $j_2$. This gives rise to a \textit{rectangular structure} in $\B$ in which a row's importance, $\beta_{1,j_1}$, is fixed across all columns (and similarly for $\beta_{2,j_2}$).

\begin{figure}[!b]
\centering    
\subfloat[Rank 1]{\includegraphics[width=0.25\textwidth]{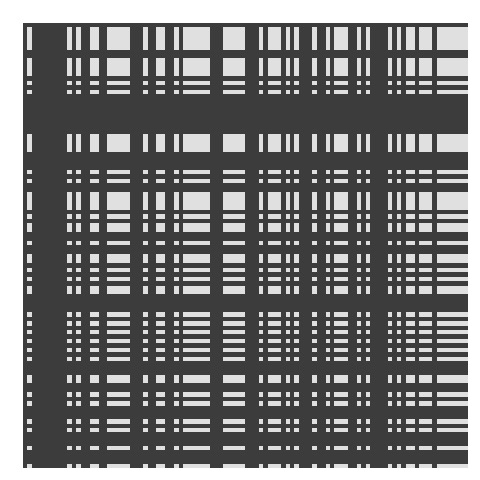} \label{fig:ordered_parafac_1a}}
\subfloat[Rank 3]{\includegraphics[width=0.25\textwidth]{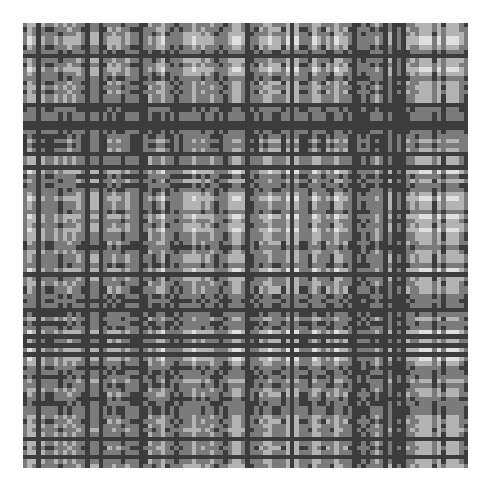} \label{fig:ordered_parafac_3}}
\subfloat[Rank 3 - ordered]{\includegraphics[width=0.25\textwidth]{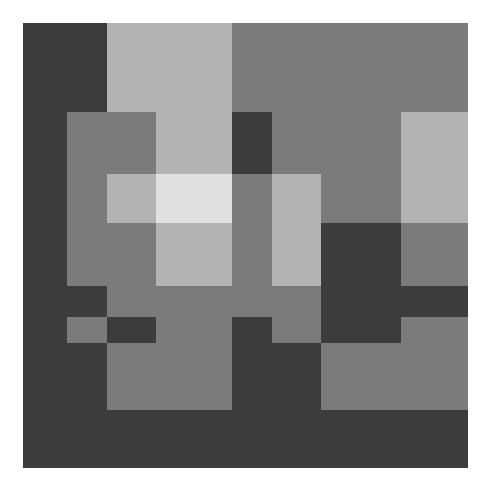} \label{fig:ordered_parafac_3b}}
\caption{Illustration of the \parafac{} Inflexibility. Panel (a) shows a rank-1 tensor of the form $\B = \beta_1 \otimes \beta_2$, for vectors $\beta_1, \beta_2 \in \{0, 1\}^{100}$. Panel (b) shows a rank-3 matrix that is the sum of the three rank-1 tensors like the one in panel (a). Panel (c) shows the same rank-3 matrix with rows and columns reordered according to their mean entry.}
\label{fig:ordered_parafac}
\end{figure}

We further illustrate this in \cref{fig:ordered_parafac_1a} where we plot $\beta_1 \otimes \beta_2$ for randomly generated vectors $\beta_1, \beta_2 \in \{0, 1\}^{100}$.
It is evident from \cref{fig:ordered_parafac_1a} that rank-1 matrices are organized in a rectangular structure where rows and columns are either uniformly important or not. Even though the generated vectors are binary for ease of illustration, the rectangular structure persists even when $\beta_1, \beta_2$ include non-binary entries.
The rectangular structure observed in rank-1 tensors indicates that a rank-1 ($D=1$) approximation to the coefficient tensor could be quite limiting. Generally, a rank-$D$ approximation for $D > 1$ is employed to estimate the coefficient tensor. \cref{fig:ordered_parafac_3} shows a matrix $\B$ of rank $D = 3$, summing over three rank-1 tensors like the one in \cref{fig:ordered_parafac_1a}. The rank-3 tensor alleviates but does not annihilate the rectangular structure observed previously. This is most obvious in \cref{fig:ordered_parafac_3b} where the rows and columns of \cref{fig:ordered_parafac_3} are re-ordered according to their mean entry.
In \cref{app_sec:hard_demonstration} we further demonstrate the inflexibility of the hard \parafac{}'s block structure.

The said block structure is also evident in the work by \cite{Zhou2013tensor, Guhaniyogi2017bayesian} and \cite{Li2018tucker} where they simulated data based on binary coefficient matrices. When these matrices represent combinations of rectangles (such as squares or crosses), the approximation performed well in estimating the true coefficient tensor. However, in situations where the true coefficient tensor was irregular, an increase in the rank was necessary in order to vaguely approximate the truth. 

\section{Soft tensor regression}
\label{sec:soft_tensor_regression}

Our approach proceeds by increasing the number of parameters in the regression model \cref{eq:tensor_regression2} and subsequently imposing sufficient structure to ensure model regularization and adaptability, simultaneously. It borrows from the low-rank structure of the hard \parafac{}, but it is more flexible than the rigid form in \cref{eq:hardB_entry} and \cref{fig:ordered_parafac}, and can adaptively expand away from low-rank. We introduce tensors $\B_k\tod$ of equal dimensions to $\B$ and write
\begin{equation}
\B = \sum_{d = 1}^D \B_1\tod \hadamard \B_2\tod \hadamard \dots \hadamard \B_K\tod.
\label{eq:softB}
\end{equation}
From \cref{eq:softB}, the coefficient with indices $\subss = (\subs)$ is written as the sum of $D$ products of $K$ parameters
\begin{equation}
\B_{\subss} = \sum_{d = 1}^D \beta_{1,\subss}\tod \beta_{2,\subss}\tod \dots \beta_{K,\subss}\tod,
\label{eq:softB_entry}
\end{equation}
where $\beta_{k,\subss}\tod$ is the $\subss^{th}$ entry of the tensor $\B_k\tod$.
For reasons that will become apparent later, the parameters $\beta_{k,\subss}\tod$ are referred to as the $j_k^{th}$ row-specific contributions along mode $k$ to the coefficient $\B_{\subss}$. Note that these row-specific contributions are allowed to depend on all indices $\subss$.
For unrestricted $\B_k\tod$s, \cref{eq:softB} does not impose any restrictions on the coefficient tensor and any tensor $\B$ can be written in this form (for example, take $D = 1$, $\B_1^{(1)} = \B$ and $\B_k^{(1)} = \mathbf{1}$, for all $k > 1$).

The representation of the coefficient tensor in \cref{eq:softB} might appear counter-productive at first since the already high-dimensional problem of estimating the $\prod_{k = 1}^K p_k$ parameters in $\B$ is translated to an {\it even higher}-dimensional problem with $DK \prod_{k = 1}^K p_k$ parameters in \cref{eq:softB}.
However, as we will see, that is not a problem if adequate structure is imposed on the tensors $\B_k\tod$. In fact, in \cref{subsec:parafac_representation} we show that the hard \parafac{} can be written in the form \cref{eq:softB} for carefully designed tensors $\B_k\tod$, which shows that sufficient structure on the tensors $\B_k\tod$ can lead to drastic dimensionality reduction. In \cref{subsec:soft_representation}, we present our approach which is based on imposing structure on the tensors $\B_k\tod$ in a careful manner such that it simultaneously allows for low-dimensional and flexible estimation of $\B$. We refer to the resulting characterization as the soft \parafac{} of $\B$.

\subsection{Representation of the hard \parafac{} motivating the soft \parafac{}}
\label{subsec:parafac_representation}

\begin{figure}[!t]
\centering
\subfloat[Hard \parafac{}]{
\includegraphics[width = 0.4\textwidth]{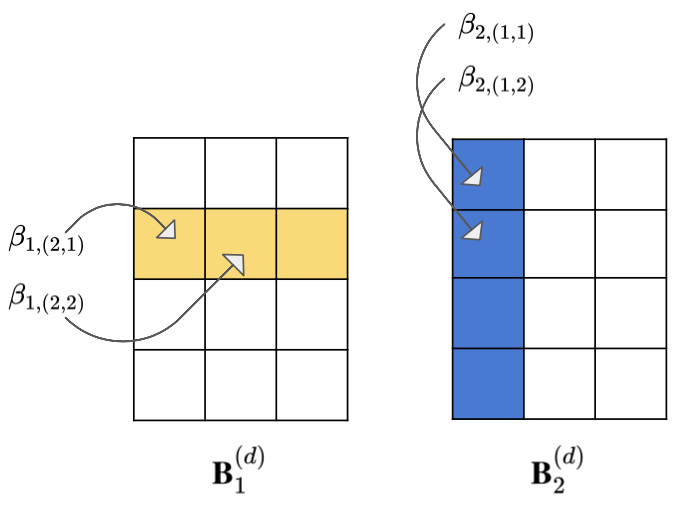} \label{fig:hard_parafac_visual}}
\hspace{0.1\textwidth}
\subfloat[Soft \parafac{}]{\includegraphics[width = 0.4\textwidth]{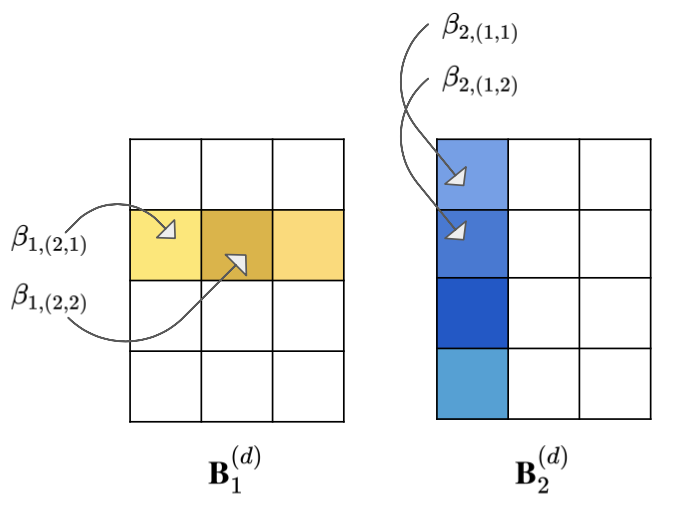} \label{fig:soft_parafac_visual}}
\caption{Row-specific Contributions for the Hard and Soft \parafac{}. Left: For the hard \parafac{}, the contributions are fixed across remaining indices. Right: For the soft \parafac{}, the contributions of a row and column are entry specific and centered around an overall mean.}
\label{fig:parafac_visual}
\end{figure}

As shown in \cref{eq:hardB_entry}, the hard \parafac{} row-specific contributions to each entry of the coefficient tensor are fixed across the remaining indices. Hence, the hard \parafac{} can be written in the form \cref{eq:softB} by specifying tensors $\B_k\tod$ of which two entries with the same $j_k$ index are equal:
$$
\big[\B_k\tod \big]_{j_1 j_2 \dots j_k \dots j_K} = \big[\B_k\tod \big]_{j_1' \dots j_{k-1}' j_k j_{k + 1}' \dots j_K'},
$$
This structure on the tensors $\B_k\tod$ can be visualized as $p_k$ \textit{constant} slices along mode $k$ representing the fixed row-specific contributions to all coefficient entries that involve it. This structure is illustrated in \cref{fig:hard_parafac_visual} for a 4-by-3 coefficient matrix. As an example, the contribution of row 2 along mode 1 is constant ($\beta_{1, (2, 1)}= \beta_{1, (2, 2)}= \beta_{1, (2, 3)}$), and the same is true for the contribution of row 1 along mode 2 ($\beta_{2, (1, 1)} = \beta_{2, (2, 1)} = \beta_{2, (3, 1)} = \beta_{2, (4, 1)}$). 
The connection between \cref{eq:softB} and the hard \parafac{} is the reason why we refer to $\beta_{k,\subss}\tod$ as row-specific contributions along mode $k$.

This demonstrates that the hard \parafac{} is one example of structure that can be imposed on the $\B_k\tod$s in order to approximate $\B$. However, the hard \parafac{} structure is quite strict, in that it imposes equalities across the $p_k$ slices of $\B_k\tod$. Furthermore, since the hard \parafac{} can only capture coefficient tensors of rank up to $D$, the hard \parafac's rigid structure on the tensors $\B_k\tod$ can severely limit the flexibility of the model in capturing a true coefficient tensor $\B$ of higher rank.

\subsection{The soft \parafac{}}
\label{subsec:soft_representation}

The soft \parafac{} builds upon the hard \parafac{}'s low-rank structure, while providing additional flexibility by introducing entry-specific variability in the row contributions. Instead of forcing all entries of $\B_k\tod$ with the same index $j_k$ to be equal to each other, the soft \parafac{} assumes that the entries of $\B_k\tod$ are centered around a $j_k$-specific value, $\gammakj\tod$.
Specifically, the soft \parafac{} specifies that, for all $k = 1, 2, \dots, K$, $j_k = 1, 2, \dots, p_k$, and $d = 1, 2, \dots D$,
\begin{equation}
\betakj\tod \sim N(\gammakj\tod, \sigma^2_k \zeta\tod),
\label{eq:soft_beta_dist}
\end{equation}
for some $\gammakj\tod \in \mathbb{R}, \sigma^2_k, \zeta\tod > 0$, where again $\subss$ is the $(\subs)$ entry of the tensor.
The connection between the soft and hard \parafac{} is evident by noticing that the mean values $\gamma$ depend only on the index $j_k$, and not on the remaining indices in $\subss$. In fact, the $\gamma$ parameters resemble the $j_k$-specific entries in the hard \parafac{}, and they specify that \softer{} is based on an underlying $\gamma$-based rank-D \parafac{}:
\[
\E[\B_\subss | \Gamma, S, Z] = \sum_{d = 1}^D \gamma_{1,j_1}\tod \gamma_{2,j_2}\tod \dots \gamma_{K,j_K}\tod,
\]
where $\Gamma, S, Z$ are the collections of the $\gamma, \sigma, \zeta$ parameters respectively.
At the same time, \cref{eq:soft_beta_dist} allows variation within the mode$-k$ slices of $\B_k\tod$ by considering them as random effects centered around an overall mean. This implies that row $j_k$'s importance is allowed to be {\it entry-specific} leading to a softening in the hard \parafac{} structure. The soft \parafac{} is illustrated in \cref{fig:soft_parafac_visual}. Here, the row-contributions are centered around a common value (a value resembling the row-contribution according to the hard \parafac{}) but are entry-specific. For example, $\beta_{1, (2,1)}$ is similar but not equal to $\beta_{1, (2, 2)}, \beta_{1, (2, 3)}$.

\begin{figure}[H]
\centering
\includegraphics[width = 0.75\textwidth, trim = 20 30 0 0, clip]{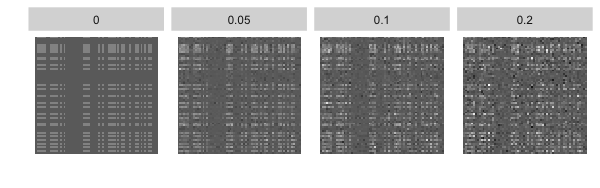}
\caption{The Softening Variance. Soft \parafac{} matrices for increasing variance of the entry-specific contributions. For variance equal to 0, the matrix corresponds to a rank-1 binary matrix generated as $\gamma_1 \otimes \gamma_2$ for $\gamma_1, \gamma_2 \in \{0, 1\}^{64}$. The remaining matrices are generated according to the soft \parafac{} and \cref{eq:soft_beta_dist} for variance equal to 0.05, 0.1, and 0.2.}
\label{fig:parafac_plus_sd}
\end{figure}

The entry-specific contributions deviate from the baseline according to a mode-specific parameter, $\sigma^2_k$, and a component-specific parameter, $\zeta\tod$. As we will discuss later, the inclusion of $\zeta\tod$ in the variance forces a larger amount of shrinkage on the entry-specific importance for components $d$ that have limited overall importance. For $\sigma^2_k \zeta\tod = 0$ the soft \parafac{} reverts back to the hard \parafac{}, with row-specific contributions fixed at $\gammakj\tod$. However, larger values of $\sigma^2_k \zeta\tod$ allow for a \parafac{}-based approximation that deviates from its hard underlying structure and can be used to represent {\it any} true tensor $\B$.
This is further illustrated in \cref{fig:parafac_plus_sd} where $\gamma_1, \gamma_2 \in \{0, 1\}^{64}$, and entry-specific contributions are generated according to \cref{eq:soft_beta_dist} with $\sigma^2_k \zeta\tod \in \{0, 0.05, 0.1, 0.2\}$. The soft \parafac{} resembles a structured matrix though higher values of the conditional variance lead to further deviations from a low-rank structure.

The structure imposed by the soft \parafac{} has an interesting interpretation. The parameters $\gammakj\tod$ represent the baseline importance of row $j_k$ along the tensor's $k^{th}$ mode. In contrast to the hard \parafac{}, the soft \parafac{} allows for structured, graph-based deviations of a row's importance, by allowing row $j_k$'s contribution to manifest differently based on the rows of the other modes that participate with it in a coefficient entry, $\subss \setminus \{j_k\}$, through $\betakj \tod$.
This interpretation of the soft \parafac{} structure is coherent in network settings like the one in our brain connectomics study, where we expect a brain region to have some baseline value for its connections, but the magnitude of this importance might slightly vary depending on the other region with which these connections are made. In this sense, defining deviations from the hard \parafac{} through deviations in the row-specific contributions as specified in \cref{eq:soft_beta_dist} represents a {\it tensor-based} relaxation of the hard \parafac{} structure, which is itself tensor-based.

\subsection{Bayesian inference in the soft tensor regression framework}
\label{subsec:bayesian}

\softer{} is placed within the Bayesian paradigm, which allows for straightforward uncertainty quantification.
We consider the structure on $\B_k\tod$ expressed in \cref{eq:soft_beta_dist} as part of the prior specification on the model parameters of \cref{eq:tensor_regression2}. Since $\gammakj\tod$ are the key building blocks for the mean of $\B$ representing the underlying hard \parafac{}, we borrow from \cite{Guhaniyogi2017bayesian} and specify
\begin{align*}
\gammakj \tod & \sim N(0, \tau_\gamma \zeta\tod \wkj \tod) \\
\tau_\gamma & \sim \Gamma(a_\tau, b_\tau) \\
\wkj \tod & \sim Exp((\lambda_k\tod)^2 / 2),\\
\lambda_k \tod & \sim \Gamma(a_\lambda, b_\lambda) \\
\bm \zeta & \sim \text{Dirichlet}(\alpha/D, \alpha/D, \dots, \alpha/D),
\end{align*}
where $\bm \zeta = (\zeta^{(1)}, \zeta^{(2)}, \dots, \zeta^{(D)})$. Therefore, the parameters $\gammakj\tod$ vary around 0 with variance that depends on an overall parameter $\tau_\gamma$, and component and row-specific parameters $\zeta\tod$ and $w_{k,j_k}\tod$. As discussed in \cite{Guhaniyogi2017bayesian}, the row-specific components $\wkj\tod$ lead to an adaptive Lasso type penalty on $\gammakj\tod$ \citep{Armagan2013}, and $\gammakj\tod | \tau_\gamma, \zeta\tod, \lambda_k\tod$ follows a double exponential (Laplace) distribution centered at 0 with scale $\tau_\gamma \zeta\tod / \lambda_k\tod$ \citep{Park2008}.

The component-specific variance parameter $\zeta\tod$ is included in the prior of $\gammakj\tod$ to encourage only a subset of the $D$ components to contribute substantially in the tensor's low-rank \parafac{} approximation. This is because parameters $\gammakj\tod$ for $d$ with small $\zeta\tod$ are shrunk closer to zero. For that reason, we also include $\zeta\tod$ in the conditional variance of $\betakj\tod$ in \cref{eq:soft_beta_dist} to ensure that penalization of the baseline row contributions $\gammakj\tod$ is accompanied with penalization of the row contributions $\betakj\tod$, and that a reduction in the variance of $\gammakj\tod$ is not overcompensated by an increase in the conditional variance of $\betakj\tod$.

We assume normal prior distributions on the intercept and scalar covariates' coefficients $(\mu, \bm \delta) \sim N(0, \Sigma_0)$, and inverse gamma priors on the residual variance $\tau^2 \sim IG(a_\tau, b_\tau)$ and the mode-specific variance components $\sigma^2_k \sim \Gamma(a_\sigma, b_\sigma)$.
Specific choices for the hyperparameter values are discussed in \cref{subsec:hyperparameters}.

\subsection{Choosing hyperparameters to achieve desirable characteristics of the induced prior}
\label{subsec:hyperparameters}

The prior distribution on the coefficient tensor $\B \sim \pi_{\B}$ is induced by our prior specification on the remaining parameters. The choice of hyperparameters can have a large effect on model performance, and the use of diffuse, non-informative priors can perform poorly in some situations \citep{Gelman2008}. For that reason and in order to assist default choices of hyperparameters leading to weakly informative prior distributions with interpretable flexibility, we study the properties of the induced prior on $\B$.

We do so in the following way. First, in \cref{theory:variance_B} we provide expressions for the induced prior expectation, variance and covariance for the entries in $\B$. These expressions illustrate the importance of certain hyperparameters in how the soft \parafac{} transitions away from its underlying, low-rank, hard version.
Then, in \cref{theory:prior_targets} we provide default values for hyperparameters for a standardized 2-mode tensor predictor such that, a priori, $\Var(\B_\subss) = V^*$, and the proportion of the variance that arises due to the proposed \parafac{} softening is equal to $AV^*$. Studying the proportion of prior variability due to the softening is motivated by \cref{fig:parafac_plus_sd} in that hyperparameters should be chosen such that most of the coefficient tensor's \textit{prior} variability arises from a low-rank structure. Therefore, \cref{theory:prior_targets} permits us to choose hyperparameter values that directly specify the amount of a priori \parafac{} softening. Finally, in \cref{subsec:theory_consistency} we study statistical properties of the induced prior on $\B$, and we show full support over the class of coefficient tensors, irrespective of the base rank used in the soft \parafac{}, leading to consistent posterior distributions.
All proofs are in \cref{app_sec:proofs}.


\begin{prop}
For $\subss, \subss' \in \otimes_{k = 1}^K \{1, 2, \dots, p_k\}$ such that $\subss \neq \subss'$,
we have that $\E(\B_\subss) = 0$, $\Cov(\B_\subss, \B_{\subss'}) = 0$, and for $a_\lambda > 2$,
$$\Var(\B_\subss) = \Big\{ D \prod_{r = 0}^{K - 1} \frac{\alpha/D + r}{\alpha + r} \Big\}
\Big[ \sum_{l = 0}^K \rho_l \binom{K}{l} \Big\{ \lambdafrac[T] \Big\}^l
\Big( \addvar \Big)^{K -l} \Big],
$$
where $\rho_0 = 1$ and $\rho_l = a_\tau (a_\tau + 1) \dots (a_\tau + l - 1)$ for $l \geq 1$.
\label{theory:variance_B}
\end{prop}

\begin{remark} \textit{The hyperparameters of the softening variance, $a_\sigma, b_\sigma$}. \label{remark:add_var}
Remember that $\sigma^2_k$ is the parameter driving the \parafac{} softening by allowing row-specific contributions to vary.
From \cref{theory:variance_B}, it is evident that the prior of $\sigma^2_k$ is only influential on the first two moments of $\B_\subss$ through its mean, $\addvar$, with higher prior expectation of $\sigma^2_k$ leading to higher prior variance of $\B_\subss$. Therefore, prior elicitation for $a_\sigma, b_\sigma$ could be decided based on the ratio $\addvar$.
\end{remark}


\begin{remark} \textit{Variance of coefficient entries for the hard \parafac{}.}
For $\E(\sigma^2_k) = 0$, the prior variance of the coefficient tensor entries is equal to the prior variance of the hard \parafac{},
$$
\Var^{hard}(\B_\subss) = \Big\{ D \prod_{r = 0}^{K - 1} \frac{\alpha/D + r}{\alpha + r} \Big\} \frac{\rho_K}{b_\tau^K} \Big\{ \lambdafrac \Big\}^K.
$$
\end{remark}

Comparing the variance of $\B_\subss$ based on the soft and hard \parafac{} allows us to quantify the amount of additional flexibility that is provided by the \parafac{} softening, expressed as
$$
AV = \frac{\Var(\B_\subss) - \Var^{hard}(\B_\subss)}{\Var(\B_\subss)} \in [0, 1).
$$
We refer to this quantity as the \textit{additional variance}. Motivated by \cref{fig:parafac_plus_sd}, we would like to ensure that chosen hyperparameters assign more prior weight to coefficient matrices that resemble low-rank factorizations. At the same time, choice of hyperparameters should ensure a sufficiently but not overly large prior variance of the regression coefficients.
\cref{theory:prior_targets} provides conditions on the hyperparameters for matrix predictors ($K = 2$), for which $\Var(\B_\subss) = V^*$, and $AV = AV^*$, for values $V^*>0$ and $AV^* \in [0, 1)$. For a tensor predictor with $K > 2$, conditions on the hyperparameters that ensure a given target variance and target additional variance can be acquired by following similar steps.

\begin{prop}
For a matrix predictor, target variance $V^* \in (0, \infty)$, target additional variance $AV^* \in [0,1)$, and hyperparameters satisfying $a_\lambda > 2$,
\begin{equation}
\lambdafrac = \frac{b_\tau}{a_\tau} \sqrt{\frac{V^*(1 - AV^*)a_\tau}{C (a_\tau + 1)}}
\label{eq:condition_lambdafrac}
\end{equation}
and
\begin{equation}
\addvar = \sqrt{\frac{V^*(1 - AV^*)a_\tau}{C(a_\tau + 1)}} \bigg\{ \sqrt{  1 - \frac{a_\tau + 1}{a_\tau} \big\{1 - \big(1 - AV^*)^{-1} \big\}} - 1 \bigg\}, 
\label{eq:condition_addvar}    
\end{equation}
where $C = (\alpha / D + 1)/(\alpha + 1)$, we have that a priori $\Var(\B_\subss) = V^*$, and $AV = AV^*$.
\label{theory:prior_targets}
\end{prop}

\cref{theory:prior_targets} is used in our simulations and study of brain connectomics to choose hyperparameters such that, a priori, $\Var(\B_\subss) = 1$ and $AV = 10\%$, assuming a tensor predictor with standardized entries.
Specifically, we set $a_\tau = 3$, $a_\sigma = 0.5$ and calculate the values of $b_\tau, b_\sigma$ for which $V^* = 1$ and $AV^* = 10\%$. These values correspond to $b_\tau \approx 6.3 \sqrt{C} $ and $b_\sigma \approx 8.5 \sqrt{C} $.
We specify $\alpha_\sigma = 0.5 < 1$ to encourage, a priori, smaller values of $\sigma^2_k$. Throughout, we use $\alpha = 1$ and $D = 3$.
For the hyperparameters controlling the underlying hard \parafac{}, we specify $a_\lambda = 3$ and $b_\lambda = \sqrt[2K]{a_\lambda}$. Lastly, assuming centered and scaled outcome and scalar covariates, we specify $(\mu, \bm \delta^T)^T \sim N(0, \mathbb{I}_{p +1})$, and residual variance $\tau^2 \sim IG(2, 0.35)$ which specifies $P(\tau^2 < 1) \approx 0.99$.

\begin{remark}
\textit{Interplay between variance hyperparameters}.
From \cref{eq:condition_addvar}, it is evident that the prior mean of $\sigma^2_k$, the variance component in the \parafac{} softening, depends on the target variance and the proportion of that variance that is attributable to the \parafac{} softening, and does {\it not} depend on the remaining hyperparameters (considering that $a_\tau / (a_\tau + 1) \approx 1$).
Furthermore, note that \cref{eq:condition_lambdafrac} only includes hyperparameters which drive the underlying hard \parafac{}, and it depends on $V^*$ and $AV^*$ only through $V^*(1 - AV^*)$, which expresses the prior variability in $\B$ that is {\it not} attributed to the \parafac{} softening.
Therefore, in \softer{} there is a clear and desirable separation between the hard and soft \parafac{} variance hyperparameters.
Lastly, since $\lambdafrac$ in \cref{eq:condition_lambdafrac} is the prior mean of $\wkj\tod$, these results illustrate the interplay between the two components in the variance $V^*(1 - AV^*)$ of the underlying hard \parafac{} structure: when the prior mean of $\tau_\gamma$ increases, the prior mean of $\wkj$ has to decrease in order to maintain the target variance due to the underlying \parafac{} at $V^*(1 - AV^*)$.
\end{remark}

\subsection{Posterior consistency and \softer's robustness to the underlying rank}
\label{subsec:theory_consistency}

In this section, we focus on the choice of the rank $D$ for \softer{}. We discuss that results based on \softer{} are likely robust to small changes in the choice of the underlying rank. To support this claim, we first provide two intuition-based arguments, and then a theoretical one. Finally, \softer{}'s robustness to the choice of $D$ is empirically illustrated in simulated examples in \cref{sec:simulations}.

When employing the hard \parafac{}, \cite{Guhaniyogi2017bayesian} recommended using $D = 10$ for a predictor of dimension $64 \times 64$. The reason is that the prior on $\bm \zeta$ permits some amount of flexibility in the number of components that contribute to the coefficient matrix approximation in that 
\begin{enumerate*}[label = (\alph*)]
\item if the matrix can be well-approximated by a rank lower than $D$, the prior leads to a reduction in the approximation's effective rank, and
\item if all $D$ components are useful in estimation, then all of them acquire sufficient weight.
\end{enumerate*}
Since \softer{} employs the same prior on $\bm \zeta$, it also allows for sparsity in the effective components in the underlying hard \parafac{}, in that if the true coefficient tensor is of rank lower than $D$, higher order components will be heavily shrunk towards zero, and softening will be minimal. This intuition will be further supported by simulations in \textsection\ref{subsec:sims_400} where we illustrate that \softer{} reverts back to the hard \parafac{} when the underlying low-rank structure is true.

\begin{figure}[!b]
\centering
\includegraphics[width = 0.45\textwidth]{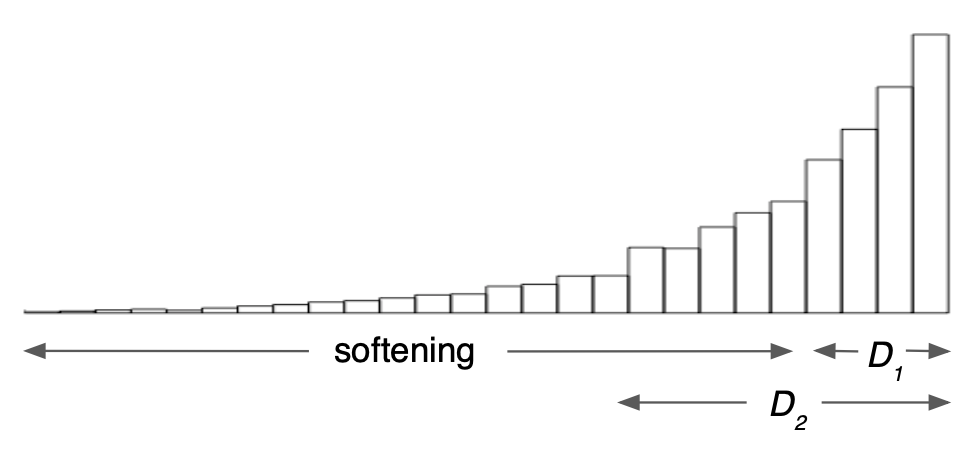}
\caption{Hypothetical histogram of singular values of a coefficient matrix. A rank $D_1$ \parafac{} approximation would incorporate components corresponding to the $D_1$ largest singular values. Increasing the rank to $D_2$ allows for $D_2-D_1$ additional components, whereas \softer{} allows for the incorporation of deviations corresponding to any singular value.}
\label{fig:rank_choice}
\end{figure}

However, if the true coefficient tensor is not of low rank, \softer{} can expand away from a low-rank structure in two ways: it can use additional components in the underlying hard \parafac{}, or it can soften away from the rigid underlying low-rank structure. The deviations based on the \parafac{} softening can effectively capture components corresponding to singular values of any magnitude. In \cref{fig:rank_choice} we illustrate the range of singular values that would be accommodated when expanding away from a rank-$D_1$ hard \parafac{} approximation by (1) increasing the hard \parafac{} rank, and (2) softening the \parafac{}. Increasing the hard \parafac{} rank would include components corresponding to some small singular values, but softening the \parafac{} accommodates deviations from the underlying $D_1$-rank structure across all singular values. Since \softer{} based on an underlying $D$-rank can capture these arbitrary deviations from a $D$-rank structure, increasing the rank $D$ in \softer{}'s underlying \parafac{} is not expected to drastically alter the results, making \softer{} {\it robust} to the choice of rank $D$. This intuition is empirically illustrated in \textsection\ref{subsec:sims_rank}.

\softer{}'s robustness to the choice of underlying rank $D$ and its ability to capture {\it any} true coefficient tensor is also evident in the following theoretical results. In \cref{prop:prior_support} we show that the prior on $\B$, $\pi_{\B}$, has full $L_\infty$ prior support in that it assigns positive prior weight to any $\epsilon$-sized neighborhood of any true coefficient tensor $\B^0$. This indicates that, even if the value of $D$ is lower than the coefficient tensor's true rank, \softer{} assigns positive prior weight to a neighborhood of the true tensor. Full prior support is a key element for establishing sufficient flexibility of a Bayesian procedure. In turn, the posterior distribution of the coefficient tensor is consistent, irrespective of the true coefficient tensor's rank, or the rank used in \softer{}'s underlying \parafac{}.

\begin{prop}[Full prior support]
\label{prop:prior_support}
Let $\epsilon > 0$. Then, $\pi_{\B}\big(\neigh\big) > 0$, where
$\neigh = \{ \B: \max_{\subss} | \B_\subss^0 - \B_\subss| < \epsilon \}.$
\end{prop}

\noindent
We assume that the true data generating model is \cref{eq:tensor_regression2} with true coefficient tensor $\B^0$, and that the tensor predictor $\tensor$ has bounded entries. Since our interest is in estimating $\B^0$, we assume that $\tau^2 = 1$, $\mu = 0$ and $\bm \delta = \bm 0$ are known. Then, we have the following.
\begin{prop}
\label{prop:consistency}
The posterior distribution of $\B$ is weakly consistent for $\B^0$.
\end{prop}

\noindent
These results indicate that softening the \parafac{} allows us to capture any truth. Since they hold irrespective of the choice of the underlying rank, they also indicate that \softer{} is robust to the choice of rank $D$, at least asymptotically.


\subsection{Approximating the posterior distribution of the coefficient tensor}
\label{subsec:MCMC}

Since there is no closed-form for the posterior distribution of $\B$, we approximate it using Markov chain Monte Carlo (MCMC). An MCMC scheme where most parameters are updated using Gibbs sampling is shown in \cref{app_sec:MCMC}. We found this approach to be sufficiently efficient when the sample size is larger or of similar order to the number of parameters. However, in very high-dimensional settings where $p \gg n$, mixing and convergence was slow under reasonable time constraints. For that reason, and in order to provide a sampling approach that performs well across $n,p$ situations, we instead rely on Hamiltonian Monte Carlo (HMC) implemented in Stan \citep{Carpenter2017stan} and on the R interface \citep{Rstan} to acquire samples from the posterior distribution. HMC is designed to improve mixing relative to Gibbs sampling by employing simultaneous updates, and relying on gradients calculated with automatic differentiation to obtain efficient proposals. Using Stan, we employ the No-U-Turn sampler (NUTS) algorithm \citep{Hoffman2014NoUTurn} which automatically tunes the HMC parameters to achieve a target acceptance rate (the default step size adaptation parameters were used). If MCMC convergence is slow, one could increase the NUTS parameter $\delta$ in RStan from 0.8, which is the default, to a value closer to 1.

MCMC convergence was assessed based on visual inspection of traceplots across chains with different starting values and the potential scale reduction factor \citep{Gelman1992} for the regression coefficients $\mu, \bm \delta, \B$ and the residual variance $\tau^2$.
Note that the remaining parameters are not individually identifiable as the underlying \parafac{} parameters are themselves non-identifiable, and therefore the corresponding softening parameters are also non-identifiable.
In simulations, we considered a model that restricts the underlying \parafac{} parameters to make them identifiable (using the constraints discussed in Section 2.3 of \cite{Guhaniyogi2017bayesian}). We found that the original and constrained models had identical estimation performance, but that the MCMC for the unconstrained model including non-identifiable parameters required a smaller number of iterations to converge than the constrained model that uses the identifiable underlying \parafac{}.


\section{Simulations}
\label{sec:simulations}

To illustrate the performance of \softer{} and compare it against alternatives, we simulated data under various scenarios. In one set of simulations (\cref{subsec:sims_400}), we considered a tensor predictor of dimension $32 \times 32$, corresponding coefficient tensors ranging from close to low-rank to full-rank, and with different degrees of sparsity, and sample size equal to 400.
In another set of simulations (\cref{subsec:sims_rank}), we considered a tensor predictor of dimension $20 \times 20$ and corresponding coefficient tensor of rank $3, 5, 7, 10$ and $20$ in order to investigate the performance of \softer{} relative to the hard \parafac{} for a true coefficient tensor that increasingly deviates from low rank form, and various choices of the algorithmic rank $D$. The sample size in this situation was 200. In all scenarios, the predictor's entries were drawn independently from a $N(0, 1)$ distribution, and the outcome was generated from a model in the form \cref{eq:tensor_regression2} with true residual variance $\tau ^ 2 = 0.5$. 

We considered the following methods:
\begin{enumerate*}[label=(\alph*)]
\item \softer{},
\item the Bayesian hard \parafac{} approach of \cite{Guhaniyogi2017bayesian}, and 
\item estimating the coefficient tensor by vectorizing the predictor and performing Lasso. We considered these two competing approaches because they represent the two extremes of how much prioritization is given to the predictor's array structure (the hard \parafac{} directly depends on it, the Lasso completely ignores it), whereas \softer{} is designed to exploit the predictor's structure while allowing deviations from it. We also considered
\item the Bayesian tensor regression approach of \cite{Spencer2020infer} which is based on the Tucker decomposition, a generalization of the \parafac{} decomposition.
\end{enumerate*}

Methods were evaluated in terms of how well they estimated the true coefficient tensor $\B$ by calculating
\begin{enumerate*}
\item[(1)] the entry-specific bias and mean squared error of the posterior mean (for the Bayesian approaches) and the penalized likelihood estimate (for the Lasso), and
\item[(2)] the frequentist coverage of the 95\% credible intervals.
\end{enumerate*}
In order to evaluate the methods' performance in accurately identifying important entries (entries with non-zero coefficients), we calculated methods'
\begin{enumerate*}
\item[(3a)] sensitivity (percentage of important entries that were identified),
\item[(3b)] specificity (percentage of non-important entries that were correctly deemed non-important),
\item[(3c)] false positive rate (percentage of identified entries that are truly not important), and
\item[(3d)] false negative rates (percentage of non-identified entries that are important).
\end{enumerate*}
For the Bayesian methods, an entry was flagged as important if its corresponding 95\% credible interval did not overlap with zero. Hierarchical Bayesian models have been shown to automatically perform adjustment for multiple testing error \citep{Scott2010bayes, muller2006fdr}. Confidence intervals and entry selection for the Lasso were not considered.
We also evaluated the models' predictive performance by estimating
\begin{enumerate*}
\item[(4)] the predictive mean squared error defined as the mean of the squared difference between the true outcome and the predictive mean over 1,000 new data points.
\end{enumerate*}
Lastly, 
\begin{enumerate*}
\item[(5)] we compared the hard \parafac{} and \softer{} in terms of their computational time, for various ranks $D$.
\end{enumerate*}

Additional simulations are shown in \cref{app_sec:more_sims} and are summarized below, where applicable.

\subsection{Simulation results for tensor predictor of dimensions 32$\times$32}
\label{subsec:sims_400}

\begin{figure}[!t]
\begin{center}
\begin{minipage}{0.21\textwidth}
\hspace{2pt} \large \it Truth \phantom{f} \\
\includegraphics[height=0.135\textheight]{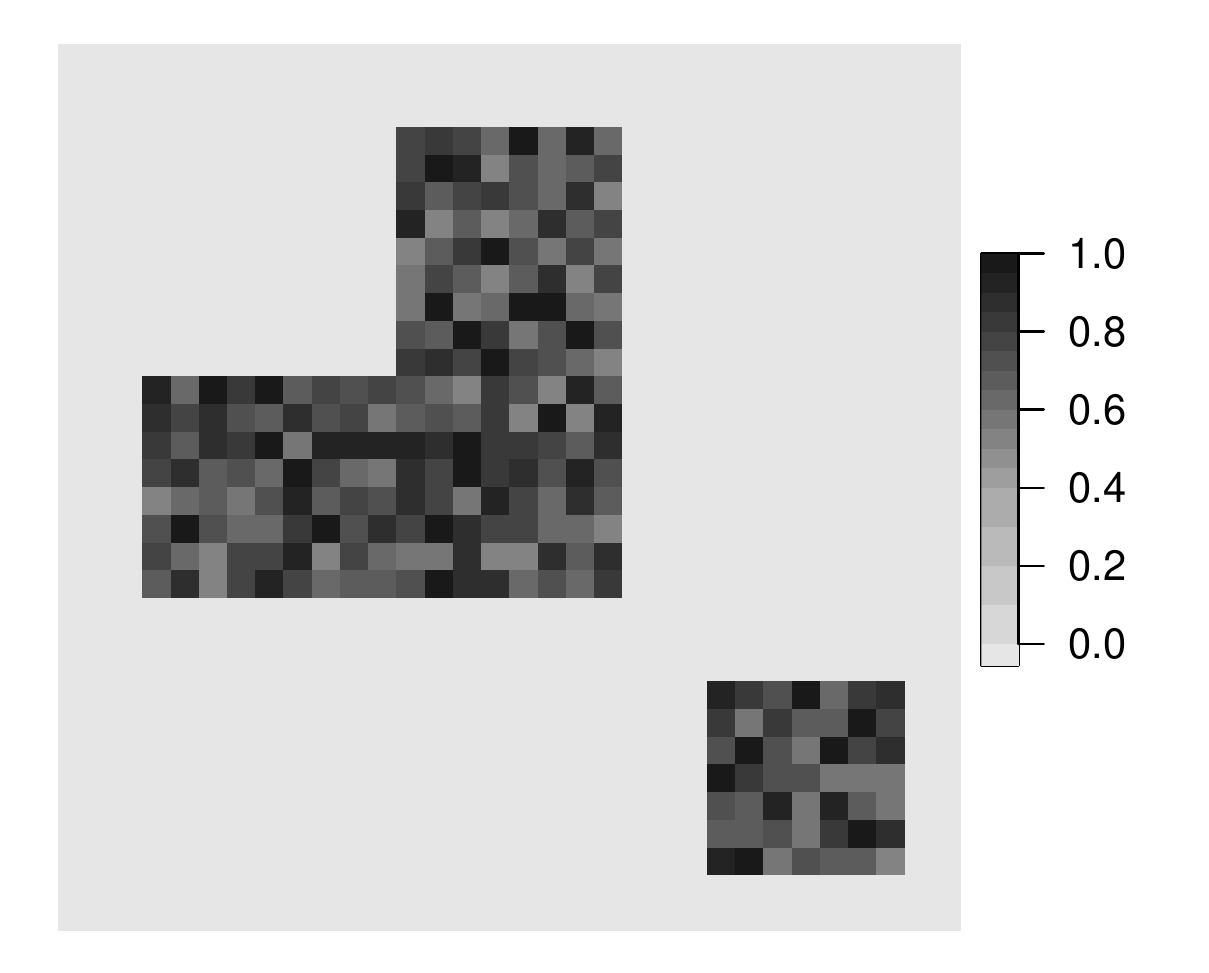} \\
\includegraphics[height=0.135\textheight]{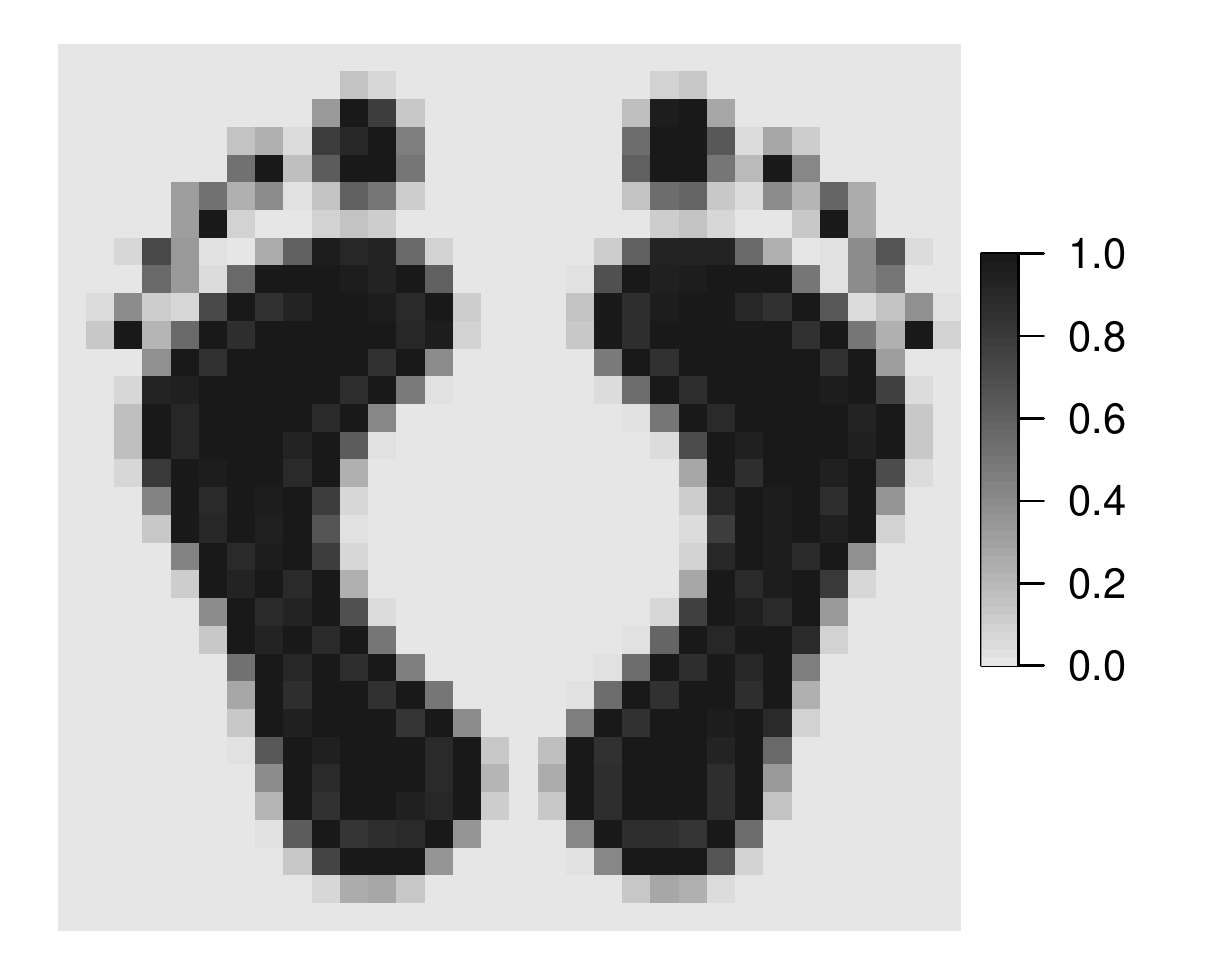} \\
\includegraphics[height=0.135\textheight]{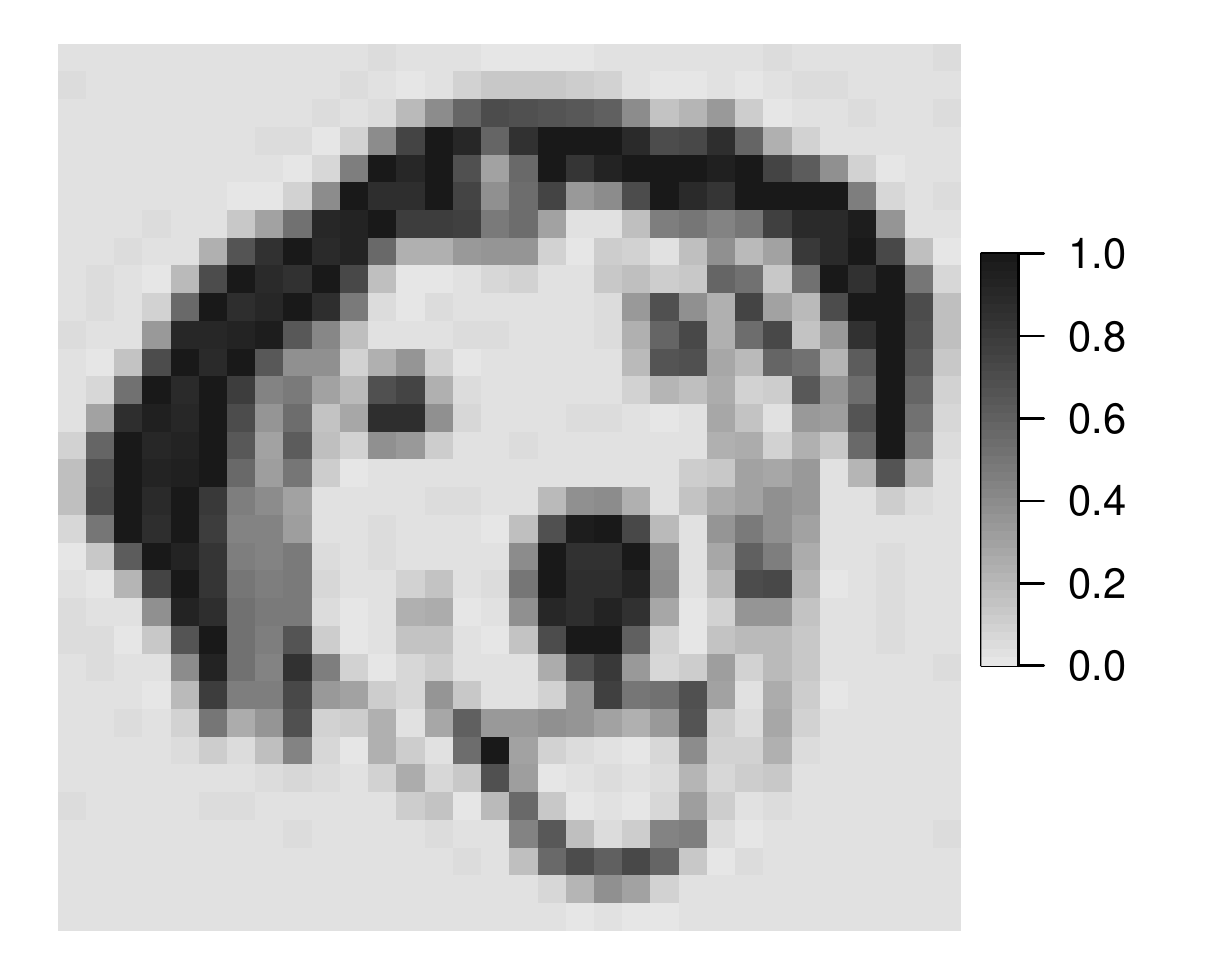} \\
\includegraphics[height=0.135\textheight]{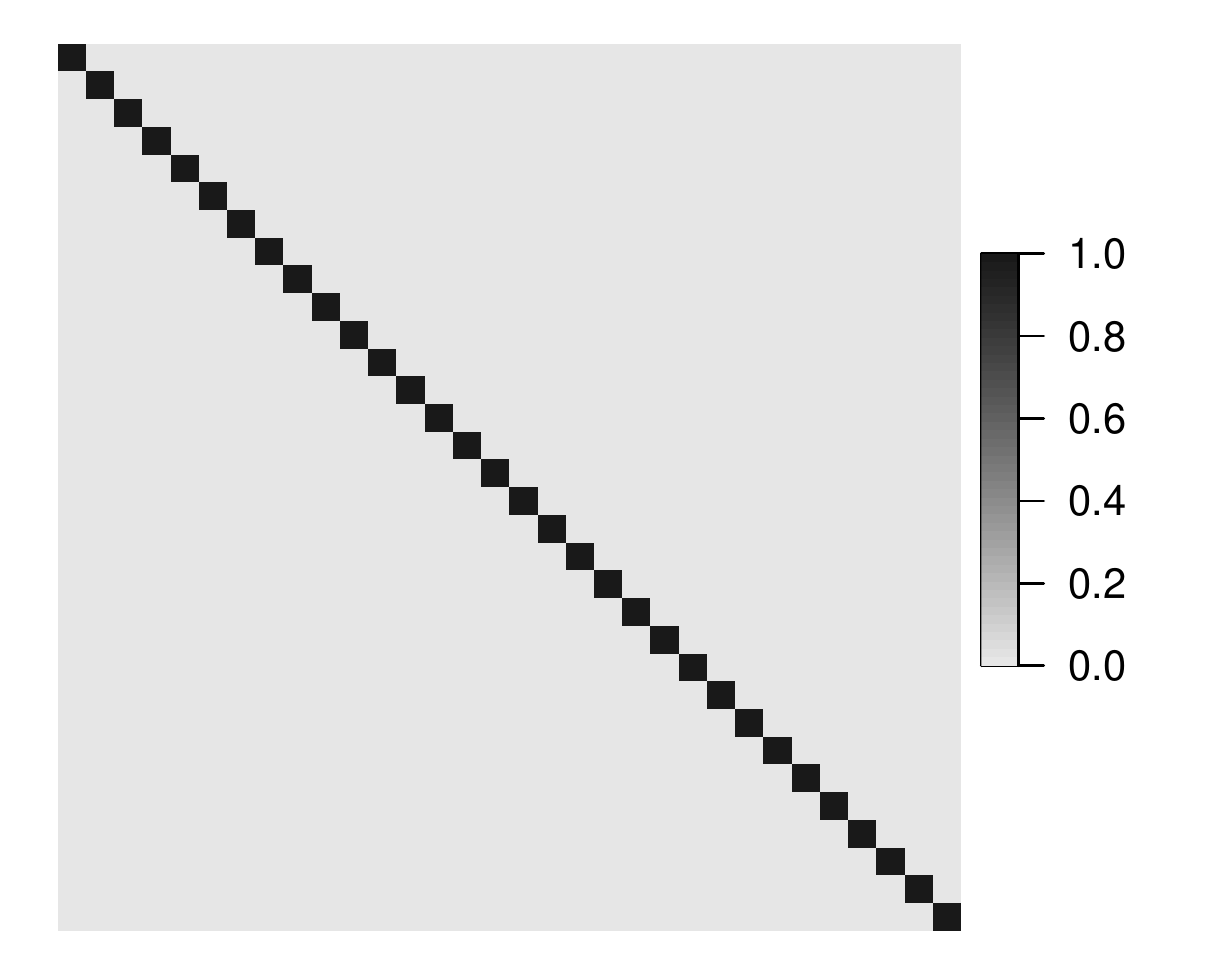}
\end{minipage}
\begin{minipage}{0.17\textwidth}
\hspace{2pt} \large \it \softer{} \\
\includegraphics[height=0.135\textheight,trim=0 0 70 0, clip]{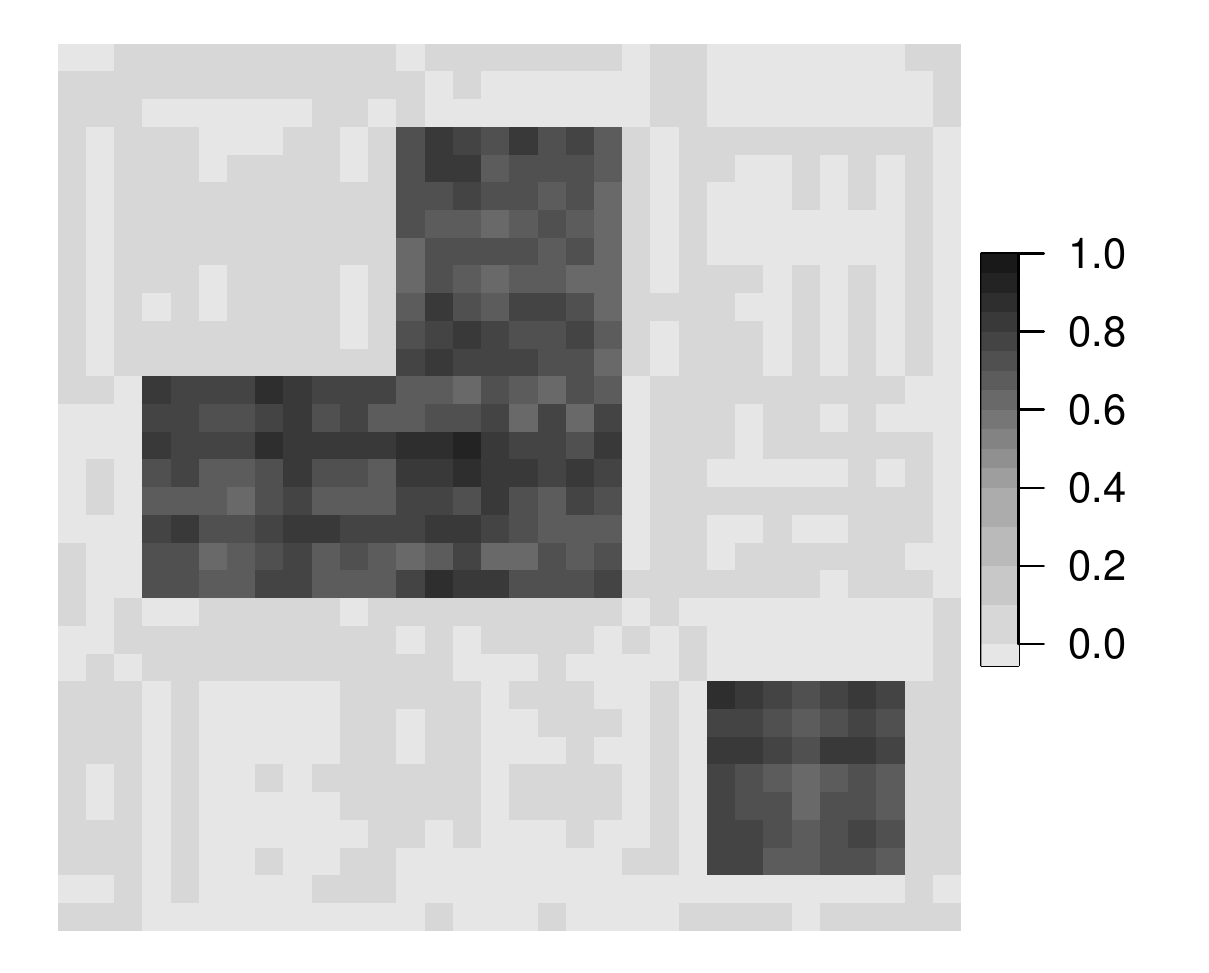} \\
\includegraphics[height=0.135\textheight,trim=0 0 70 0, clip]{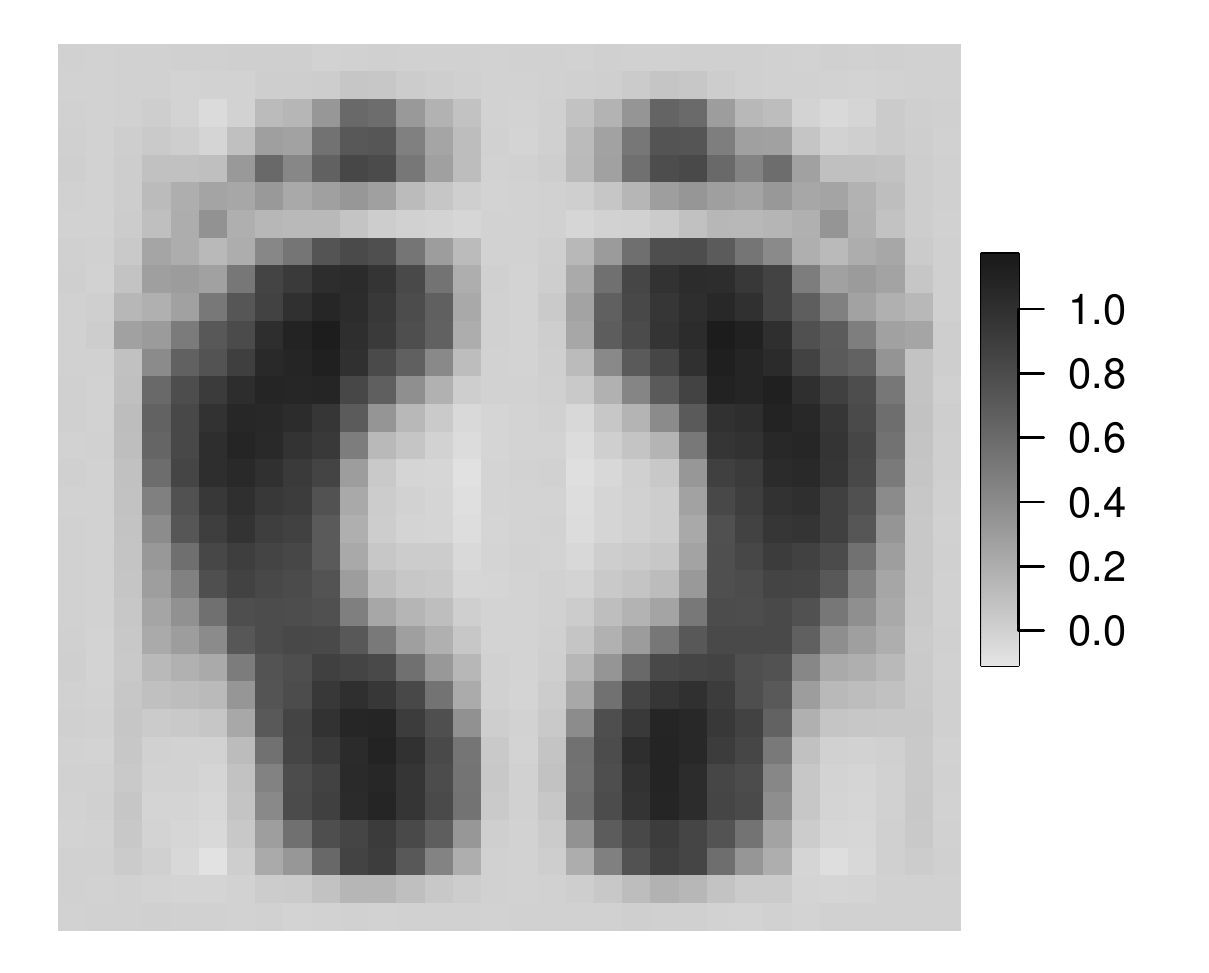} \\
\includegraphics[height=0.135\textheight,trim=0 0 70 0, clip]{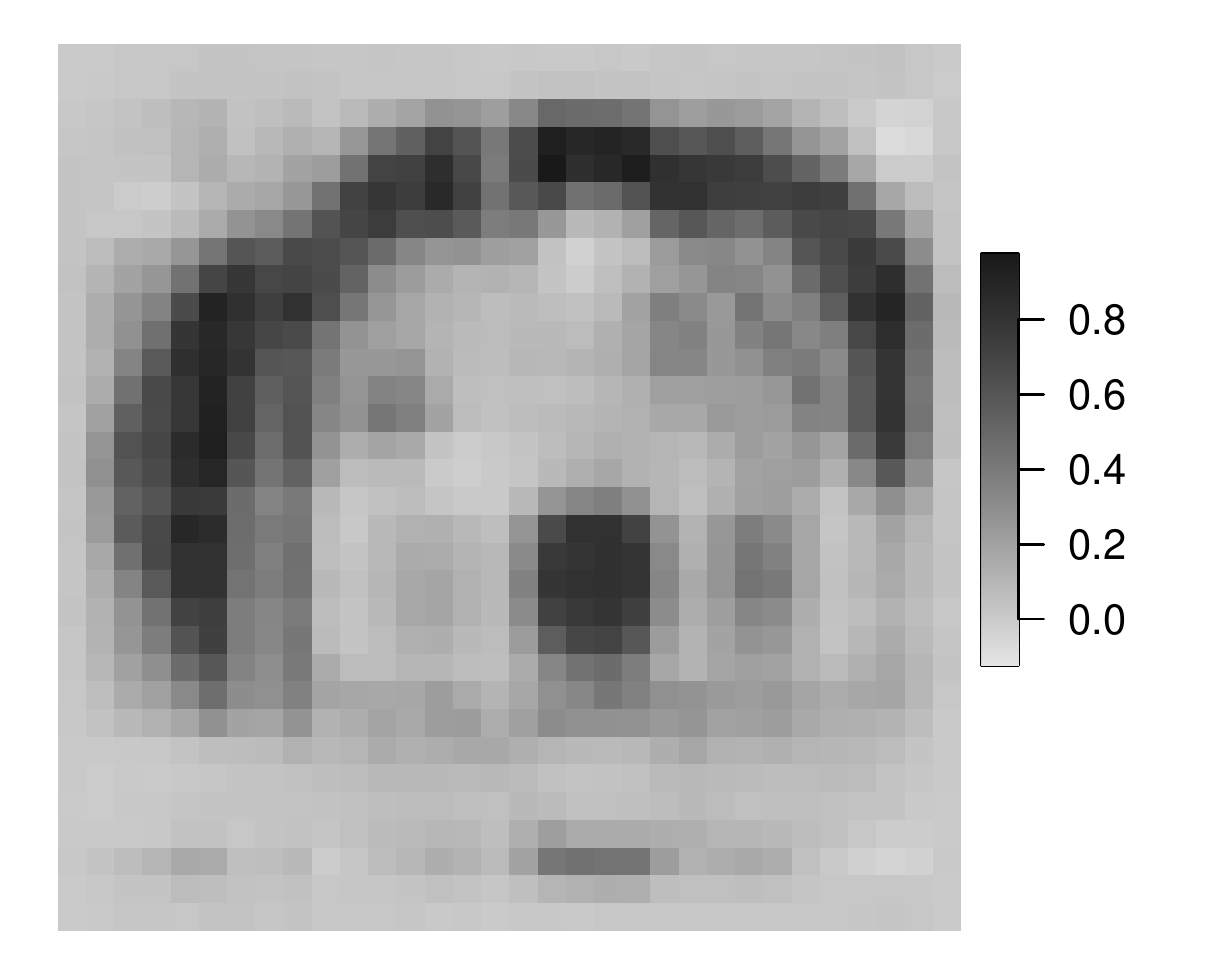} \\
\includegraphics[height=0.135\textheight,trim=0 0 70 0, clip]{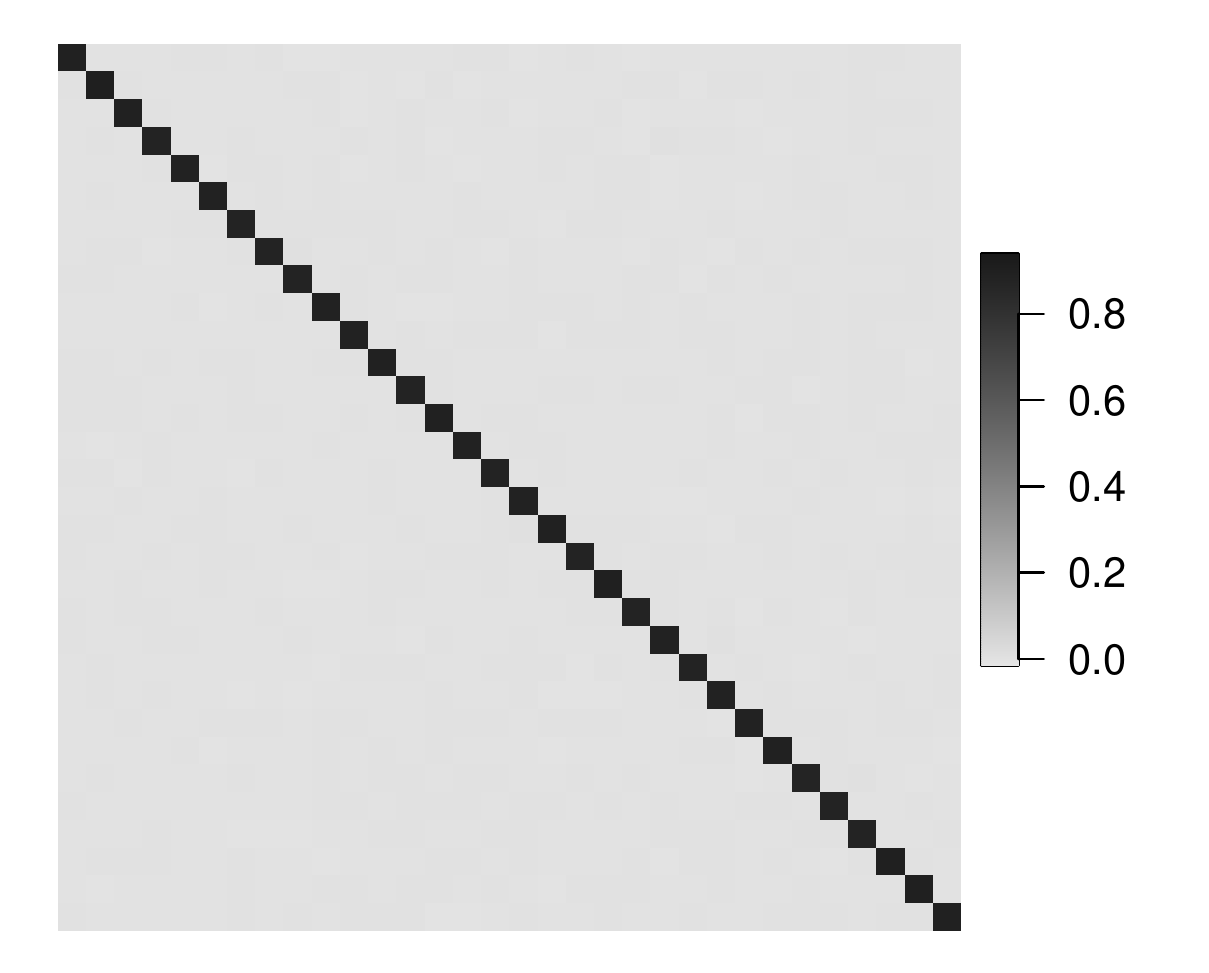}
\end{minipage}
\begin{minipage}{0.17\textwidth}
\hspace{2pt} \large \it \parafac{} \phantom{f} \\
\includegraphics[height=0.135\textheight,trim=0 0 70 0, clip]{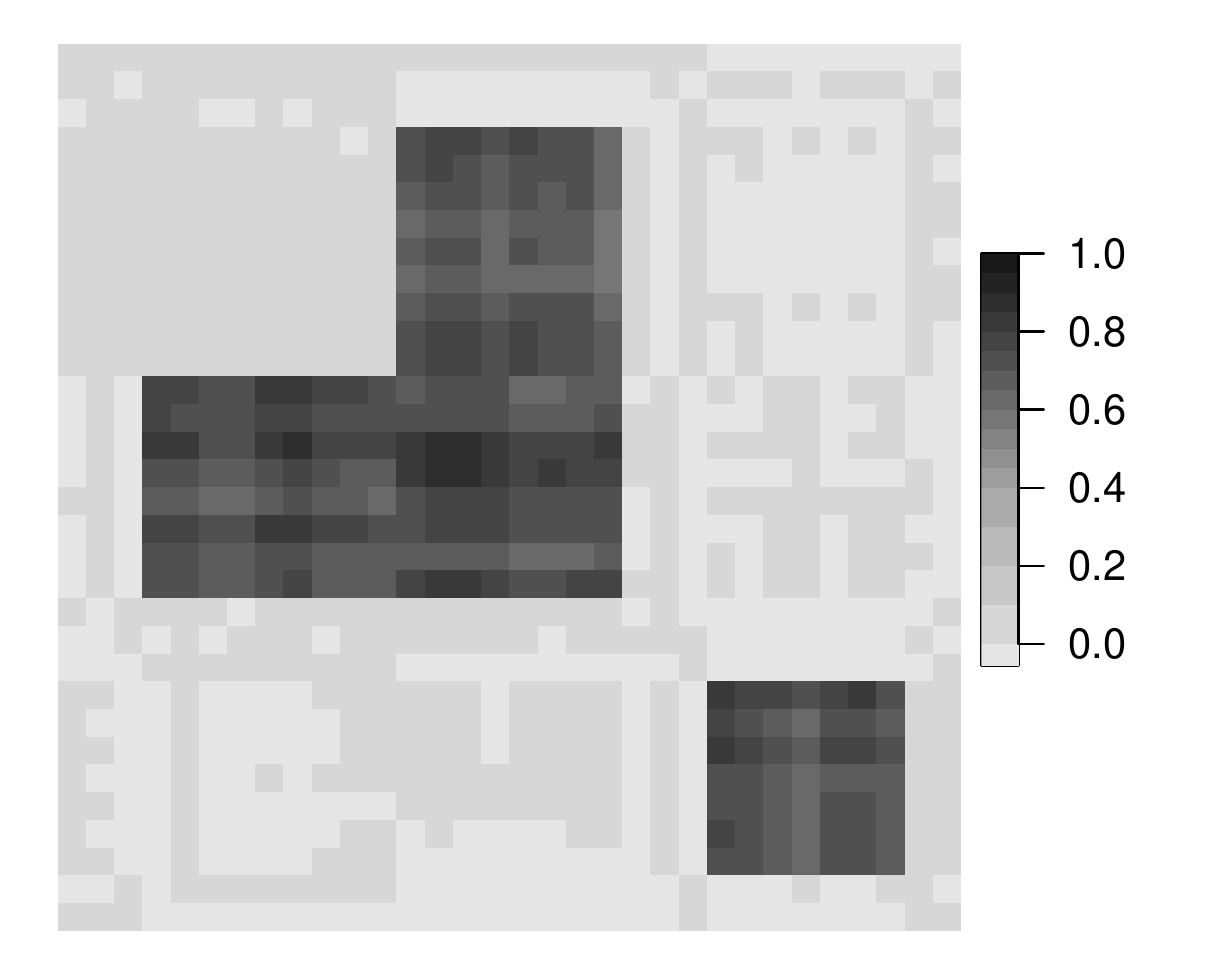} \\
\includegraphics[height=0.135\textheight,trim=0 0 70 0, clip]{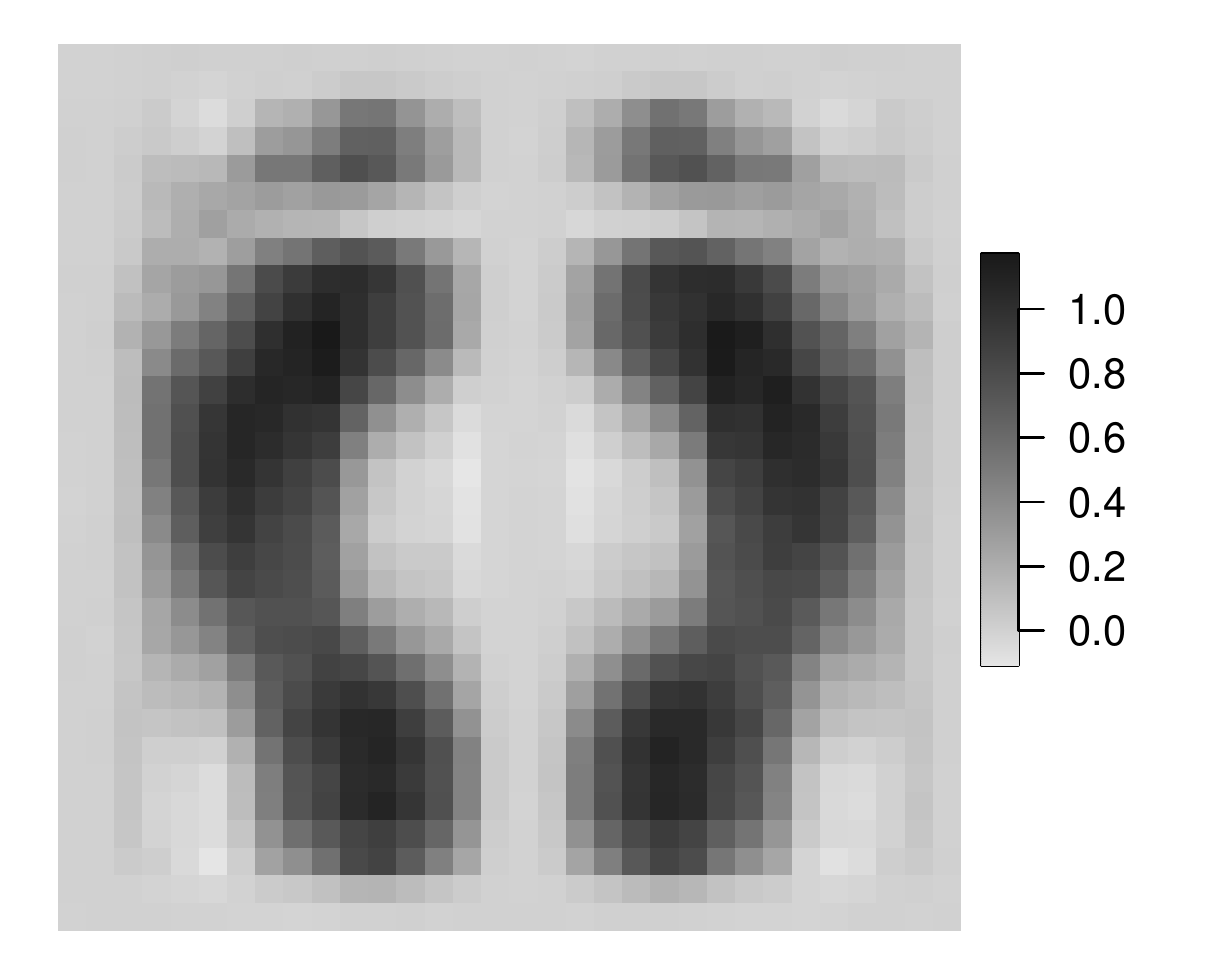} \\
\includegraphics[height=0.135\textheight,trim=0 0 70 0, clip]{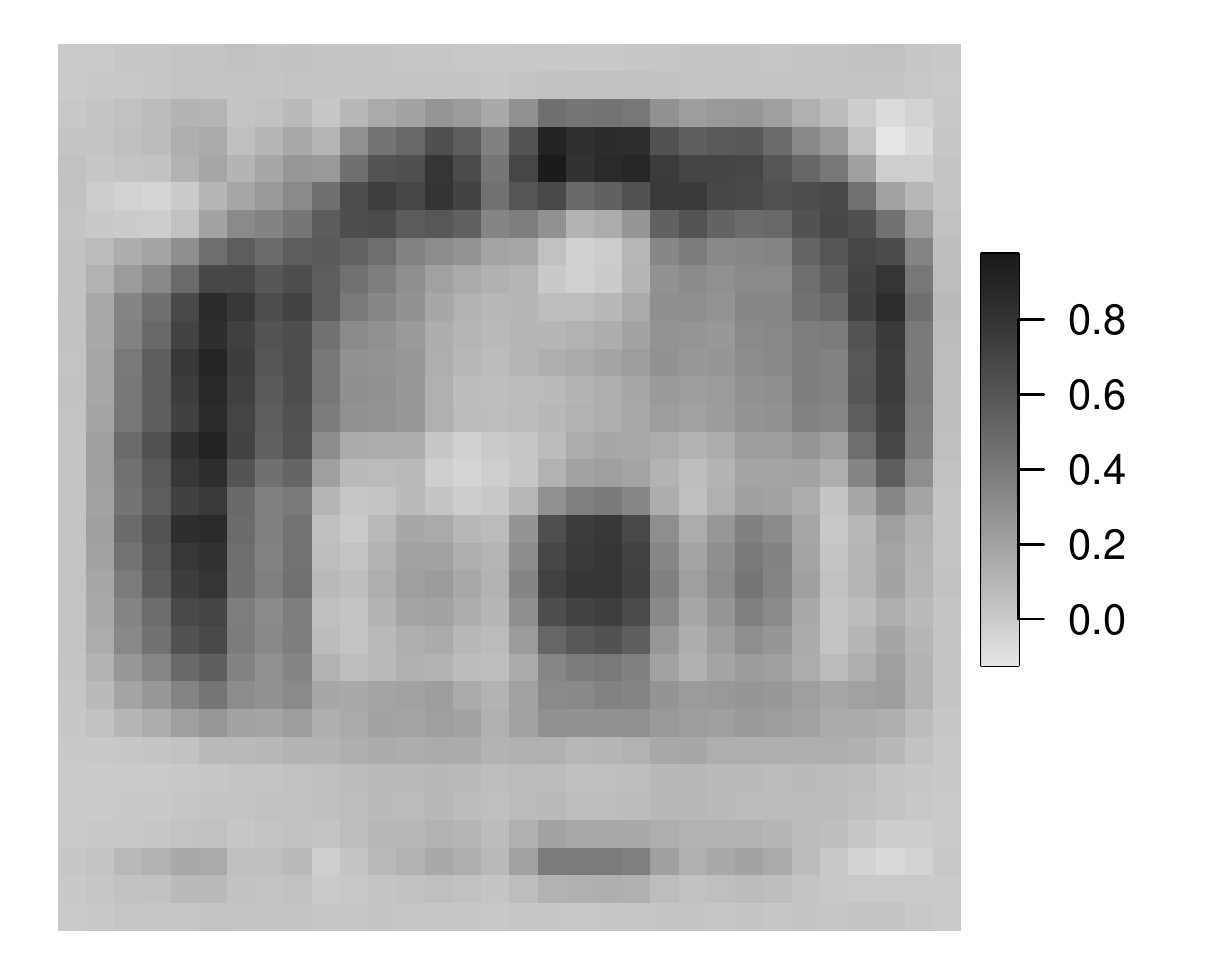} \\
\includegraphics[height=0.135\textheight,trim=0 0 70 0, clip]{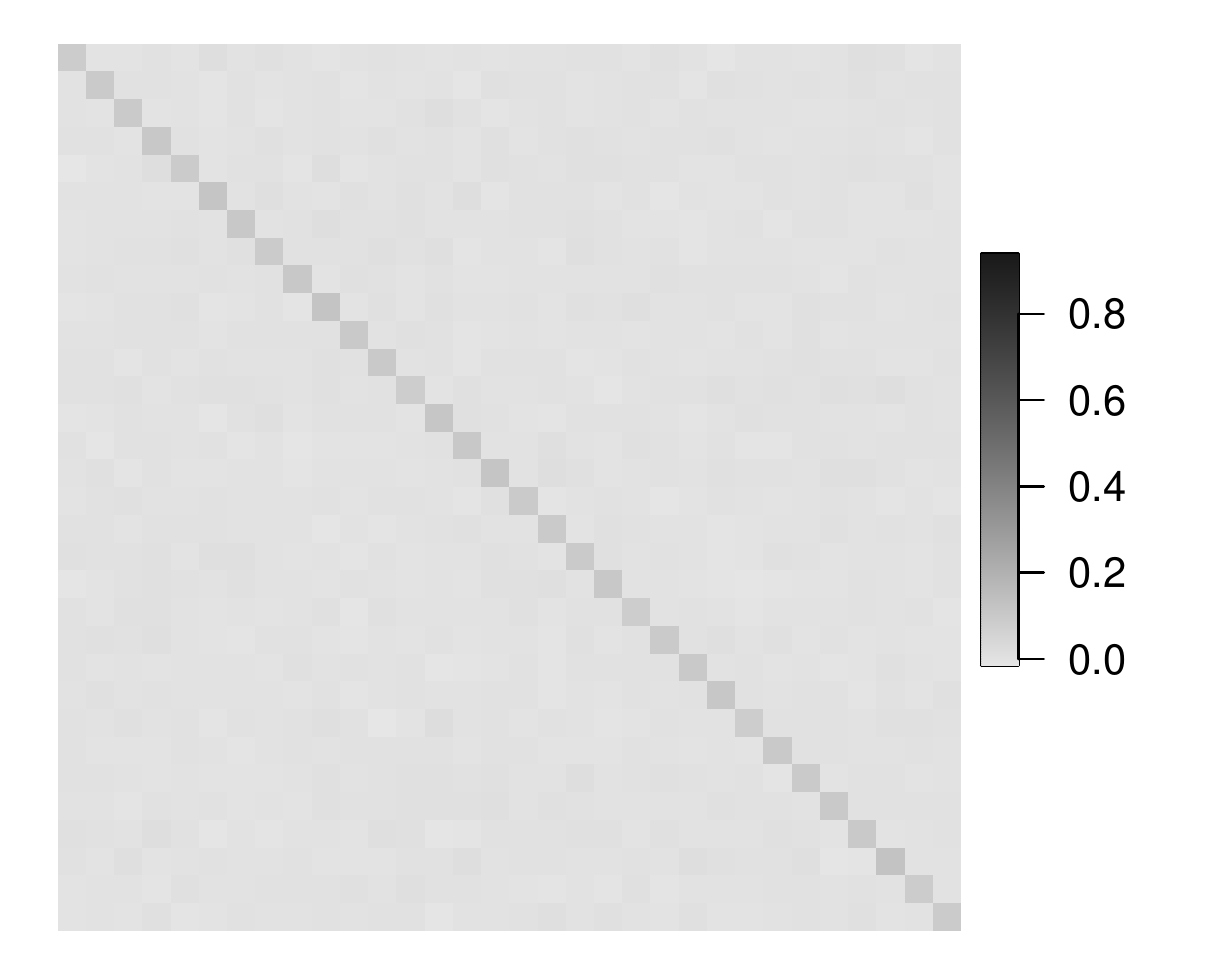}
\end{minipage}
%
\begin{minipage}{0.17\textwidth}
\hspace{2pt} \large \it Tucker \phantom{f} \\
\includegraphics[height=0.135\textheight,trim=0 0 70 0, clip]{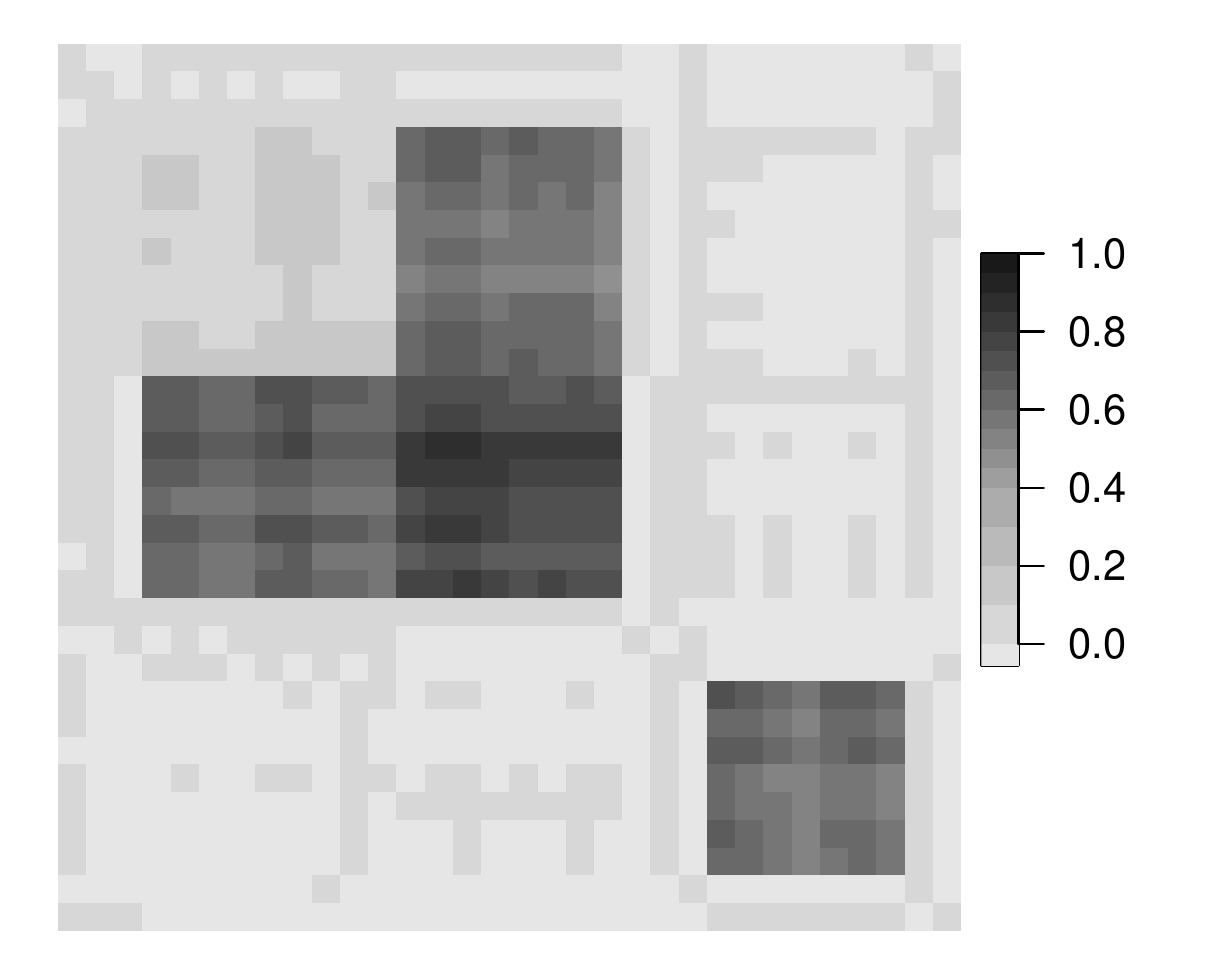} \\
\includegraphics[height=0.135\textheight,trim=0 0 70 0, clip]{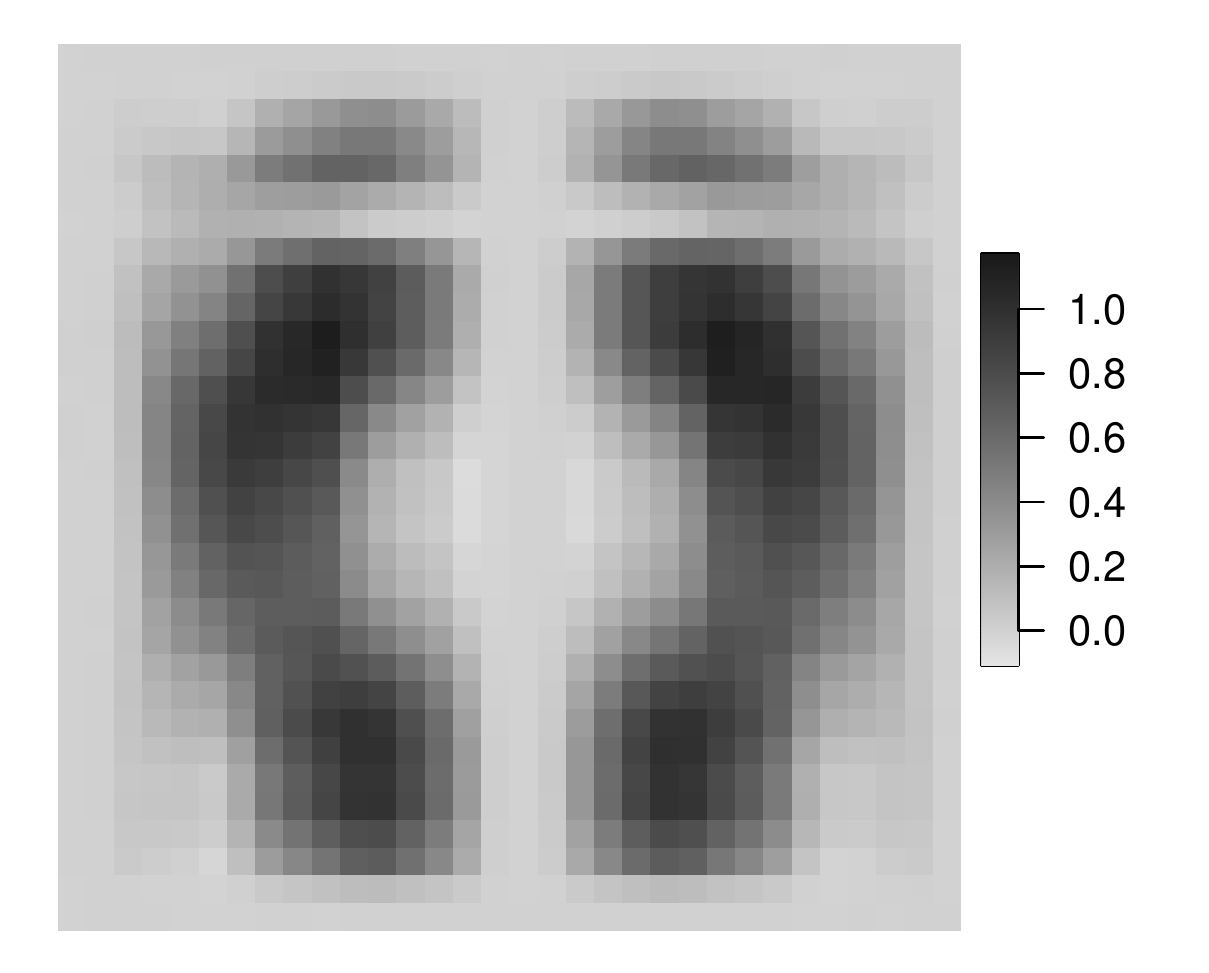} \\
\includegraphics[height=0.135\textheight,trim=0 0 70 0, clip]{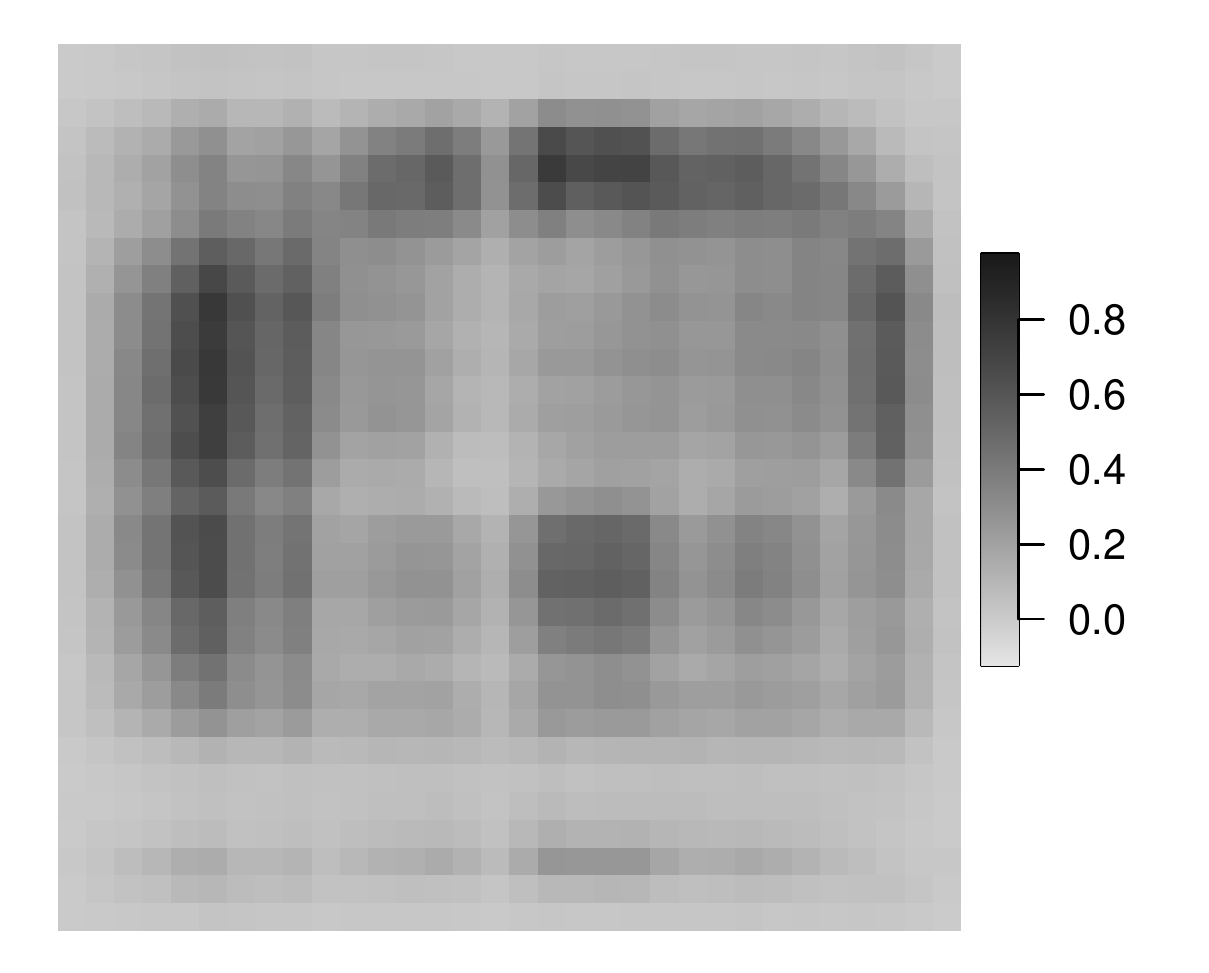} \\
\includegraphics[height=0.135\textheight,trim=0 0 70 0, clip]{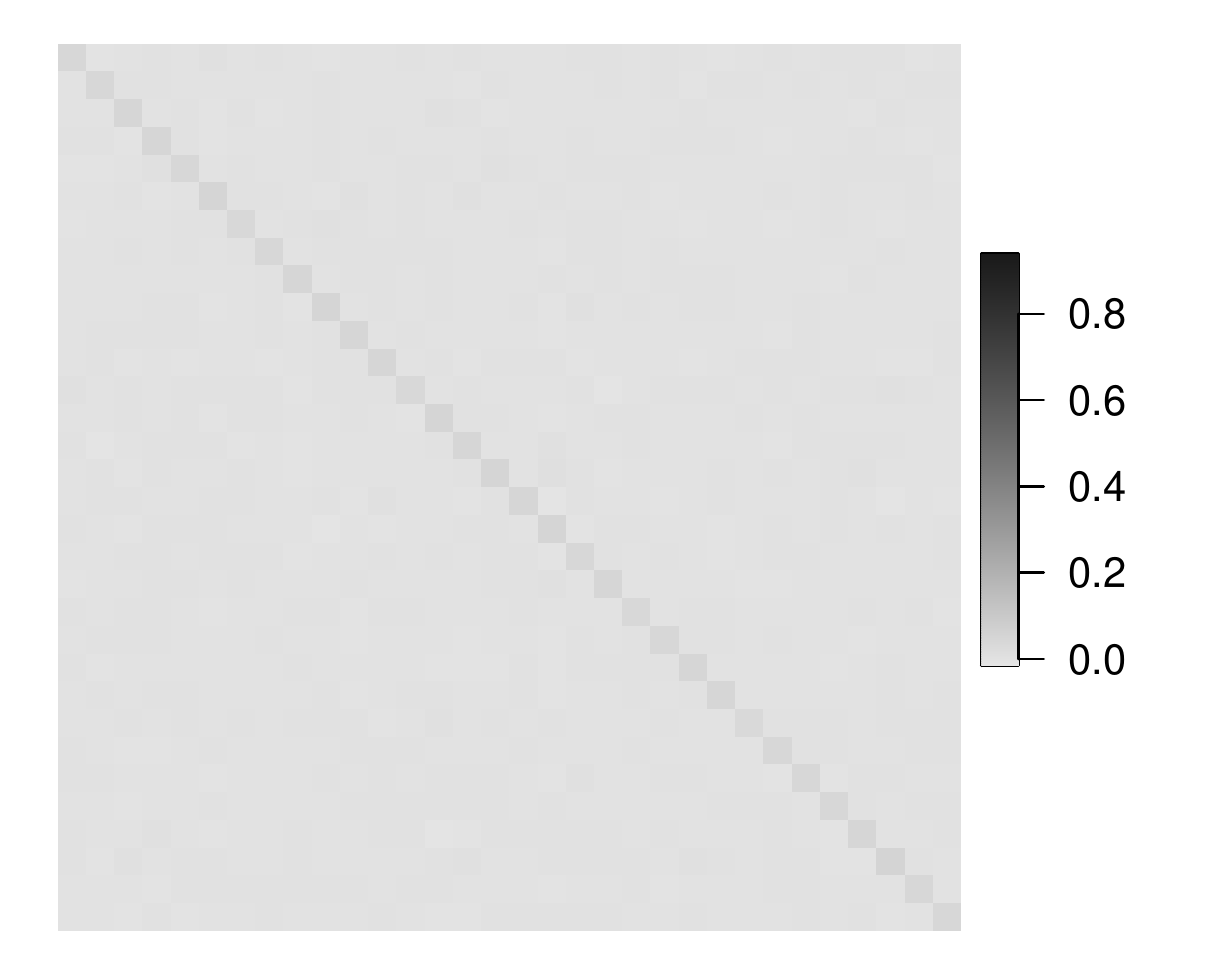}
\end{minipage}
\begin{minipage}{0.21\textwidth}
\hspace{2pt} \large \it Lasso \phantom{f} \\
\includegraphics[height=0.135\textheight]{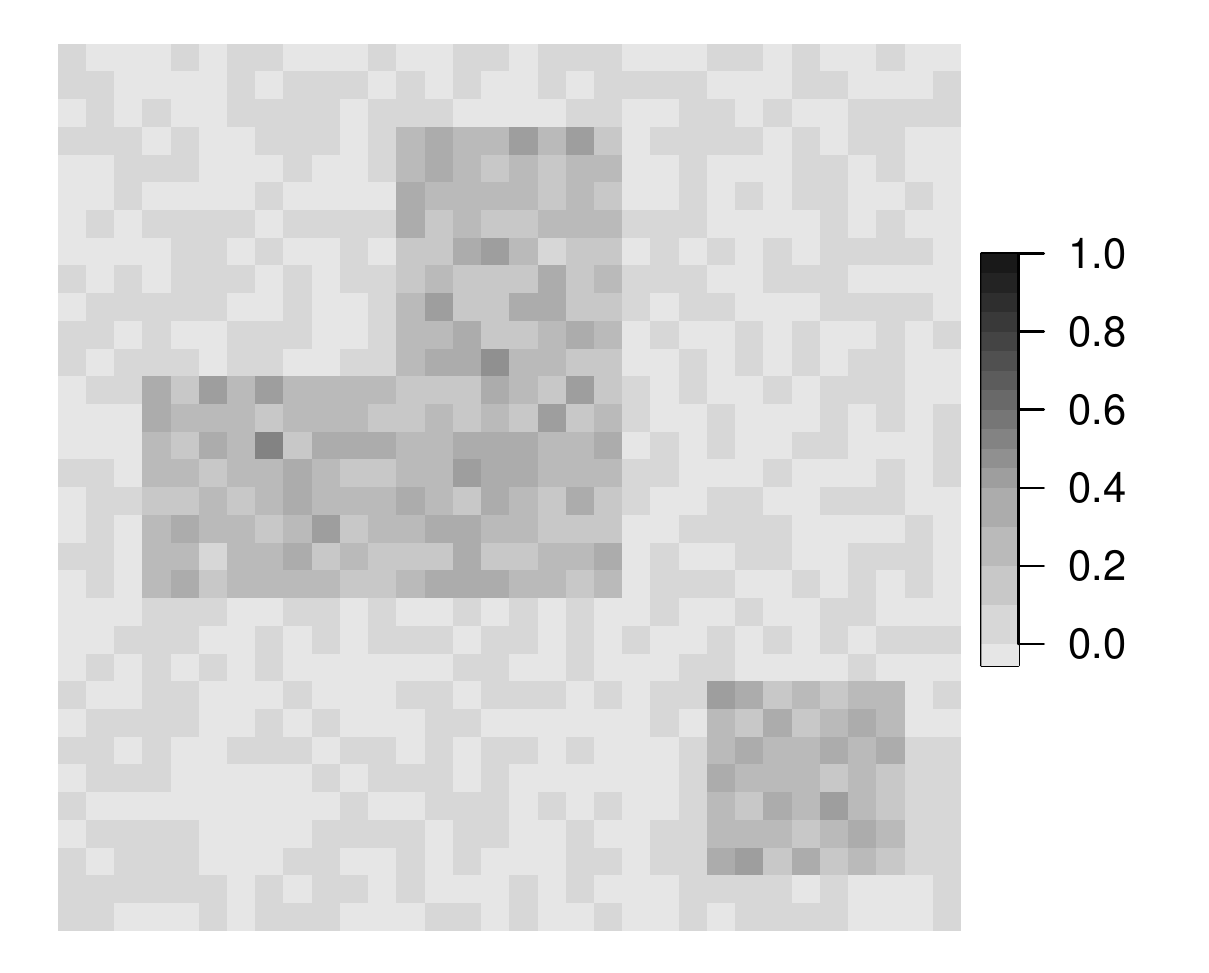} \\
\includegraphics[height=0.135\textheight]{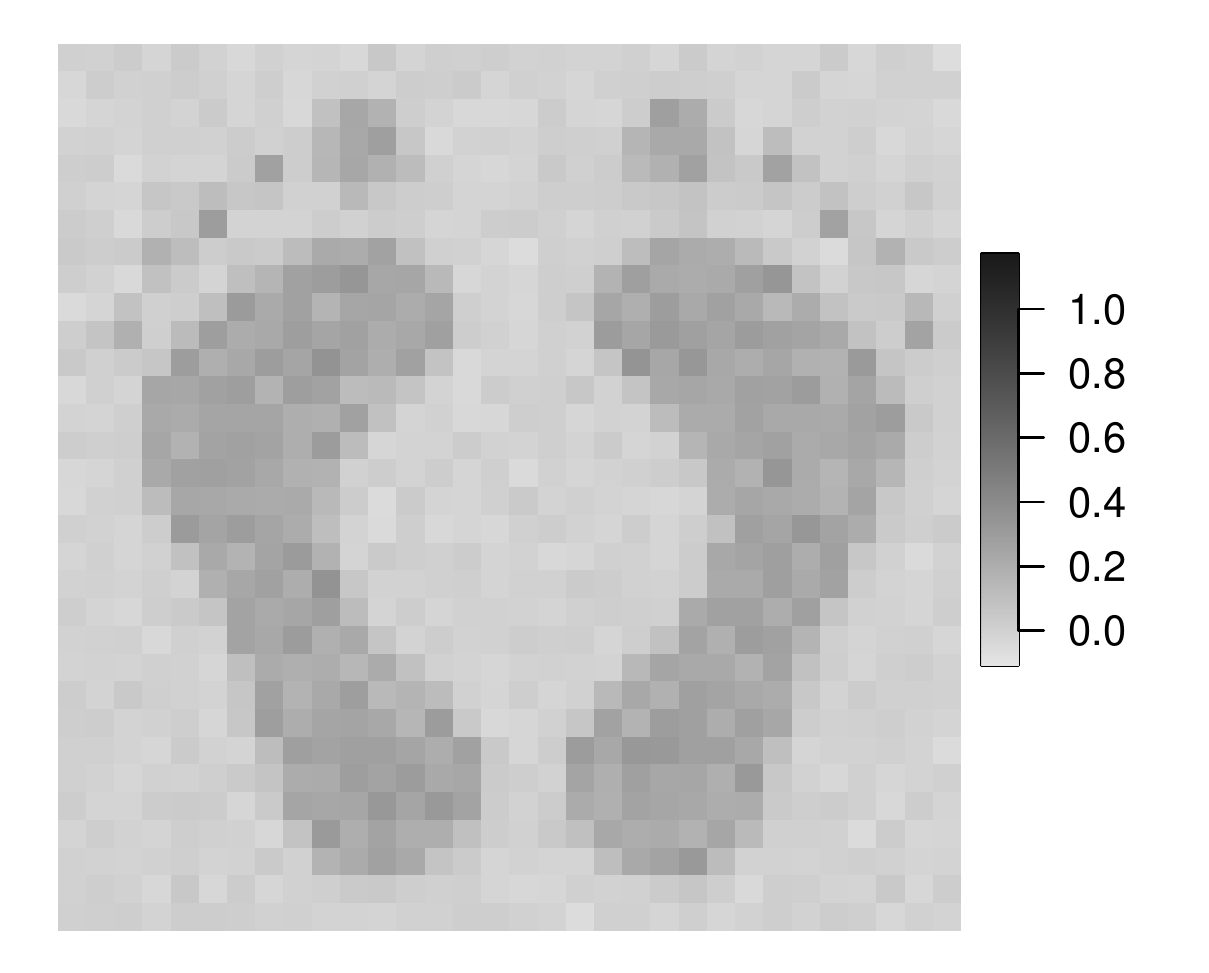} \\
\includegraphics[height=0.135\textheight]{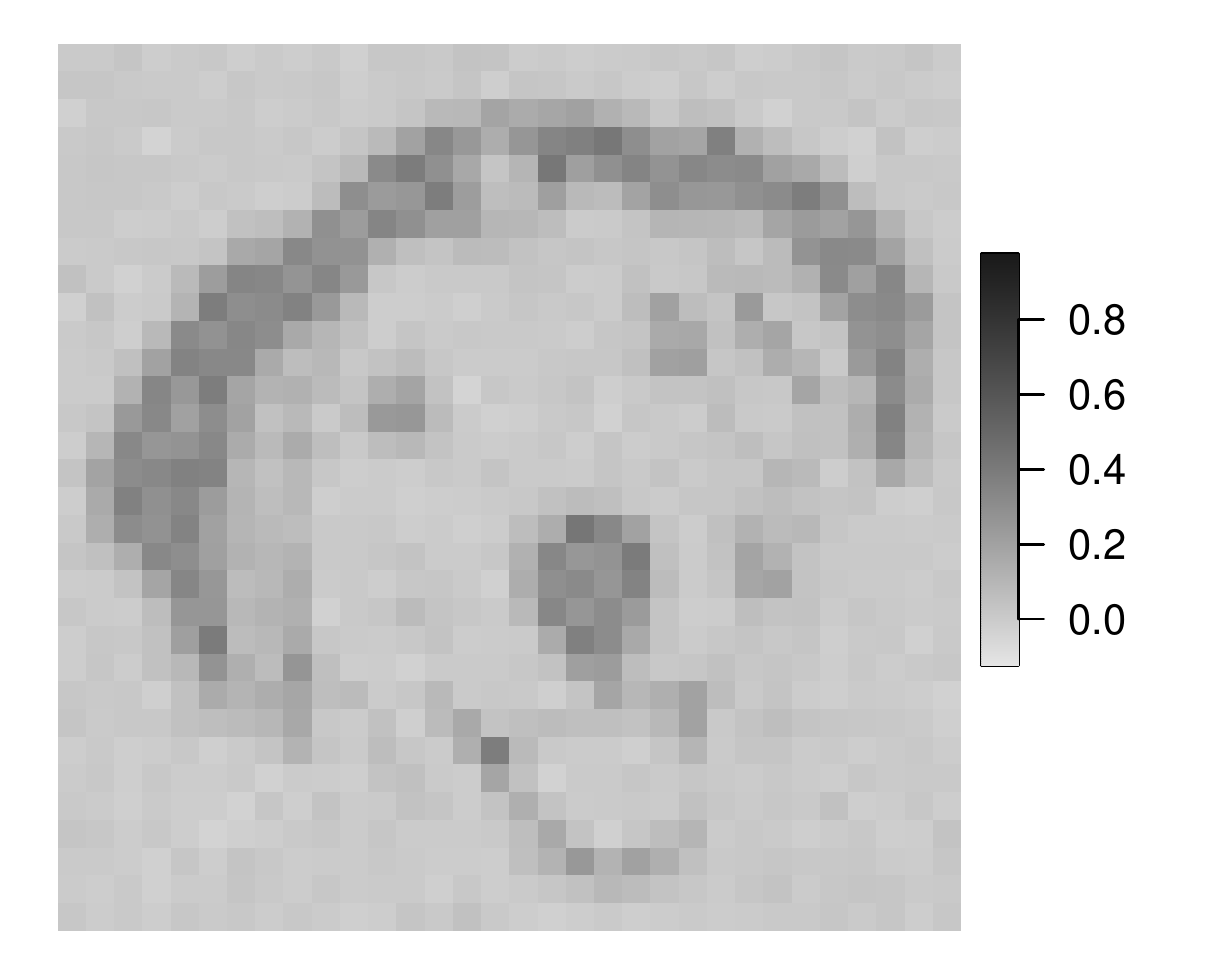}\\
\includegraphics[height=0.135\textheight]{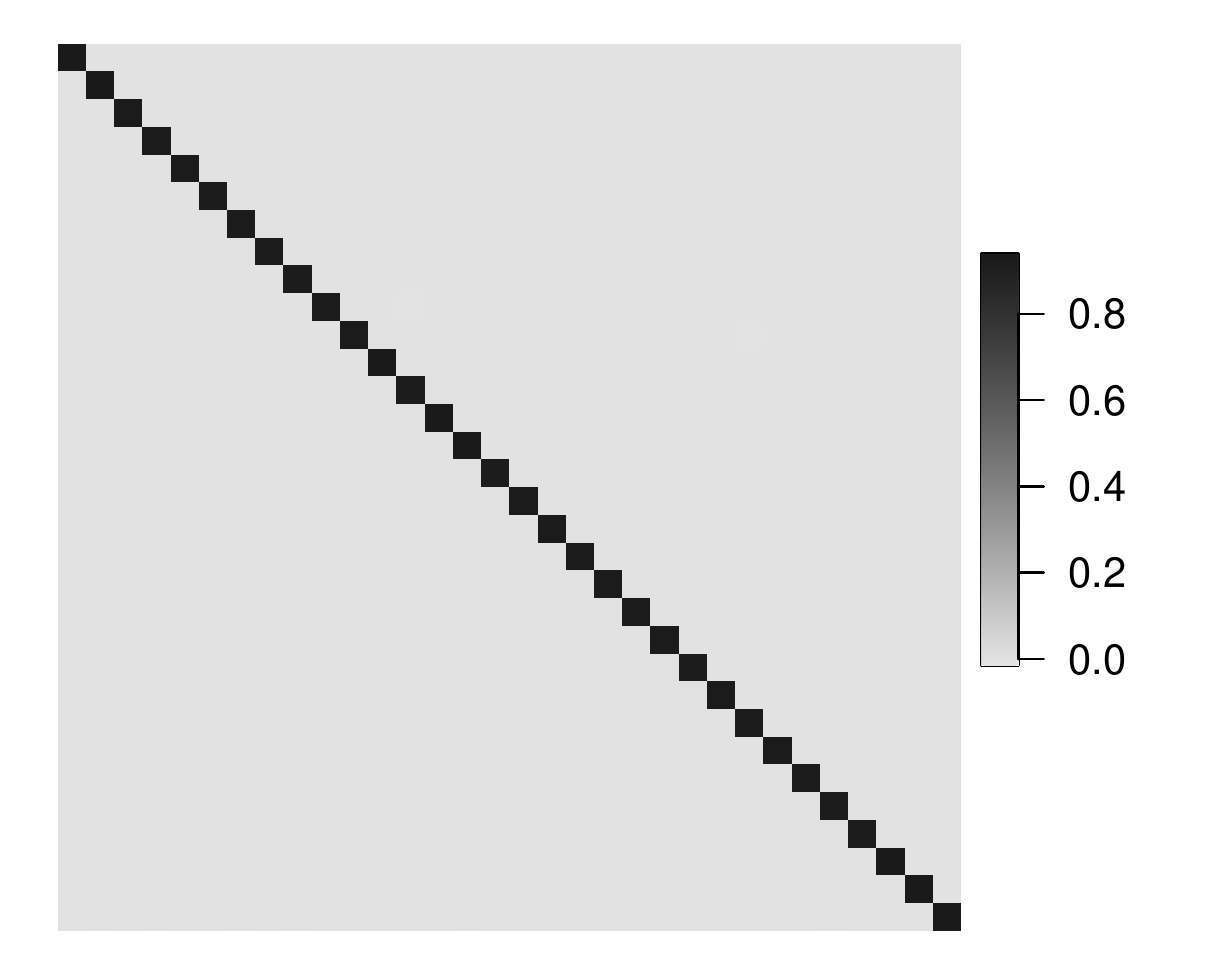}
\end{minipage}
\end{center}
\caption{True coefficient matrix and average across simulated data sets of the coefficient matrix posterior mean for \softer{}, hard \parafac{} and Tucker regression, and the penalized estimator for the Lasso.}
\label{fig:sim_pictures}
\end{figure}

The scenarios we considered represent settings of varying complexity and sparsity.
The first column of \cref{fig:sim_pictures} shows the true coefficient tensors (squares, feet, dog, diagonal). The squares coefficient matrix is used as a scenario where the true coefficient matrix is rectangular, but not low rank, and not sparse. The next two scenarios represent situations where the underlying structure is not of low-rank form, but could be potentially approximated by a low rank matrix up to a certain degree, hence representing scenarios somewhat favorable to the hard \parafac{}. In the last case, the diagonal coefficient matrix is used to represent a sparse coefficient matrix of full-rank without a network-structure, a scenario that is favorable for the Lasso, but is expected to be difficult for the hard \parafac{}.
Therefore, the scenarios we consider here represent a wide range of rank and sparsity specifications, and we investigate \softer's ability to use a low rank structure when such structure is useful (feet, dog), and expand away from it when it is not (squares, diagonal). Even though none of these coefficient tensors is low-rank, we considered a low-rank example in the Appendix, and we discuss it briefly below.

The remaining columns of \cref{fig:sim_pictures} show the average posterior mean or penalized estimate across simulated data sets. Results from \softer{} and the hard \parafac{} correspond to $D = 3$, though the methods are also considered with rank $D = 7$ (the results are shown in the Appendix and are discussed below). In the squares, feet and dog scenarios, the hard \parafac{} performs decently in providing a low-rank approximation to the true coefficient matrix. However, certain coefficient entries are estimated poorly to fit its rectangular structure. This is most evident in the squares scenario where the non-zero coefficients are obscured in order to fit in a low-rank form.
Results for the approach based on the Tucker decomposition are worse, with estimated coefficient matrices that deviate from the truth more.
In the diagonal scenario, the hard \parafac{} and Tucker approaches totally miss the diagonal structure and estimate (on average) a coefficient matrix that is very close to zero. In contrast, the Lasso performs best in the sparse, diagonal scenario and identifies on average the correct coefficient matrix structure. However, in the squares, dog and feet settings, it underestimates the coefficient matrix since it is based on assumed sparsity which is not true, and does not borrow any information across coefficient matrix entries.
In all situations, \softer{} closely identifies the structure of the underlying coefficient matrix, providing a compromise between tensor-based and unstructured estimation, and having small biases across all simulated scenarios (average bias also reported in \cref{tab:sims_n400}). In the squares, feet and dog scenarios, \softer{} bases estimation on the low-rank structure of the underlying hard \parafac{}, but it expands from it to better describe the true coefficient matrix's details. At the same time, \softer{} also performs well in the diagonal scenario, where the true coefficient tensor is full-rank. Therefore, the strength of \softer{} is found in its ability to use the low-rank structure of the \parafac{} when necessary, and diverge from it when needed.

In \cref{tab:sims_n400}, we report the methods' bias and root mean squared error (rMSE), predictive mean squared error, and frequentist coverage of the 95\% credible intervals. Conclusions remain unchanged, with \softer{} performing similarly to the hard \parafac{} when its underlying structure is close to true, and has the ability to diverge from it and estimate a coefficient tensor that is not low-rank in other scenarios. This is evident by an average coverage of 95\% posterior credible intervals that is 94.7\% in the diagonal scenario.
In terms of their predictive ability, the pattern observed for the bias and mean squared error persists, with \softer{} having the smallest mean squared predictive error in the squares, feet and dog scenarios, and the Lasso for the diagonal scenario, followed closely by \softer. Across all scenarios, the approach based on the Tucker decomposition performs worst among the Bayesian methods in terms of both estimation and predictive accuracy.

\begin{table}[!t]
\centering \small
\begin{tabular}{lll cccc}
& & & \softer{} & \parafac{} & Tucker & Lasso \\ \hline

squares & Truly zero & bias &  \textbf{0.003} & 0.005 & 0.013 & 0.012  \\
& & rMSE & \textbf{0.033} & 0.05 & 0.05 & 0.143  \\
& & coverage & 99.7\% & 98.3\% & 99.8\% & -- \\[5pt]

& Truly non-zero & bias & \textbf{0.084} & 0.106 & 0.136 & 0.501  \\
& & rMSE & \textbf{0.11} & 0.148 & 0.166 & 0.601  \\
& & coverage & 80.1\% & 68.4\% & \textbf{90.7\%}  & -- \\[5pt]

& Prediction & MSE & \textbf{5.05} & 8.99 & 12.16 & 111.8 \\
\hline

feet & Truly zero & bias & 0.037 & 0.046 & 0.057 & \textbf{0.016} \\
& & rMSE & \textbf{0.092} & 0.109 & 0.102 & 0.198  \\
& & coverage & 96.9\% & 94.2\% & 95.5\% & -- \\[5pt]

& Truly non-zero & bias & \textbf{0.116} & 0.138 & 0.175 & 0.43  \\
& & rMSE & \textbf{0.184} & 0.21 & 0.226 & 0.558  \\
& & coverage & \textbf{89.2\%} & 80\% & 84.8\% & --  \\[5pt]

& Prediction & MSE & \textbf{31.9} & 41.8 & 50.6 & 264.6\\ \hline

dog & Truly zero & bias & 0.047 & 0.059 & 0.083 & \textbf{0.029} \\
& & rMSE & \textbf{0.111} & 0.129 & 0.129 & 0.151 \\
& & coverage & 98.0\% & 92.8\% & {95.5\%} & -- \\[5pt]

& Truly non-zero & bias & \textbf{0.129} & 0.159 & 0.214 & 0.350 \\
& & rMSE & \textbf{0.205} & 0.230 & 0.261 & 0.435 \\
& & coverage & \textbf{88.2\%} & 76.9\% & 78.5\% & -- \\[5pt]

& Prediction & MSE & \textbf{35.0} & 45.4 & 61.7 & 138.4 \\ \hline

diagonal & Truly zero & bias & 0.002 & 0.004 & 0.002 & \textbf{$<$0.001} \\
& & rMSE & 0.02\phantom{0} & 0.051 & 0.024 & \textbf{0.009} \\
& & coverage & 100\% & 100\% & 100\% & -- \\[5pt]

& Truly non-zero & bias & 0.111 & 0.899 & 0.954 & \textbf{0.07} \\
& & rMSE & 0.126 & 0.906  & 0.955 & \textbf{0.084} \\
& & coverage & \textbf{94.7\%} & 3\% & 0.8\% & -- \\[5pt]

& Prediction & MSE & 1.41 & 29.7 & 30.6 & \textbf{0.81} \\ \hline
\end{tabular}

\caption{Simulation Results. Average bias, root mean squared error, frequentist coverage of 95\% credible intervals among truly zero and truly non-zero coefficient entries, and predictive mean squared error for \softer{} (with $D = 3$), the hard \parafac{} (with $D = 3$), Tucker regression, and Lasso for the simulation scenario with tensor predictor of dimensions $32 \times 32$ and sample size $n = 400$. Bold text is used for the approach performing best in each scenario and for each metric. If no entry is bold, no conclusive argument can be made.}
\label{tab:sims_n400}
\end{table}

\begin{table}[!b]
\centering
\begin{tabular}{l l cc l l}
& & Sensitivity & Specificity & FPR & FNR \\ \hline
squares & \softer{}  & 100 & 99.7 & 0.9 (0, 2.5) & 0 \\
        & \parafac{} & 100 & 98.3 & 4.7 (1.2, 8.7) & 0 \\
        & Tucker & 100 & 99.9 & 0.3 (0, 0.4) & 0 \\
        \hline
feet & \softer{} & 64.5 & 96.9 & 2.9 (1.7, 4.1) & 36.3 \\
    & \parafac{} & 68.4 & 94.1 & 5.2 (3.3, 7.1) & 34.3 \\
    & Tucker & 63.6 & 95.5 & 4.4 (3.3, 5.5) & 37.2 \\
    \hline
dog$^{**}$ & \softer{} & 52.9 & 96.7 & 5.2 (2.7, 8.1) & 34.7 \\
    & \parafac{} & 63.1 & 90.1 & 12.4 (8.4, 15.9) & 30.9  \\ 
    & Tucker & 46.3 & 93.1 & 11.5 (5.2, 17.6) & 38.6 \\
    \hline
diagonal & \softer & 100 & 100 & 0 (0, 0) & 0 \\
& \parafac{} & 3 & 100 & 28.8 (0, 70) & 3 \\
& Tucker & $<1$ & 100 & 33.3 (10, 50) & 3.1 \\
\hline
\end{tabular}
\caption{Methods' performance in identifying important entries. For sensitivity, specificity and false negative rate (FNR), results are shown as average across simulated data sets ($\times 100$), and for false positive rate$^*$ (FPR) as average (10$^{th}$, 90$^{th}$ percentile) ($\times 100$).}
\label{tab:sims_significance}
\begin{flushleft}
{\footnotesize $^*$The average FPR is taken over simulated data sets for which at least one entry was identified as important.} \\
{\footnotesize $^{**}$Most coefficients in the dog simulation were non-zero. Results are presented considering coefficients smaller than 0.05 as effectively zero.}
\end{flushleft}
\end{table}

\cref{tab:sims_significance} shows the performance of \softer{}, the hard \parafac{}, and Tucker regression approaches for identifying the important entries of the tensor predictor (entries with non-zero coefficients). Perfect performance would imply specificity and sensitivity equal to 100, and false positive and negative rates (FPR, FNR) equal to 0. The methods perform comparably in terms of specificity, sensitivity, and FNR, except for the diagonal scenario where the sensitivity of the hard \parafac{} and Tucker approaches is dramatically lower. However, the big difference is in the FPR. Even though \softer{}'s FPR is at times higher than 5\%, it remains at much lower levels than the hard \parafac{} and Tucker approaches for which FPR is on average over 10\% in the dog and approximately 30\% in the diagonal scenario. In \cref{app_sec:significance_disagree} we investigate the cases where \softer{} and hard \parafac{} return contradicting results related to an entry's importance. We illustrate that, when \softer{} disagrees with \parafac{} and identifies an entry as significant, the entry's true coefficient varies uniformly over the range of coefficients. On the other hand, when \parafac{} identifies entries as important and \softer{} does not, it is most likely for entries with small (or zero) coefficient values, in an effort to fit the estimated coefficient matrix into a low-rank form. These results indicate that identifying important entries based on the hard \parafac{} could lead to false discovery rates much higher than the desired one, whereas \softer{} alleviates this issue.

\cref{app_sec:more_sims} shows additional simulation results. \cref{app_subsec:sims_altern} shows results for alternative coefficient matrices, including a coefficient matrix of rank 3, a scenario favorable to the hard \parafac{}. There, we see that \softer{} collapses to the underlying hard \parafac{} structure when such a structure is true, and does not use its allowed deviations. These simulation results are in agreement with our first intuition-based argument of \cref{subsec:theory_consistency} with regards to the choice of rank $D$ for \softer{}.
\cref{app_sec:sims_n200} shows results for a subset of the coefficient matrices and for sample size $n = 200$. Simulations with a smaller $n$ to $p$ ratio show that \softer{} performs comparably to the hard \parafac{} for the dog and feet scenarios and has substantially smaller bias and rMSE for the truly non-zero coefficients in the diagonal scenario. The most notable conclusion is that \softer{} results are closer to the hard \parafac{} results when the sample size is small. This indicates that the data inform the amount of \parafac{} softening which depends on the sample size. Finally, \cref{app_subsec:D_7} shows results for \softer{} and the hard \parafac{} when $D = 7$. Even though the hard \parafac{} shows substantial improvements using the higher rank, \softer{} performs almost identically for rank 3 or 7, illustrating its robustness to the choice of the underlying rank. This last result is further investigated in \cref{subsec:sims_rank}.

\subsection{Simulation results for coefficient tensor of increasing rank}
\label{subsec:sims_rank}

We evaluated the performance of the hard and soft \parafac{} tensor regression with various values of the algorithmic rank $D$ and the coefficient tensor's true rank. We considered tensor predictor of dimension $20 \times 20$, rank of the true coefficient matrix equal to $3, 5, 7, 10$ and $20$, and we evaluated the hard \parafac{} and \softer{} with $D \in \{1, 3, 5, 7, 10\}$. For every value of the true rank, we generated 100 data sets.

\begin{figure}[!b]
\centering
\includegraphics[width = 0.9\textwidth]{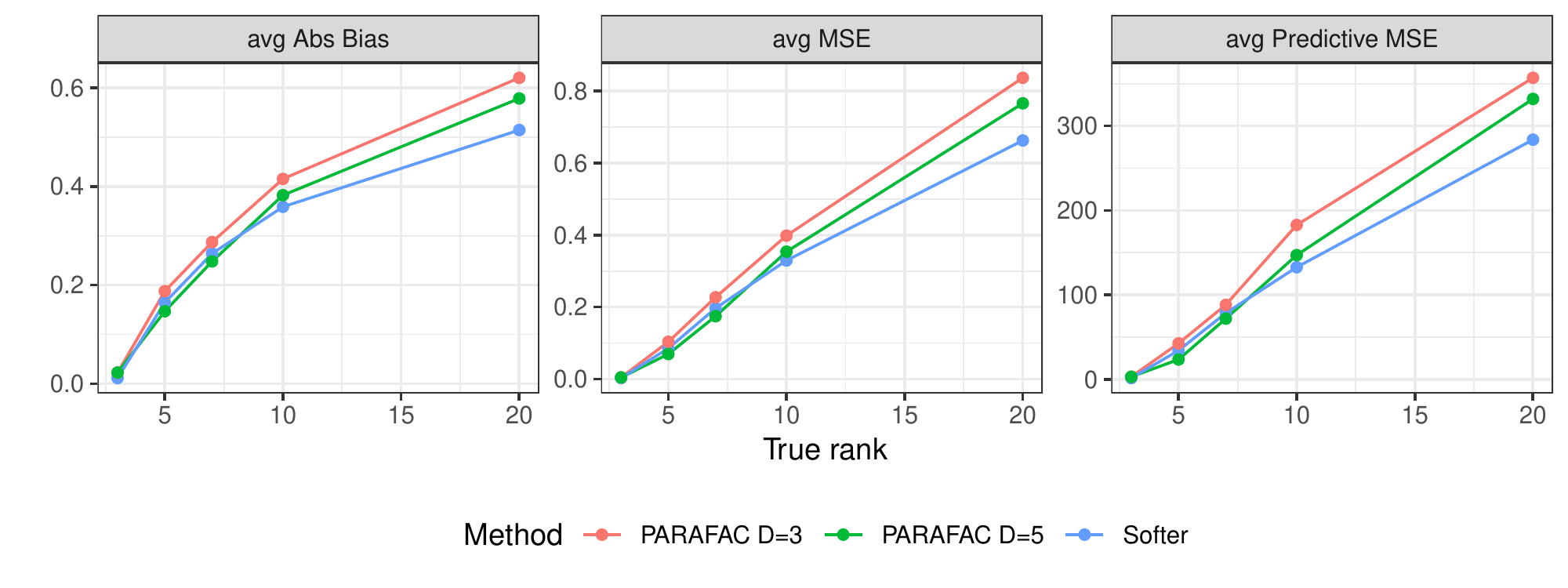}
\caption{Average absolute bias, estimation mean squared error and predictive mean squared error (y-axis) for tensor predictor of dimensions $20 \times 20$ and true coefficient matrix of increasing rank (x-axis). Results are shown for the hard \parafac{} with $D = 3$ (red) and $D = 5$ (green), and \softer{} with $D = 3$ (blue).}
\label{fig:sims_increasing_rank}
\end{figure}

In \cref{fig:sims_increasing_rank}, we show the average across entries of the coefficient matrix of the absolute bias and mean squared error, and the predictive mean squared error. For illustrative simplicity, we include a subset of the results: \softer{} with $D = 3$, and the hard \parafac{} with $D = 3$ and $D = 5$, though the complete results are discussed below.
When the true rank of the coefficient matrix is 3, the three approaches perform similarly. This indicates that both the hard \parafac{} with $D = 5$ and \softer{} are able to convert back to low ranks when this is true. For true rank equal to 5 or 7, the hard \parafac{} with $D = 5$ slightly outperforms \softer{}. However, for $D > 7$, \softer{} based on a rank-3 underlying structure performs best both in estimation and in prediction. These results indicate that, in realistic situations where the coefficient tensor is not of low-rank form, \softer{} with a low rank has the ability to capture the coefficient tensor's complex structure more accurately than the hard \parafac{}.

In \cref{app_subsec:sims_rank}, we show the performance of \softer{} and the hard \parafac{} across all ranks $D$ considered, $D \in \{1, 3, 5, 7, 10\}$. The conclusions about the methods' relative performance remain unchanged: even though \softer{} and the hard \parafac{} perform similarly for coefficient tensors with small true rank, \softer{} with $D > 1$ outperforms the hard \parafac{} in terms of both estimation and prediction when the coefficient tensor is of high rank. These results illustrate another important point. The performance of the hard \parafac{} depends heavily on the choice of rank $D$. In contrast, \softer{}'s performance is strikingly similar across all values of $D$, illustrating again that the results from \softer{} are robust to the choice of $D$, in agreement with the second intuition-based argument on \softer{}'s robustness in \cref{subsec:theory_consistency}.

Finally, \cref{tab:comp_time} shows the average time across simulated data sets for 15,000 MCMC iterations for the two methods. As expected, the computational time for both approaches increases as the value of $D$ increases, though the true rank of the coefficient tensor seems to not play a role. Here, we see that \softer{} generally requires two to three times as much time as the hard \parafac{} {\it for the same value of $D$}, which might seem computationally prohibitive at first. However, our results in \cref{fig:sims_increasing_rank} and \cref{app_subsec:sims_rank} show that \softer{} requires {\it a smaller value of $D$} in order to perform similarly to or better than the hard \parafac{} in terms of both estimation and prediction. Therefore, \softer{}'s computational burden can be drastically alleviated by fitting the method for a value of $D$ which is substantially lower than the value of $D$ for the hard \parafac{}. In \cref{sec:discussion}, we also discuss frequentist counterparts to our model that might also reduce the computational load.

\begin{table}[ht]
\centering
\begin{tabular}{rlccccc}
  \hline
 & & \multicolumn{5}{c}{True rank of coefficient tensor} \\
 & \hspace{30pt} &
 \hspace{10pt}3\phantom{0}\hspace{10pt} & \hspace{10pt}5\phantom{0}\hspace{10pt} &
 \hspace{10pt}7\phantom{0}\hspace{10pt} &
 \hspace{10pt}10\hspace{10pt} &
 \hspace{10pt}20\hspace{10pt} \\ 
  \hline \hline
\softer{} & $D = 1$ & 28 & 29  & 30  & 37  & 39  \\ 
 &  $D = 3$  & 166  & 110  & 101  & 104  & 117  \\ 
 &  $D = 5$  & 181  & 149  & 158  & 159  & 152  \\ 
 &  $D = 7$  & 212  & 189  & 205  & 216  & 220  \\ 
 &  $D = 10$ & 275  & 236  & 255  & 288  & 259  \\ 
   \hline
Hard \parafac{} &   $D = 1$ & 16 & 16 & 15 & 15 & 15 \\ 
 &  $D = 3$  & 30 & 39 & 46 & 35 & 36 \\ 
 &  $D = 5$  & 69 & 54 & 57 & 53 & 56 \\ 
 &  $D = 7$  & 100 & 90 & 82 & 72 & 76 \\ 
 &  $D = 10$ & 104 & 117 & 108 & 104 & 102 \\ 
   \hline
\end{tabular}
\caption{Computational time (in minutes) for \softer{} and the hard \parafac{}.}
\label{tab:comp_time}
\end{table}


\section{Estimating the relationship between brain connectomics and human traits}
\label{sec:application}

Data from the Human Connectome Project (HCP) contain information on about 1,200 healthy young adults including age, gender, various brain imaging data, and a collection of measures assessing cognition, personality, substance intake and so on (referred to as traits here). We are interested in studying the brain structural connectome, referring to anatomical connections of brain regions via white matter fibers tracts. The white matter fiber tracts can be indirectly inferred from diffusion MRI data. Two brain regions are considered connected if there is at least one fiber tract running between them. However, there can be thousands of fiber tracts connecting a pair of regions. Properties of the white matter tracts in a connection, such as number of tracts, and patterns of entering the regions, might be informative about an individual's traits.
Using data from the HCP and based on the soft tensor regression framework, we investigate the relationship between different connectome descriptors and human traits.

Structural connectivity data were extracted using state-of-the-art pipelines in \cite{Zhang2018mapping}. In total, about 20 connectome descriptors (adjacency matrices) describing different aspects of white matter fiber tract connections were generated (see \cite{Zhang2018mapping} for more information on the extracted descriptors). Each adjacency matrix has a dimension of $68 \times 68$, representing $R = 68$ regions' connection pattern. The $68$ regions  were defined using the Desikan-Killiany atlas \citep{Desikan2006automated}. Of the 20 extracted connectome features, we consider two in this analysis: (a) count, describing the number of streamlines, and (b) connected surface area (CSA), describing the area covered by small circles at the interactions of fiber tracts and brain regions, since they are the most predictive features according to results in \cite{Zhang2019tensor}.

We examine the relationship between these descriptors of structural brain connections and \numtraits{} traits, covering domains such as cognition, motor, substance intake, psychiatric and life function, emotion, personality and health. The full list of outcomes we analyze is presented in \cref{app_tab:app_outcomes} and includes both binary and continuous traits. For binary traits, a logistic link function is assumed.

\subsection{Adapting \softer{} for (semi-)symmetric brain connectomics analysis}

The nature of the brain connectivity data implies that the $R \times R$-dimensional tensor predictor including a specific connectivity feature among $R$ regions of interest (ROIs) is symmetric and the diagonal elements can be ignored since self-loops are not considered. Further, considering $p$ features simultaneously would lead to an $R \times R \times p$ tensor predictor which is semi-symmetric (symmetric along its first two modes). The (semi-)symmetry encountered in the predictor allows us to slightly modify \softer{} and reduce the number of parameters by imposing that the estimated coefficient matrix $\B$ is also (semi-)symmetric. We provide the technical details for the (semi-)symmetric \softer{} in \cref{app_sec:symmetric_softer}.

\subsection{Analyses of the brain connectomics data}

For the purpose of this paper, we investigate the relationship between features of brain connections and human traits by regressing each outcome on each of the two predictors (count and CSA) separately. Even though analyzing the relationship between the traits and multiple features simultaneously is possible, we avoid doing so here for simplicity. We analyze the data employing the following methods: (1) symmetric \softer{} with $D = 6$, (2) the hard \parafac{} approach of \cite{Guhaniyogi2017bayesian} which does not impose symmetry of the coefficient tensor with $D = 10$, and (3) Lasso on the vectorized lower triangular part of the tensor predictor. Since publicly available code for non-continuous outcomes is not available for the hard \parafac{} approach, we only consider it when predicting continuous outcomes.

We compare methods relative to their predictive performance. For each approach, we estimate the out-of-sample prediction error by performing 15-fold cross validation, fitting the method on 90\% of the data and predicting the outcome on the remaining 10\%. In the case of \softer{} and the hard \parafac{} we also investigate the presence of specific brain connections that are important in predicting any of the outcomes by checking whether their coefficients' 95\% posterior credible intervals include zero.
Additional results based on \softer{} for a different choice of baseline rank or when symmetry is ignored are included in \cref{app_sec:additional_application} and are summarized below.


\subsection{Using features of brain connections for predicting human traits}

For continuous outcomes, methods' predictive performance was evaluated by calculating the percentage of the marginal variance explained by the model defined as $1 - (\text{CV MSE}) / (\text{marginal variance})$. For binary outcomes, we used the model's estimated linear predictor to estimate the optimal cutoff for classification based on Youden's index \citep{youden1950index} and calculated the average percentage of correctly classified observations in the held-out data.

\begin{figure}[!b]
\centering
\includegraphics[width = \textwidth]{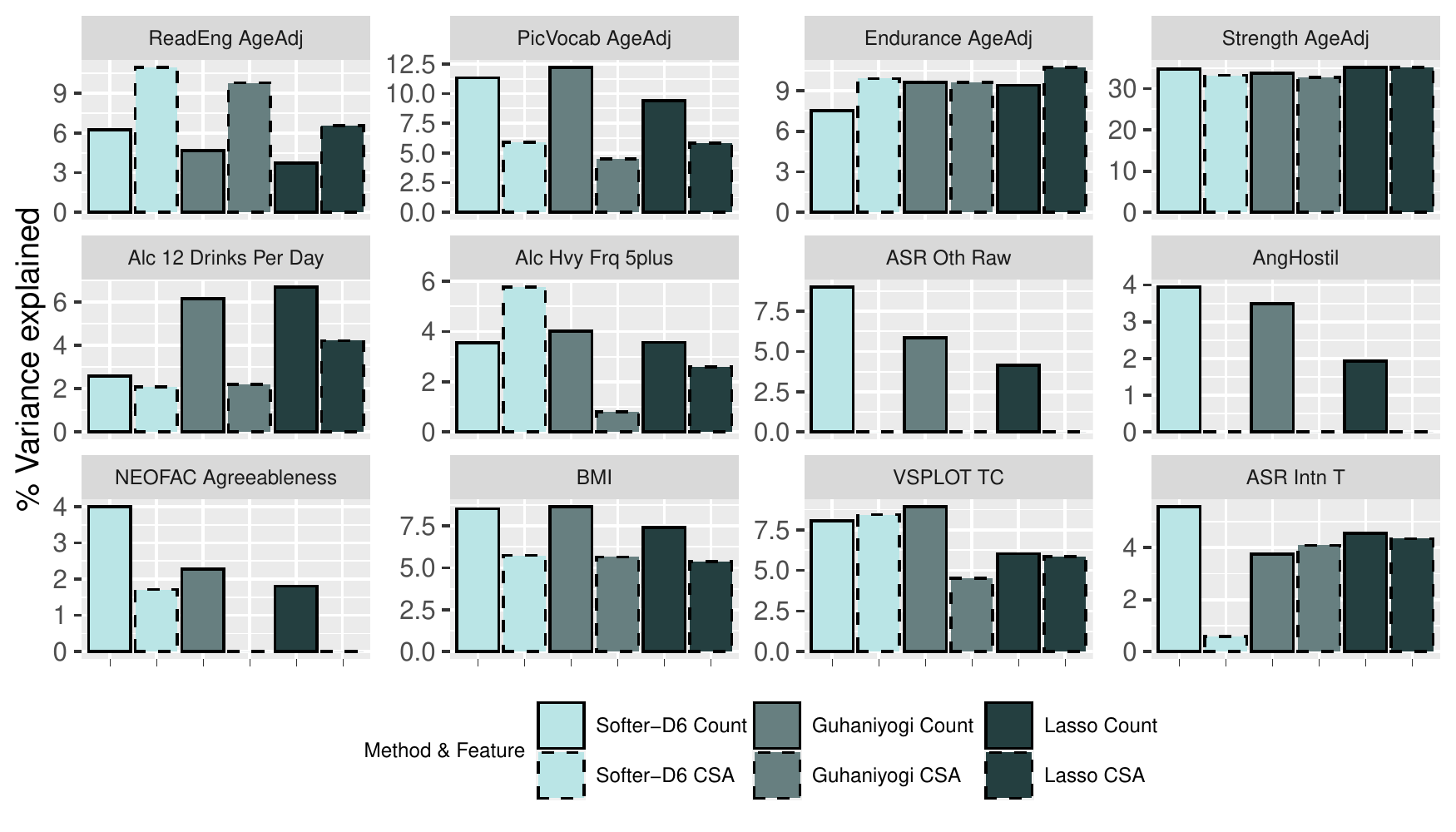} \\
\includegraphics[width = 0.75\textwidth]{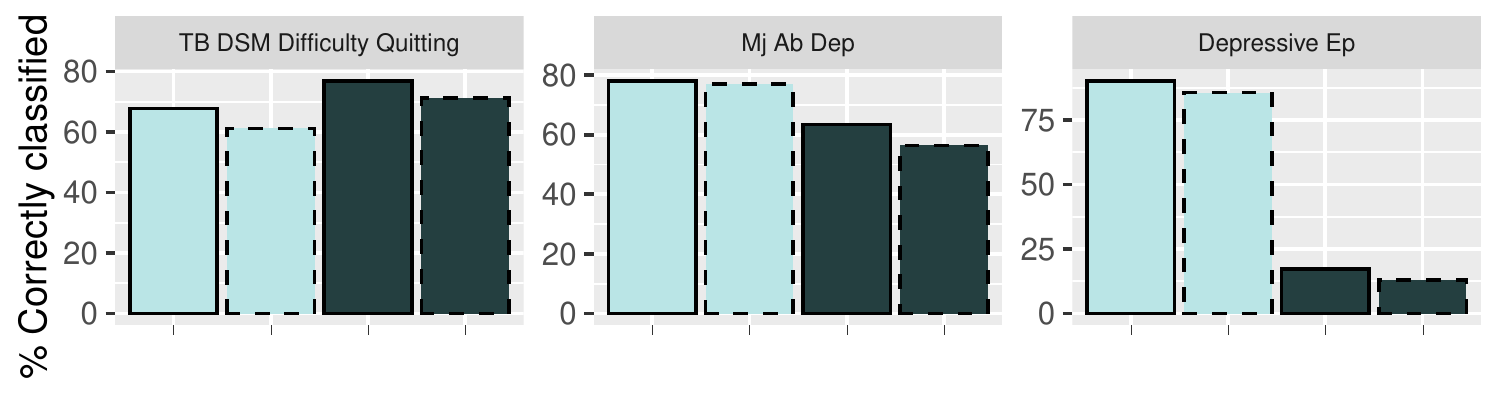}
\caption{Top: Percentage of outcome variance explained by the tensor predictor for continuous outcomes calculated as [1 - MSE / (marginal variance)] $\times 100$. Bottom: Average percentage of units correctly classified for binary outcomes. Results are presented using different color for each method, and different line-type for each feature of brain connections.}
\label{fig:app_predict}
\end{figure}

\cref{fig:app_predict} shows these results for the three approaches considered, and for each feature separately. For most outcomes, one of the two features appeared to be most predictive of the outcome across approaches. For example, the count of streamlines was more predictive than CSA of an individual's anger level (\texttt{AngHostil}), independent of the modeling approach used. By examining the methods' predictive performance, it is evident that features of brain connectomics are, in some cases, highly predictive of outcomes. Specifically, over 30\% of the variance in an individual's strength level, and over 10\% of the variance in endurance, reading comprehension, and picture vocabulary ability can be explained by the count or CSA of streamlines of their brain connections.

Not one approach outperformed the others in prediction across all features and outcomes. However, approaches that accommodate the network structure perform better than Lasso in most situations. One example is \softer{}'s performance relative to Lasso when predicting individuals' previous depressive episode (\texttt{Depressive Ep}). Here, Lasso performs worse than the random classifier, whereas \softer{} has over 90\% accuracy. Even when the number of observations is less than 300 (difficulty quitting tobacco, \texttt{TB DSM Difficulty Quitting}), \softer{} performs only slightly worse than Lasso.
For continuous outcomes, \softer{} and hard \parafac{} perform comparably.
As we saw in the simulations in \cref{sec:simulations} and in \cref{app_sec:sims_n200}, the similar predictive performance of \softer{} and hard \parafac{} could be due to the limited sample size that forces \softer{} to heavily rely on the underlying low-rank structure for estimation, essentially reverting back to the hard \parafac{}.

The low signal in predicting some outcomes implies low power in identifying pairs of brain regions whose connection's features are important. In fact, 95\% credible intervals for all coefficients using the hard \parafac{} overlapped with zero. In contrast, \softer{} identified seven important connections:
five of them were for predicting whether an individual has had a depressive episode (three using count of streamlines as the predictor, and two using CSA), one in predicting an individual's strength, and one in predicting the variable short penn line orientation (VSPLOT).
The identified connections are listed in \cref{tab:app_connections} and agree with the literature in neuroscience.
All identified connections in predicting a depressive episode involve the parahippocampal, which is the posterior limit of the amygdala and hippocampus and is located in the temporal lobe, and ROIs located in the frontal lobe (paracentral, lateral orbitofrontal, pars orbitalis). Dysfunction of the parahippocampal (as well as the amygdala and hippocampus) has been identified in various studies as an important factor in major depression and emotion-related memory observed in depression \citep{Mayberg2003modulating, Seminowicz2004limbic, LaBar2006cognitive, Zeng2012identifying}. Further, dysregulation of the pathways between the frontal and temporal lobes has been identified as predictive of depression \citep{mayberg1994frontal,Steingard2002smaller}, even when explicitly focusing on the cortical regions \softer{} identified as important \citep{Liao2013depression}. 
Two of the three connections were identified as important irrespective of the tensor predictor used (count or CSA). Even though these predictors are related, they describe different aspects of brain connectivity. Therefore, \softer{} identified the same connections as important based on two separate measures of brain connectivity.
The identified connection in predicting strength involves the precuneus and superior parietal regions in the parietal lobe. Precuneus' connectivity has been associated with a variety of human functions, including motor-related traits \citep{Cavanna2006precuneus, Wenderoth2005role,Simon2002topographical}, and the parietal lobe in general is believed to control humans' motor  system \citep{Fogassi2005motor}.

\begin{table}[!b]
\centering
\begin{tabular}{llll}
  Outcome & Feature & ROI 1 & ROI 2 \\ \hline
Depressive & Count &  (lh) Parahippocampal   &  (rh) Lateral Orbitofrontal    \\
Episode & & & (lh) Paracentral  \\
        & & &  (lh) Pars Orbitalis  \\ \hline
& CSA & &  (rh) Lateral Orbitofrontal  \\
& & & (lh) Paracentral  \\ \hline
  
Strength & Count & (rh) Precuneus  & (rh) Superior Parietal \\ \hline


VSPLOT & CSA & (lh) Banks of Superior Temporal Sulcus & (lh) Rostral Anterior Cingulate \\ \hline

\end{tabular}
\caption{Brain connections with important features in predicting human traits.}
\label{tab:app_connections}
\end{table}

\cref{app_sec:application_additional_results} includes results from the symmetric \softer{} using a smaller rank ($D=3$), and from \softer{} using the same rank as the results in this section ($D=6$) but ignoring the known symmetry of the predictor. All three versions of \softer{} perform similarly in terms of prediction, with potentially slightly lower predictive power for the symmetric \softer{} with rank 3. The entries identified as important by \softer{} with $D = 3$ are similar, though not identical to the ones in \cref{tab:app_connections}. When symmetry is not accounted for, \softer{} does not identify any important connections. Finally, in \cref{app_subsec:identified_reproducibility} we investigate the reproducibility of identified connections across random subsamples including 90\% of the observations. We find that for strength, when connections are identified in the subsample, they generally include the ones in \cref{tab:app_connections}. Finally, when predicting whether an individual has has a depressive episode, {\it any} of the connections identified across {\it any} subsample and for {\it either} predictor involved at least one of Parahippocampal (in the left hemisphere) or Lateral Orbitofrontal (in the right hemisphere) ROIs, implying that the importance of these two regions is strikingly reproducible.

\section{Discussion}
\label{sec:discussion}

In this paper, we considered modeling a scalar outcome as a function of a tensor predictor within a regression framework.
Estimation of regression models in high dimensions is generally based on some type of assumed sparsity of the true underlying model: sparsity directly on covariates, or ``latent sparsity'' by assuming a low-dimensional structure. When the assumed sparsity is not true, the model's predictive ability and estimation can suffer.
Our approach is positioned within the class of latent sparsity, since it exploits a low-dimensional underlying structure. We focused on adequately relaxing the assumed structure by softening the low-dimensional \parafac{} approximation and allowing for interpretable deviations of row-specific contributions.
We show that softening the \parafac{} leads to improved estimation of coefficient tensors, better performance in identifying important entries, consistent estimation irrespective of the underlying rank used for estimation, and more accurate predictions.
The approach is applicable to both continuous and binary outcomes, and was adapted to (semi-)symmetric tensor predictors, which is common in settings where the predictor is measured on a network of nodes. \softer{} was used to study the relationship between brain connectomics and human traits, and identified several important connections in predicting depression.

Combining the two types of assumed sparsity for low-rank \textit{and} sparse matrix estimation has received some attention in the literature, especially in machine learning with matrix data. \cite{Candes2011robust},
\cite{Waters2011recovering} and \cite{Zhou2011randomized} decomposed matrices as the sum of a low-rank and a sparse component. \cite{Zhang2016lowrank} employed such decomposition in anomaly detection by studying the Mahalanobis distance between the observed data and the low-rank component. \cite{Richard2012estimation} developed a penalization-based approach to estimating matrices that are simultaneously sparse and low-rank by adopting one type of penalty for sparsity and one for rank. All these approaches have been formulated as optimization problems and algorithms are generally based on iterative procedures.

Within a Bayesian regression framework, \cite{Guha2018bayesian} combined the two types of sparsity and proposed a network-based spike and slab prior on the nodes' importance. Under that model, a node is either active or inactive, and active nodes are expected to contribute to the outcome based on a low-rank coefficient tensor. In that sense, this approach has important commonalities to estimating a coefficient tensor that is simultaneously sparse and low-rank. Even though we find that approach to be promising, we find that node selection in itself can be too restrictive in some settings, and future work could incorporate hierarchical or parallel node and entry selection.

\softer{} has similarities but also differences from the methods discussed above. On one hand, \softer{} provides a relaxation of an assumed low-rank form. However, this relaxation (or softening) is not sparse in any way, and \textit{every} element of the tensor is allowed to deviate from the low-rank structure. We find that an exciting line of research would combine low-rank and sparse approaches while allowing for sufficient flexibility to deviate from both of them.
Important questions remain on the interplay between entry selection and the assumed structure on the coefficient tensor. We illustrated in simulations that tensor regression based on the hard \parafac{} can perform poorly for entry selection. Future work could focus on studying the properties of multiplicity control in structured settings, and forming principled variable selection approaches with desirable properties within the context of structured data.

Even though we showed that the posterior distribution of the coefficient tensor based on \softer{} is consistent irrespective of the tensor's true rank or the algorithmic rank used, additional theoretical results would help illuminate \softer{}'s performance. For example, it would be interesting to understand how \softer{} performs under the asymptotic regime where the tensor predictor is allowed to grow with the sample size.
Furthermore, even though our simulation results indicate that \softer{} reverts back to the hard \parafac{} when the underlying low-rank structure is true, future approaches to ``soft'' regression can investigate whether the soft and hard approaches asymptotically converge if the hard structure is true. If this result was derived, then it would imply a new form of robustness: robustness in terms of the method chosen, soft or hard.


Finally, a criticism of Bayesian methods is often their computational burden. One way forward in alleviating this issue for \softer{} is to re-formulate our model within the frequentist paradigm. For example, one could impose a penalty term on the magnitude of the $\gammakj\tod$ parameters using
\(
\text{penalty}_{\gamma}(d) = \sum_k \sum_{j_k} (\gammakj\tod)^2,
\)
substitute the softening structure in \cref{eq:soft_beta_dist} with a penalty term such as
\(
\text{penalty}_{\beta}(d) = \sum_k \sum_\subss (\betakj\tod - \gammakj\tod)^2,
\)
and maximize the penalized likelihood with appropriate tuning parameters. One might need to link the tuning parameter for $\text{penalty}_\gamma(d)$ with that of $\text{penalty}_\beta(d)$ to ensure that penalization of one does not lead to overcompensation from the other (similarly to our discussion in \cref{subsec:bayesian} for the inclusion of $\zeta\tod$ in the distribution for $\betakj\tod$).
Such $L2$-norm penalties with a continuous outcome would imply that the maximization problem is convex and optimization procedures can be developed relatively easily. Even though this reformulation would not reduce the number of parameters, it is expected to require far fewer iterations than an MCMC procedure. Alternative norms could also be considered to impose more sparsity, such as using a group-Lasso penalty instead of $\text{penalty}_\beta(d)$.

\section*{Acknowledgements}
This work was supported by grant R01MH118927 from the National Institutes of Health (NIH) and R01-ES027498-01A1 from the National Institute of Environmental Health Sciences (NIEHS) of the National Institutes of Health (NIH).

\newpage
\section*{Appendices}

\appendix
\setcounter{table}{0}
\renewcommand{\thetable}{\thesection.\arabic{table}}
\setcounter{equation}{0}
\renewcommand{\theequation}{\thesection.\arabic{equation}}
\setcounter{figure}{0}
\renewcommand{\thefigure}{\thesection.\arabic{figure}}

\numberwithin{figure}{section}
\numberwithin{table}{section}
\numberwithin{equation}{section}

\crefalias{section}{appsec}

\section{Proofs}
\label{app_sec:proofs}

\renewcommand*{\proofname}{\textbf{Proof of \cref{theory:variance_B}}}
\begin{proof} $\ $\\
\textbf{Expectation.}
\noindent
We use $S, Z$ and $W$ to denote the collection of $\sigma^2_k, \zeta\tod$ and $\wkj\tod$, over $k$, $d$, and $(k,j_k,d)$ accordingly. We start by noting that
$$\betakj\tod | \sigma^2_k, \zeta\tod, \tau_\gamma, \wkj\tod \sim N(0, \sigma^2_k\zeta\tod + \tau_\gamma\zeta\tod\wkj\tod),$$
and, if $(k,j_k,d) \neq (k',j_k', d')$
\begin{equation}
\betakj\tod \independent \betakj[F][F]\tod[F] | S, Z, W, \tau_\gamma.
\label{app_eq:cond_independece}    
\end{equation}
Note that $\betakj\tod$ is \textit{not} independent of $\betakj[T][F]\tod$ conditional on $(S,Z,W,\tau_\gamma)$ when $j_k = j_k'$ due to their shared dependence on $\gammakj\tod$.
Then,
\begin{align*}
\E(\B_\subss | S, Z, W, \tau_\gamma) &=
\E\Big(\sum_{d = 1}^D \prod_{k = 1}^K \betakj\tod | S, Z, W, \tau_\gamma \Big)
= \sum_{d = 1}^D \prod_{k = 1}^K  \E\Big(\betakj\tod | S, Z, W, \tau_\gamma \Big)
= 0
\end{align*}
So, a priori, all elements of the coefficient tensor have mean 0, $\E(\B_\subss) = 0$.\\

\noindent
\textbf{Variance.}\\
\noindent
Furthermore, we have
%
$$
\Var(\B_\subss) = \E \Big\{ \Var \Big( \B_\subss | S, Z, W, \tau_\gamma \Big)\Big\} 
= \E \Big\{ \Var \Big( \sum_{d = 1}^D \prod_{k = 1}^K \betakj\tod | S, Z, W, \tau_\gamma \Big) \Big\}.
$$
Since the $\betakj\tod$ are conditionally independent across $d$, $\prod_{k = 1}^K \betakj\tod$ are also conditionally independent across $d$. Moreover, the terms of the product $\betakj\tod$ are independent across $k$ and are mean-zero random variables, implying that $\prod_{k = 1}^K \betakj\tod$ are mean zero variables. Note here that two independent mean-zero random variables $A, B$ satisfy that $\Var(AB) = \Var(A)\Var(B) $. Then,
\begin{align*}
\Var(\B_\subss) &= \E \Big\{ \sum_{d = 1}^D \prod_{k = 1}^K \Var \Big( \betakj\tod | S, Z, W, \tau_\gamma \Big) \Big\} \\
&= \E \Big\{ \sum_{d = 1}^D \prod_{k = 1}^K \zeta\tod (\sigma^2_k + \tau_\gamma \wkj\tod)\Big\} \\
&= \E_Z \Big\{ \sum_{d = 1}^D (\zeta\tod)^K  \Big\}  \E_{S,W, \tau_\gamma} \Big\{ \prod_{k = 1}^K (\sigma^2_k + \tau_\gamma \wkj\tod)\Big\},
\end{align*}
where in the last equation we used that, a priori, $Z \independent (S, W, \tau_\gamma)$ to write $\E_{S,W,\tau_\gamma|Z}$ as $\E_{S,W,\tau_\gamma}$, and separate the two expectations.

However, $\sigma^2_k + \tau_\gamma \wkj\tod$ are not independent of each other for different values of $k$ since they all involve the same parameter $\tau_\gamma$. We overcome this difficulty in calculating the expectation of the product by writing
$\prod_{k = 1}^K (\sigma^2_k + \tau_\gamma \wkj\tod) = \sum_{l = 0}^K c_l \tau_\gamma^l$, where
$$
c_l = \sum_{\mathcal{K} \subset \{1, 2, \dots, K\} : |\mathcal{K}| = l} \left( \prod_{k \in \mathcal{K}} \wkj\tod \prod_{k \not\in \mathcal{K}} \sigma^2_k \right).
$$
So, for every power of $\tau_\gamma$, $\tau_\gamma^l$, $l \in \{0, 1, \dots, K\}$, the corresponding coefficient is a sum of all terms involving $l$ distinct $w$'s and $K-l$ distinct $\sigma^2$'s. For example, for $K = 2$, $c_1 = w_{1,j_1}\tod \sigma^2_2 + \sigma^2_1 w_{2,j_2}\tod$. Writing the product in this way, separates the terms $(\wkj\tod, \sigma^2_k)$ from $\tau_\gamma$, which are a priori independent. Then,
$$
\Var(\B_\subss) = \E_Z \Big\{ \sum_{d = 1}^D (\zeta\tod)^K \Big\}
\E_{S,W, \tau_\gamma}  \Big( \sum_{l = 0}^K c_l \tau_\gamma^l \Big)
= \E_Z \Big\{ \sum_{d = 1}^D (\zeta\tod)^K  \Big\}  \Big\{ \sum_{l = 0}^K \E(\tau_\gamma^l)\E_{S,W} (c_l) \Big\}.
$$

We continue by studying $\E_{S,W}(c_l) = \sum_{\mathcal{K}: |\mathcal{K}| = l} \E_{S,W} \Big( \prod_{k\in\mathcal{K}} \wkj\tod \prod_{k \not\in\mathcal{K}} \sigma^2_k \Big) $. Note that since all parameters $\{\wkj\tod,\sigma^2_k\}_k$ for fixed $j_k$ are a priori independent (any dependence in the $\wkj\tod$ exists across $j_k$ of the same mode due to the common value $\lambda_k\tod$),
$\E_{S,W}(c_l) = \sum_{\mathcal{K}: |\mathcal{K}| = l} \Big( \prod_{k\in\mathcal{K}} \E_W (\wkj\tod) \prod_{k \not\in\mathcal{K}} \E_S (\sigma^2_k) \Big) $. Now, note that $\E(\sigma^2_k) = \addvar $, and
%
$$
\E_W (\wkj\tod) = \E_\Lambda\{\E_{W|\Lambda}[\wkj\tod]\} = 2 \E_{\Lambda}\{ (\lambda_k\tod)^{-2} \}.
$$
Since $\lambda_k\tod \sim \Gamma(a_\lambda, b_\lambda)$, $1 / \lambda_k\tod \sim IG(a_\lambda, b_\lambda)$, we have that
$$\E\{(1 / \lambda_k\tod)^2\} =
\Var(1 / \lambda_k\tod) + \E^2(1 / \lambda_k\tod) =
\frac{b_\lambda^2}{(a_\lambda - 1)(a_\lambda - 2)}, \ a_\lambda > 2.
$$
Putting this together, we have that, for $a_\lambda > 2$,
$$ \E_{S,W}(c_l) = {K \choose l}
\Big\{ \lambdafrac \Big\}^l
\Big( \addvar \Big)^{K -l}.$$
%
Further, since $\tau_\gamma \sim \Gamma(a_\tau, b_\tau)$, we have that
$$
\E(\tau_\gamma^l) = \frac{b_\tau^{a_\tau}}{\Gamma(a_\tau)} \int \tau_\gamma^{a_\tau + l - 1} \exp\{ - b_\tau \tau_\gamma \}\ \mathrm{d}\tau_\gamma = \frac{\Gamma(a_\tau + l)}{\Gamma(a_\tau)b_\tau^l} = \frac{\rho_l}{b_\tau^l},
$$
for $\rho_l = 1$ if $l = 0$, and $\rho_l = a_\tau (a_\tau + 1) \dots (a_\tau + l - 1)$ if $l \geq 1$.
%
Lastly, since $\bm \zeta \sim Dir(\alpha/D, \alpha/D, \dots, \alpha/D)$, we have that $\zeta\tod \sim Beta(\alpha/D, (D-1)\alpha/D)$, and
$$E\{(\zeta\tod)^K\} = \prod_{r = 0}^{K - 1} \frac{\alpha/D + r}{\alpha + r}
$$

\noindent
Combining all of these, we can write the prior variance for entries $\B_\subss$ of the coefficient tensor as
\begin{align*}
\Var(\B_\subss) &=
\E_Z \Big[ \sum_{d = 1}^D (\zeta\tod)^K \Big] 
\sum_{l = 0}^K \rho_l {K \choose l} \Big\{ \lambdafrac[T] \Big\}^l
\Big( \addvar \Big)^{K -l}  \\
&= \Big\{ D \prod_{r = 0}^{K - 1} \frac{\alpha/D + r}{\alpha + r} \Big\}
\Big[ \sum_{l = 0}^K \rho_l {K \choose l} \Big\{ \lambdafrac[T] \Big\}^l
\Big( \addvar \Big)^{K -l} \Big].
\end{align*}
%


\noindent
\textbf{Covariance.}\\
\noindent
Since $\E(\B_\subss|S,Z,W,\tau_\gamma) = 0$, we have that $\Cov(\B_\subss, \B_{\subss'}) = \E \Big\{ \Cov(\B_\subss, \B_{\subss'} | S, Z, W, \tau_\gamma ) \Big\}$. 
Remember from \cref{app_eq:cond_independece} that, when at least one of $k,j_k,d$ are different, $\betakj\tod \independent \betakj[F][F]\tod[F] | S, Z, W, \tau_\gamma$. However, that is not true when $(k,j_k,d) = (k',j_k',d')$, even if $\subss \neq \subss$. We write
\begin{align*}
\E \Big\{ \Cov(\B_\subss, \B_{\subss'} | S, Z, W, \tau_\gamma ) \Big\}
&= \E\Big\{ \Cov \Big(\sum_{d = 1}^D \prod_{k = 1}^K \betakj\tod, \sum_{d = 1}^D \prod_{k = 1}^K \betakj[T][F]\tod \Big | S, Z, W, \tau_\gamma) \Big\} \\
&= \sum_{d,d'=1}^D \E\Big\{ \Cov \Big( \prod_{k = 1}^K \betakj\tod, \prod_{k = 1}^K \betakj[T][F]\tod[F] | S, Z, W, \tau_\gamma \Big) \Big\}.
\end{align*}
However,
\begin{align*}
\Cov\Big(\prod_{k = 1}^K & \betakj\tod, \prod_{k = 1}^K \betakj[T][F]\tod[F] | S,Z,W,\tau_\gamma \Big) = \\ & \E\Big(\prod_{k = 1}^K \betakj\tod \betakj[T][F]\tod[F] | S,Z,W,\tau_\gamma \Big) -
\E\Big(\prod_{k = 1}^K \betakj\tod | S,Z,W,\tau_\gamma \Big)
\E\Big(\prod_{k = 1}^K \betakj[T][F]\tod[F] | S,Z,W,\tau_\gamma \Big) = \\
& \E\Big(\prod_{k = 1}^K \betakj\tod \betakj[T][F]\tod[F] | S,Z,W,\tau_\gamma \Big),
\end{align*}
where the last equation holds because the $\betakj\tod$ are independent of each other across $k$ conditional on $S, Z, W, \tau_\gamma$ and have mean zero. Furthermore, since the $\betakj\tod$ are conditionally independent across $d$, we have that for $d \neq d'$, $\Cov\Big(\prod_{k = 1}^K \betakj\tod, \prod_{k = 1}^K \betakj[T][F]\tod[F] | S,Z,W,\tau_\gamma \Big) = 0$. So we only need to study the conditional covariance for $d = d'$. For $\Gamma$ representing the set of all $\gammakj\tod$, we write
$$
\E\Big(\prod_{k = 1}^K \betakj\tod \betakj[T][F]\tod | S,Z,W,\tau_\gamma \Big)
= \E \Big\{ \E \Big( \prod_{k = 1}^K \betakj\tod \betakj[T][F]\tod | \Gamma, S,Z,W,\tau_\gamma \Big) | S,Z,W,\tau_\gamma \Big\}.
$$
Conditional on $\Gamma, S,Z,W,\tau_\gamma$, and as long as $\subss \neq \subss'$, the $\betakj\tod$ are independent across all indices, even if they have all of $k,j_k,d$ common, leading to
\begin{align*}
\E\Big(\prod_{k = 1}^K \betakj\tod \betakj[T][F]\tod | S,Z,W,\tau_\gamma \Big)
&= \E \Big\{\prod_{k = 1}^K \E \Big( \betakj\tod | \Gamma, S,Z,W,\tau_\gamma \Big) \E \Big( \betakj[T][F]\tod | \Gamma, S,Z,W,\tau_\gamma \Big) | S,Z,W,\tau_\gamma \Big\} \\
&= \E \Big( \prod_{k = 1}^K \gammakj\tod \gammakj[T][F]\tod | S,Z,W,\tau_\gamma \Big) \\
&= \prod_{k = 1}^K \E \big( \gammakj\tod \gammakj[T][F]\tod | S,Z,W,\tau_\gamma \big) \\
&= \prod_{k: j_k = j_k'} \E \Big( \big(\gammakj\tod \big)^2 | S,Z,W,\tau_\gamma \Big)  \\
& \hspace{40pt} \times \prod_{k:j_k \neq j_k'}  \E \big( \gammakj\tod | S,Z,W,\tau_\gamma \big) \E \big( \gammakj[T][F]\tod | S,Z,W,\tau_\gamma \big) \\
&= 0
\end{align*}
where the first equality holds because $\subss \neq \subss'$, the third equality holds because the $\gammakj\tod$ are conditionally independent across $k$, and the fourth equality holds because they are conditionally independent across $j_k$.

\end{proof}

\renewcommand*{\proofname}{\textbf{Proof of \cref{theory:prior_targets}}}
\begin{proof}
We want $\Var(B_\subss) = V^*$ and $AV = AV^*$. The second target will be achieved when $\Var(\B_\subss) / \Var^{hard}(\B_\subss) = (1 - AV^*)^{-1}$. Since $\addvar$ is the quantity driving the soft \parafac's additional variability we use this condition to acquire a form for $\addvar$ as a function of the remaining hyperparameters.
\begin{align*}
& \frac{\Var(\B_\subss)}{\Var^{hard}(\B_\subss)} \\
&=
\frac{\sum_{l = 0}^2 \frac{\rho_l}{b_\tau^l} \binom{2}{l} \big\{ \lambdafrac \big\}^l \big( \addvar \big)^{2 -l}}
{ \frac{\rho_2}{b_\tau^2} \big\{ \lambdafrac \big\}^2} \\
&= \sum_{l = 0}^2 \binom{2}{l} \frac{\rho_l}{\rho_2} \Big\{ \lambdafrac[T] \Big\}^{l - 2} \Big( \addvar \Big)^{2 - l} \\
&= \frac1{\rho_2} \Big\{ \lambdafrac[T] \Big\}^{- 2} \Big( \addvar \Big)^2 + 2 \frac{\rho_1}{\rho_2} \Big\{ \lambdafrac[T] \Big\}^{-1} \addvar + 1 \\
&= \frac1{a_\tau(a_\tau + 1)} \Big\{ \lambdafrac[T] \Big\}^{- 2} \Big( \addvar \Big)^2 + \frac2{a_\tau + 1} \Big\{ \lambdafrac[T] \Big\}^{-1} \addvar + 1
\end{align*}
Therefore, in order for $\Var(\B_\subss) / \Var^{hard}(\B_\subss) = (1 - AV^*)^{-1}$, $\addvar$ is the solution to a second degree polynomial. We calculate the \textit{positive} root of this polynomial.
\begin{align*}
\Delta & = \frac{4}{(a_\tau + 1)^2} \Big\{ \lambdafrac[T] \Big\}^{-2} - \frac4{a_\tau(a_\tau + 1)} \Big\{ \lambdafrac[T] \Big\}^{- 2} \big(1 - (1 - AV^*)^{-1}\big) \\
& = \frac{4}{(a_\tau + 1)^2} \Big\{ \lambdafrac[T] \Big\}^{-2} \Big[ 1 - \frac{a_\tau + 1}{a_\tau}\Big\{1 - \big(1 - AV^*)^{-1} \Big\} \Big] > 0.
\end{align*}
Since $\addvar$ is positive, we have that
\begin{align*}
\addvar &= \frac{-\frac2{a_\tau + 1} \Big\{ \lambdafrac[T] \Big\}^{-1} + \sqrt{\frac{4}{(a_\tau + 1)^2} \Big\{ \lambdafrac[T] \Big\}^{-2} \Big[ 1 - \frac{a_\tau + 1}{a_\tau}\Big\{1 - \big(1 - AV^*)^{-1} \Big\} \Big]}}
{\frac2{a_\tau(a_\tau + 1)} \Big\{ \lambdafrac[T] \Big\}^{- 2}} \\
&= \frac{- 1 + \sqrt{ 1 - \frac{a_\tau + 1}{a_\tau}\Big\{1 - \big(1 - AV^*)^{-1} \Big\}}}
{\frac1{a_\tau} \Big\{ \lambdafrac[T] \Big\}^{- 1}} \\
&= \frac{a_\tau}{b_\tau} \lambdafrac \bigg\{
\sqrt{  1 - \frac{a_\tau + 1}{a_\tau} \big\{1 - \big(1 - AV^*)^{-1} \big\}} - 1 \bigg\} \numberthis{} \label{proof_eq:add_var}.
\end{align*}
Denoting \( \displaystyle \xi = 1 - \frac{a_\tau + 1}{a_\tau} \big\{1 - \big(1 - AV^*)^{-1} \big\} \) and substituting the form of $\addvar$ in $\Var(\B_\subss)$ we have that
\begin{align*}
\Var(\B_\subss) &= C \sum_{l = 0}^2 \binom{2}{l} \frac{\rho_l}{b_\tau^l} \Big\{ \lambdafrac \Big\}^2 \Big(\frac{a_\tau}{b_\tau} \Big)^{2 - l} \Big( \sqrt{ \xi } - 1 \Big) ^{2 - l} \\
&= C \Big\{ \lambdafrac \Big\}^2 \Big(\frac{a_\tau}{b_\tau} \Big)^2 \sum_{l = 0}^2 \binom{2}{l} \frac{\rho_l}{a_\tau^l} \Big( \sqrt{\xi} - 1 \Big) ^{2 - l}\\
&= C \Big\{ \lambdafrac \Big\}^2 \Big(\frac{a_\tau}{b_\tau} \Big)^2 
\Big\{ \Big( \sqrt{\xi} - 1 \Big)^2 + 2 \Big(\sqrt{\xi} - 1 \Big) + 1 + \frac1{a_\tau} \Big\} \\
&= C \Big\{ \lambdafrac \Big\}^2 \Big(\frac{a_\tau}{b_\tau} \Big)^2 
\Big\{ \xi + \frac1{a_\tau} \Big\}
\end{align*}
Also,
\begin{align*}
\xi + \frac1{a_\tau} &=  1 - \frac{a_\tau + 1}{a_\tau} \big\{1 - \big(1 - AV^* \big)^{-1} \big\} + \frac1{a_\tau} =
 \frac{a_\tau + 1}{a_\tau} \big(1 - AV^*)^{-1} 
\end{align*}
leading to
\begin{align*}
\Var(\B_\subss) = V^* \iff \lambdafrac = \frac{b_\tau}{a_\tau} \sqrt{\frac{V^*(1 - AV^*)a_\tau}{C(a_\tau + 1)}}.
\numberthis{} \label{proof_eq:lambda_frac}
\end{align*}
Substituting \cref{proof_eq:lambda_frac} back into \cref{proof_eq:add_var}, we have that
$$
\addvar = \sqrt{\frac{V^*(1 - AV^*)a_\tau}{C(a_\tau + 1)}} \bigg\{
\sqrt{  1 - \frac{a_\tau + 1}{a_\tau} \big\{1 - \big(1 - AV^*)^{-1} \big\}} - 1 \bigg\}.
$$
\end{proof}

\renewcommand*{\proofname}{\textbf{Proof of \cref{prop:prior_support}}}
\begin{proof}
Start by noting that 
$$ \pi_{\B}(\neigh) = \E_{\Gamma, S, Z} \Big[ p \Big(\B : \max_{\subss} |\B_\subss^0 - \B_\subss| < \epsilon \Big| \Gamma, S, Z \Big) \Big], $$
where $\Gamma, S, Z$ are as defined in the proof of \cref{theory:variance_B}. Then, take $\epsilon^* = \sqrt[K]{\epsilon / (2(D - 1))}$ and write
\begin{align*}
p \Big(\B : \max_{\subss} |\B_\subss^0 - \B_\subss| < \epsilon \Big| \Gamma, S, Z \Big) & \geq
p \Big(\B : \max_{\subss} |\B_\subss^0 - \B_\subss| < \epsilon \Big| \Gamma, S, Z  \big\{|\betakj\tod| <\epsilon^*, \text{all } k, \subss, \text{ and }  d \geq 2 \big\} \Big) \\
& \hspace{40pt} \times p \Big( |\betakj\tod| <\epsilon^*, \text{all } k, \subss, \text{ and }  d \geq 2 \big\} | \Gamma, S, Z \Big).
\end{align*}
Conditional on $\Gamma,S,Z$, $\betakj\tod$ are independent normal variables with positive weight in an $\epsilon^*$-neighborhood of 0, implying that
$ p \Big( |\betakj\tod| <\epsilon^*, \text{all } k, \subss, \text{ and }  d \geq 2 \big\} | \Gamma, S, Z \Big) > 0. $

Remember from \cref{eq:softB} that
$\B = \sum_{d = 1}^D \B_1\tod \hadamard \B_2\tod \hadamard \dots \hadamard \B_K\tod$, and denote $\B\tod = \B_1\tod \hadamard \B_2\tod \hadamard \dots \hadamard \B_K\tod$. Then,
$\B_\subss = \B_\subss^{(1)} + \B_\subss^{(2)} + \dots + \B_\subss^{(D)}$. Note that
\begin{align*}
p \Big(\B : & \max_{\subss} |\B_\subss^0 - \B_\subss| <  \epsilon \Big| \Gamma, S, Z  \big\{|\betakj\tod| <\epsilon^*, \text{all } k, \subss, \text{ and }  d \geq 2 \big\} \Big) \\
& \geq p \Big(\B : \max_{\subss} |\B_\subss^0 - \B_\subss^{(1)}| < \epsilon / 2 \Big| \Gamma, S, Z  \big\{|\betakj\tod| <\epsilon^*, \text{all } k, \subss, \text{ and }  d \geq 2 \big\} \Big) \\
& = p \Big(\B : \max_{\subss} |\B_\subss^0 - \B_\subss^{(1)}| < \epsilon / 2 \Big| \Gamma, S, Z \Big),
\end{align*}
where the equality holds because the entries of $\B^{(1)}$ are independent of all $\betakj\tod$ for $d \geq 2$ conditional on $\Gamma,S,Z$,
and the inequality holds because $|\B_\subss^0 - \B_\subss^{(1)}| < \epsilon / 2$ and $|\betakj\tod| < \epsilon^*$ for $d \geq 2$ implies that
\begin{align*}
|\B_\subss^0 - \B_\subss| &= |\B_\subss^0 - \B_\subss^{(1)} - \B_\subss^{(2)} - \dots - \B_\subss^{(D)}| \\
&\leq |\B_\subss^0 - \B_\subss^{(1)}| + |\B_\subss^{(2)}| + \dots + |\B_\subss^{(D)}| \\
& < \epsilon / 2 + (D-1)(\epsilon^*)^K = \epsilon.
\end{align*}

Since all of $\betakj^{(1)}$ are independent conditional on $\Gamma, S,Z$, we have that
$$
p \Big(\B : \max_{\subss} |\B_\subss^0 - \B_\subss^{(1)}| < \epsilon / 2 \Big| \Gamma, S, Z \Big) =
\prod_{\subss} p \Big(|\B_\subss^0 - \B_\subss^{(1)}| < \epsilon / 2 \Big| \Gamma, S, Z \Big) > 0,
$$
since all entries $\B_\subss^{(1)}$ are products of draws from $K$ normal distributions and therefore assign positive weight in all $\mathbb{R}$, including the $\epsilon /2 -$neighborhood of $\B_\subss^0$.

Putting all of this together, we have the desired result that $\pi_{\B}(\neigh) > 0$.
\end{proof}

\renewcommand*{\proofname}{\textbf{Proof of \cref{prop:consistency}}}

\begin{proof}
We show that for any $\epsilon > 0$, there exists $\epsilon ^* > 0$ such that
$ \Big\{\B: \max_{\subss} | \B_\subss^0 - \B_\subss| < \epsilon^* \Big\} \subseteq
\Big\{\B: KL(\B^0, \B) < \epsilon \Big\}, $
where
$$
KL(\B_0, \B) = \int \log \frac{\phi(y;\B^0)}{\phi(y;\B)} \phi(y;\B^0) \mathrm{d} y,
$$
and $\phi(y;\B)$ is the density of a normal distribution with coefficient tensor $\B$ and variance 1. If we show this, consistency follows from \cite{Schwartz1965bayes}.

Assume that there exists $M$ such that $|\tensor_\subss| < M$ for all $\subss$ with probability 1.
For two normal distributions with mean $\langle \tensor,  \B^0\rangle_F$ and $\langle \tensor,  \B \rangle_F$ respectively, we have that
\begin{align*}
KL(\B^0, \B) &= \frac12 \Big(\langle \tensor,  \B^0\rangle_F - \langle \tensor,  \B^0\rangle_F \Big)^2 \\
&= \frac12 \Big[ \sum_\subss \big( \B^0_\subss - \B_\subss \big) \tensor_\subss \Big]^2 \\
& \leq \frac12 \Big[ \sum_\subss \big( \B^0_\subss - \B_\subss \big)^2 \Big] \Big[ \sum_\subss \tensor_\subss^2 \Big]
\end{align*}
Take $\epsilon^* = \sqrt{2\epsilon}/ \big(M (p_1p_2\dots p_K)^2 \big)$ and consider $\B \in \neigh[*]$. We will show that $\B$ satisfies that $KL(\B^0, \B) < \epsilon$ completing the proof. Note first that
\begin{align*}
\sum_\subss \big( \B^0_\subss - \B_\subss \big)^2
\leq p_1p_2\dots p_k \max_\subss \big( \B^0_\subss - \B_\subss \big)^2 < p_1p_2\dots p_k (\epsilon^*)^2 = \frac{2\epsilon}{M p_1p_2\dots p_K},
\end{align*}
and since $|\tensor_\subss| \leq M$ we have that $\sum_\subss \tensor_\subss^2 \leq M p_1p_2\dots p_K$. Putting these results together
$$
KL(\B^0, \B) < \frac12 \frac{2\epsilon}{M p_1p_2\dots p_K} M p_1p_2\dots p_K = \epsilon.
$$
\end{proof}

\section{Hard \parafac{} error in estimating the true matrix for an increasing rank}
\label{app_sec:hard_demonstration}

\begin{figure}[H]
\centering
\includegraphics[width=0.9\textwidth]{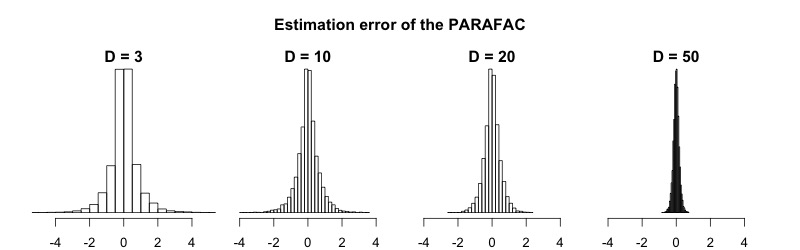}
\caption{Histogram of errors in estimating the entries of $\B$ when $\B$ resembles but is not exactly equal to a rank-1 tensor, and estimation is based on the singular value decomposition using $D\in \{3, 10, 20, 50\}$ factors.}
\label{fig:random_matrix_parafac}
\end{figure}

Due to the block structure and subsequent ``inflexibility'' of the hard \parafac{} approximation, a large number of components $D$ might be required in order to adequately approximate a coefficient tensor $\B$. 
To further demonstrate this, we considered a coefficient matrix $\B$ whose entries $\B_{ij}$ are centered around (but are not equal to) the entries of a rank-1 matrix of the form $\beta_1 \otimes \beta_2$. Therefore, even though the matrix has a somewhat rectangular structure, it is not exactly in that form. Using the singular value decomposition (which is the \parafac{} analog for matrices), we considered the quality of the approximation based on $D$ factors, for various values of $D$. \cref{fig:random_matrix_parafac} shows histograms of the difference of the true entries in $\B$ from the estimated ones. Even for $D = 20$, substantial error remains in estimating the matrix $\B$.

\section{Alternative sampling from the posterior distribution}
\label{app_sec:MCMC}

The full set of parameters is $\bm \theta = \{\mu, \bm \delta, \tau^2, \betakj\tod, \gammakj\tod, \sigma^2_k, \zeta\tod, \wkj\tod, \lambda_k\tod, \tau^2_\gamma, \text{ for all } d, k, j_k, \subss\}$. We use the notation $|\cdot$ and $|\cdot,-y$ to denote conditioning on the data and all parameters, and the data and all parameters but $y$, accordingly. Then, our MCMC updates are:
\begin{itemize}
\item $(\mu, \bm \delta) | \cdot \sim N(\bm \mu^*, \Sigma^*)$, for
$\Sigma^* = (\Sigma_0^{-1} + \widetilde{\bm C}^T \widetilde{\bm C} / \tau^2)^{-1}$, and
$\bm \mu^* = \Sigma^* \widetilde{\bm C}^T \bm R_B / \tau^2$, 
where $\widetilde{\bm C}$ is the $N\times(p+1)$ matrix with $i^{th}$ row equal to $(1, \bm C_i)$, and
$\bm R_B = (Y_1 - \langle \tensor_1, \B\rangle_F, \dots, Y_N - \langle \tensor_N, \B\rangle_F)^T$ is the vector of residuals of the outcome on the tensor predictor.
\item $\tau^2 | \cdot \sim IG(a_\tau + N / 2, b_\tau + \sum_{i = 1}^N (Y_i - \mu - \bm C_i^T \bm \delta - \langle \tensor_i, \B\rangle_F))$.
\item \( \displaystyle \sigma^2_k | \cdot \sim giG(p^*, a^*, b^*)\), for
$p^* = a_\sigma - D\prod_{k = 1}^K p_k / 2$, $a^* = 2 b_\sigma$, and $b^* = \sum_{d, \subss} (\betakj\tod - \gammakj\tod)^2 / \zeta\tod$. As a reminder, $X \sim giG(p, a, b)$ if $p(x) \propto x^{p-1}\exp\{- (ax + b/x) / 2 \} $. 
\item $\gammakj\tod | \cdot \sim N(\mu^*, \sigma^{2*})$, for {\small \( \displaystyle \sigma^{2*} = \Big\{(\tau_\gamma \wkj\tod \zeta\tod)^{-1} + \big(\sum_{l \neq k} p_l \big) / (\sigma^2_k\zeta\tod) \Big\}^{-1}, \mu^* = \sigma^{2*} \Big\{ \sum_{\subss:\subss_k = j_k} \betakj\tod / (\sigma^2_k\zeta\tod)\Big\}\)}.
\item $\tau_\gamma|\cdot \sim giG \big(a_\tau - D\sum_k p_k / 2, 2 b_\tau, \sum_{d,k,j_k} (\gammakj\tod)^2 / (\zeta\tod \wkj\tod) \big)$.
\item $\wkj\tod |\cdot \sim giG(1/2, \lambda_k^2, (\gammakj\tod)^2 / (\tau_\gamma \zeta\tod))$.
\item $[ \lambda_k\tod |\cdot, - \wkj\tod, \text{all } j_k ] \sim \Gamma(a_\lambda + p_k, b_\lambda + \sum_{j_k} |\gammakj\tod| / (\tau_\gamma\zeta\tod))$. Therefore, $\lambda_k\tod$ is updated conditional on all parameters excluding \textit{all} $\wkj\tod$, $j_k = 1, 2, \dots, p_k$. Its distribution can be acquired by noting that $\gammakj\tod | \tau_\gamma, \zeta\tod, \lambda_k\tod \sim DE(\mu = 0, b = \tau_\gamma\zeta\tod / \lambda_k\tod)$ \citep{Park2008}, where $DE$ stands for double exponential or Laplace distribution.
\item for each $k = 1, 2, \dots, K$, $d = 1, 2, \dots, D$ and $j_k = 1, 2, \dots, K$, we use $\B_{k,j_k}\tod$ to denote the $j_k^{th}$ slice of tensor $\B_k\tod$ along mode $k$, which is a $(K-1)$-mode tensor. Then,
$\vect[\B_{k,j_k}\tod] |\cdot \sim N(\bm \mu^*, \Sigma^*)$, for
$\Sigma^* = \big(\Sigma_\pi ^{-1} + \big(\sum_{i = 1}^N \Psi_i\Psi_i^T \big) / \tau^2\big)^{-1}$, and
$\bm \mu^* = \Sigma^* \big(\Sigma_\pi^{-1}\bm \mu_\pi + \big(\sum_{i = 1}^N \Psi_i R_{i,\Psi}\big) / \tau^2\big)$,
where
\begin{itemize}
    \item $\Sigma_\pi$ is a diagonal matrix of dimension $\big(\prod_{k = 1}^K p_k\big)/p_k$ with repeated entry $\sigma^2_k\zeta\tod$,
    \item $\bm \mu_\pi$ is a constant vector of length $\big(\prod_{k = 1}^K p_k\big)/p_k$ with entry $\gammakj\tod$,
    \item $\Psi_i = \vect[\B_1\tod \hadamard \dots \hadamard \B_{k-1}\tod \hadamard \B_{k+1}\tod \dots \B_K\tod \hadamard \tensor_i]$, and
    \item $R_{i,\Psi} = Y_i - \alpha - \bm C_i^T\delta - \langle \tensor_i, \sum_{r \neq d} \B_1^{(r)} \hadamard \B_2^{(r)} \hadamard \dots \hadamard \B_K^{(r)} \rangle_F$ is the residual excluding component $d$ of the coefficient tensor.
\end{itemize}
\item for each $d = 1, 2, \dots, D$, we update $\zeta\tod$ from its full conditional, and then ensure that $\bm \zeta$ sums to 1, by dividing all its entries with $\sum_{d = 1}^D\zeta\tod$. The $\zeta\tod$ update is from $\zeta\tod |\cdot \sim giG(p^*, a^*, b^*)$, where $p^* = \alpha/D - K(\prod p_k + \sum p_k) / 2$, $a^* = 0$, and $b^* = \sum_{k,\subss} (\betakj\tod - \gammakj\tod)^2 / \sigma^2_k + \sum_{k,j_k} \gammakj^2 / (\tau_\gamma \wkj\tod)$.
\end{itemize}

\section{Additional simulation results}
\label{app_sec:more_sims}

\subsection{Comparing \softer{} and \parafac{} in identifying important entries of tensor predictor}
\label{app_sec:significance_disagree}

\begin{figure}[!t]
\centering
\includegraphics[width = 0.5\textwidth]{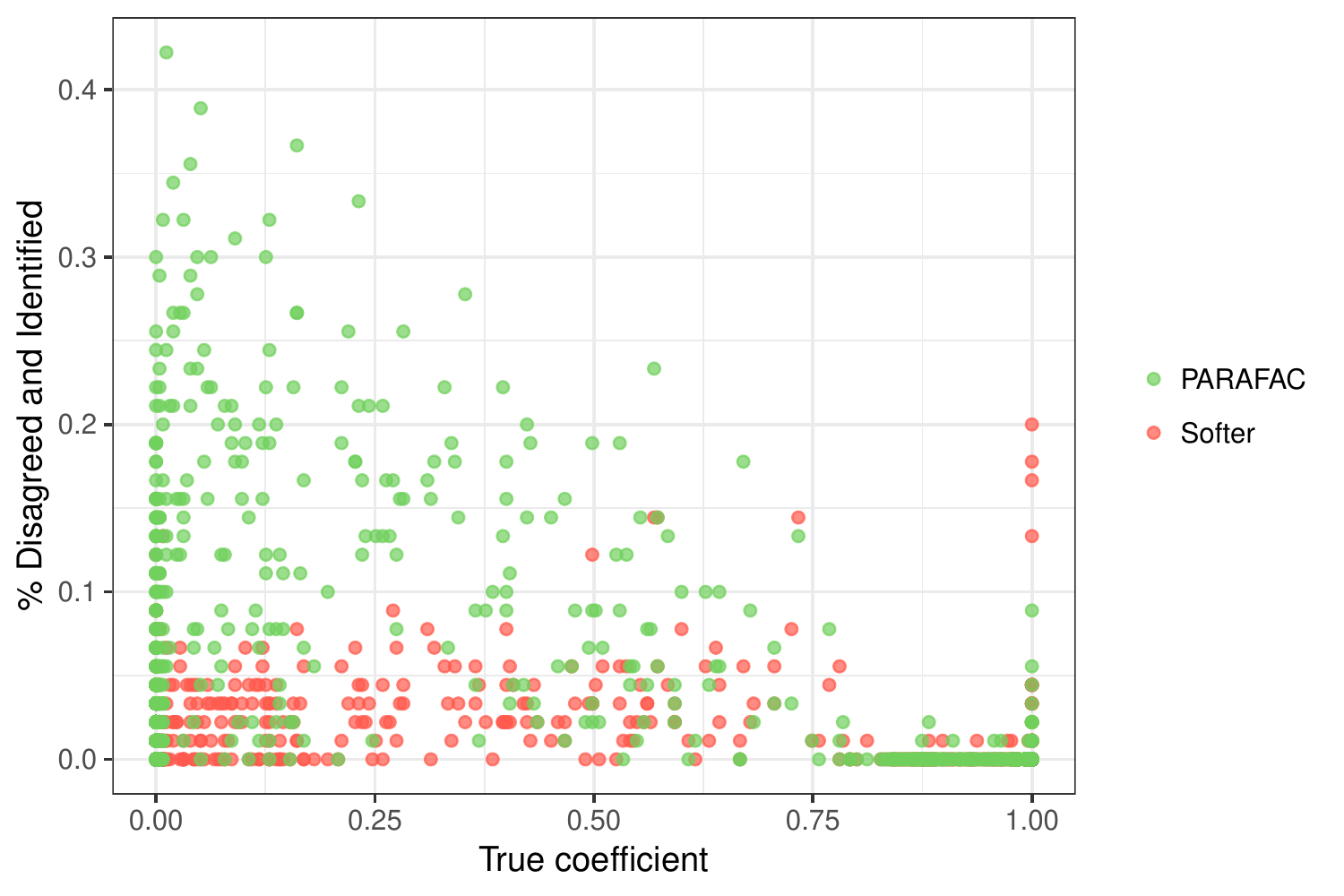}
\hspace{-50pt}
\includegraphics[width = 0.5\textwidth]{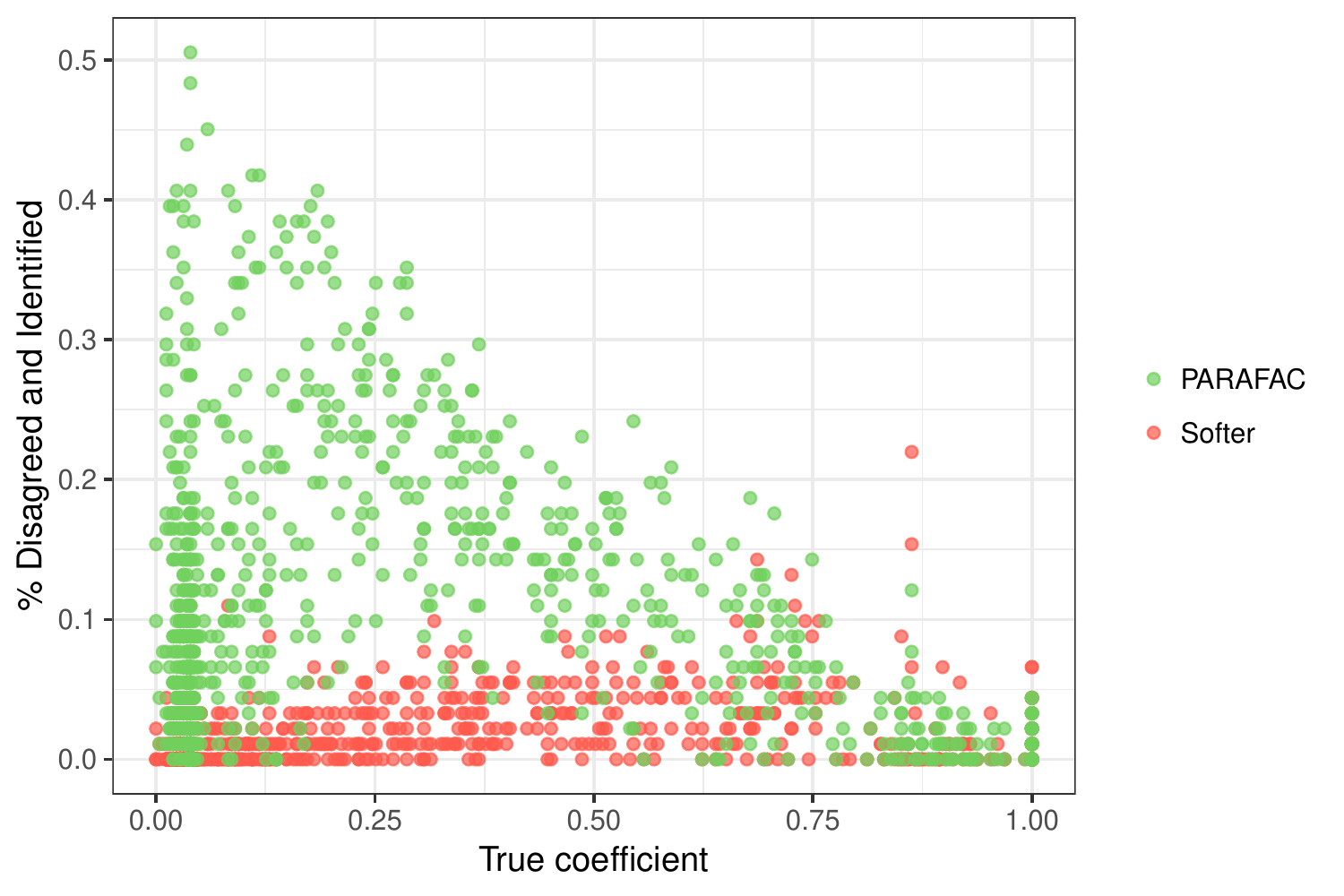}
\caption{Percentage of simulated data sets that an entry with coefficient shown on the x-axis was identified as important by the stated method but not of the other one in the feet (left) and dog (right) scenario respectively.}
\label{app_fig:disagreement}
\vspace{20pt}
\includegraphics[width = 0.3\textwidth]{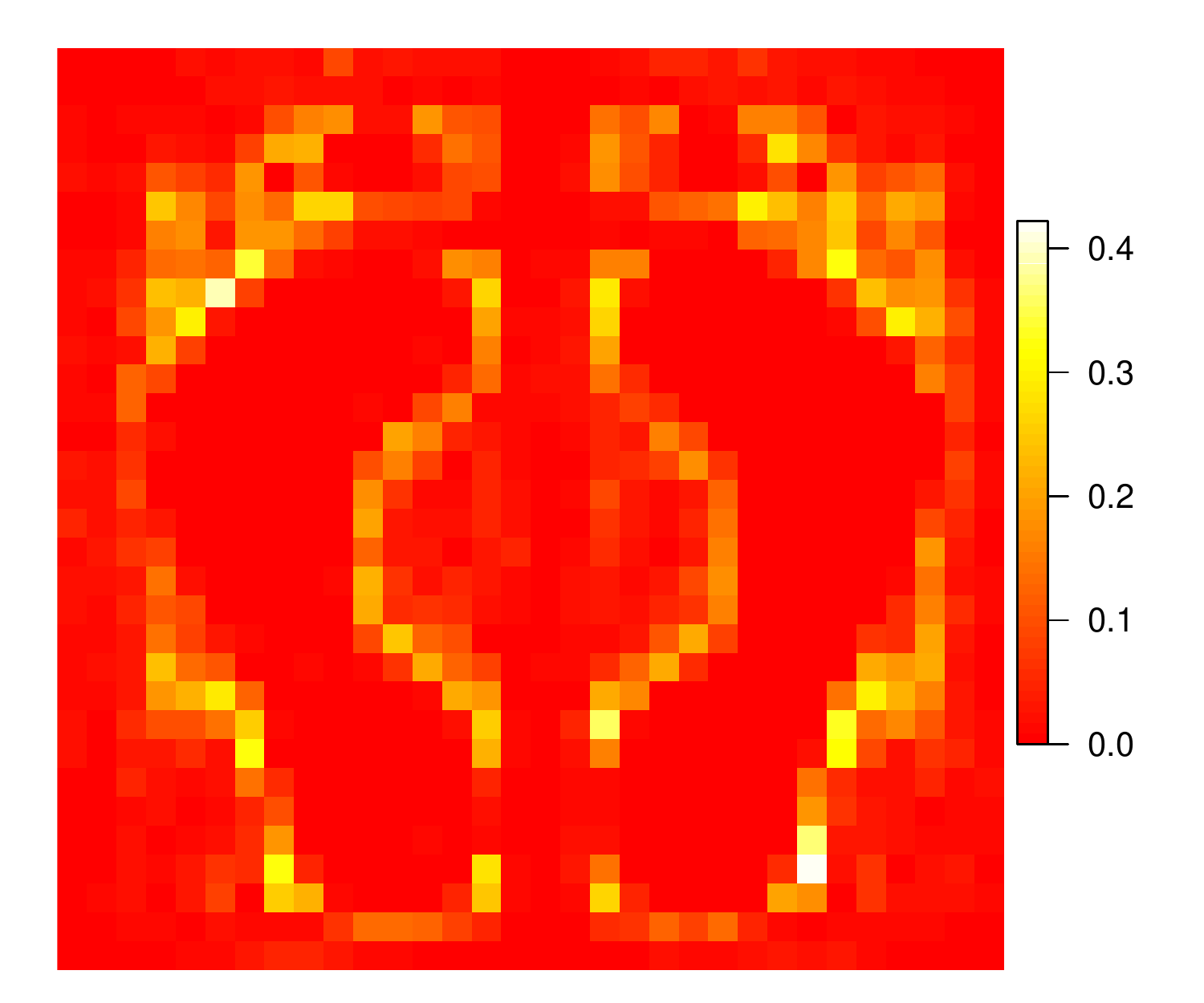}
\includegraphics[width = 0.3\textwidth]{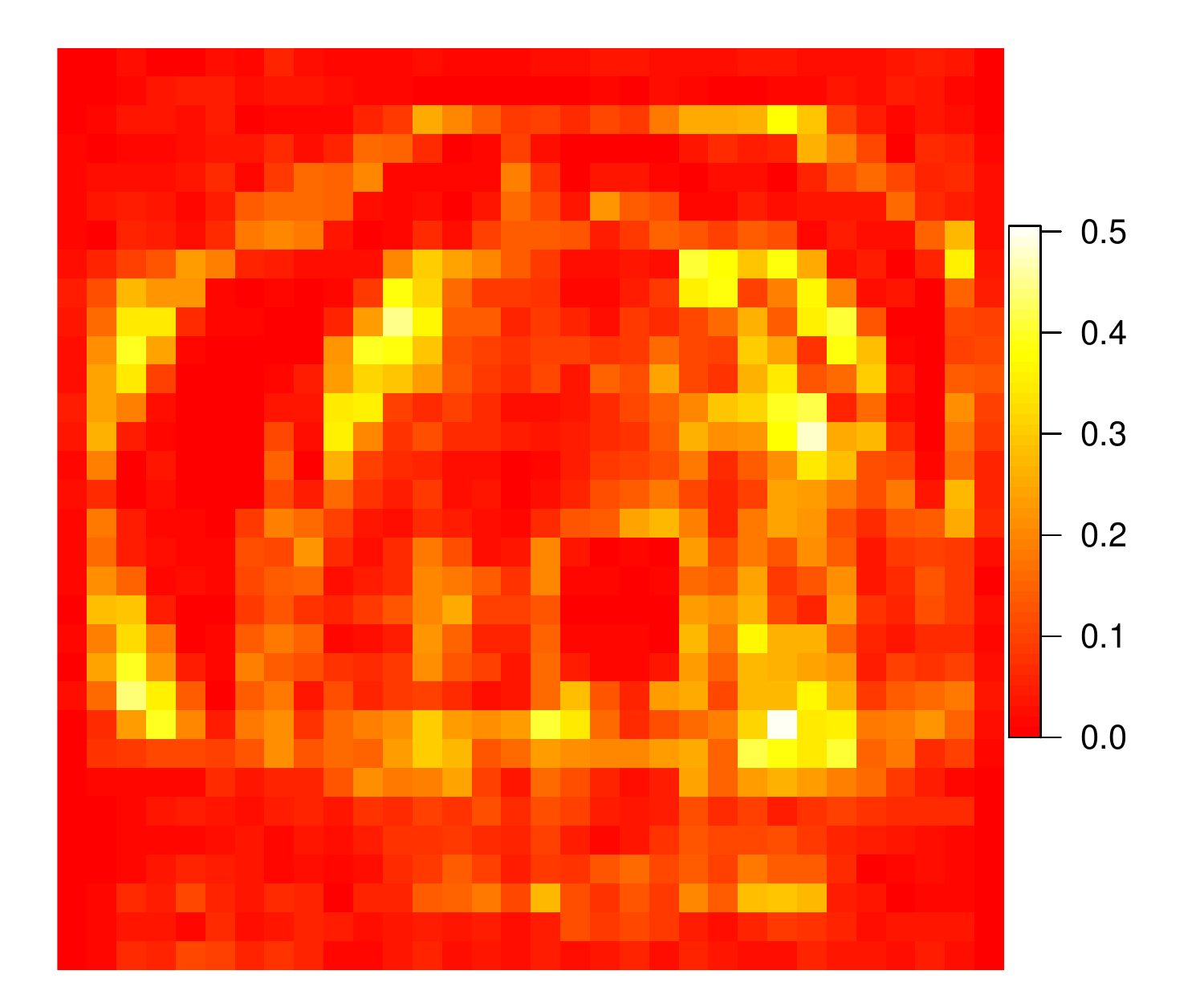}
\caption{Percentage of data sets that an entry was identified as important by hard \parafac{} and not by \softer{} in the feet (left) and dog (right) scenario respectively.}
\label{app_fig:parafac_disagree}
\end{figure}

In \cref{subsec:sims_400} we presented an evaluation of the relative performance of \softer{} and hard \parafac{} in identifying important entries. There, we shows that \softer{} has significant lower FPR indicating that the two methods systematically disagree in the entries of the tensor predictor they identify as important. In order to study their disagreement, for each entry of the tensor predictor we calculate the percentage of data sets for which \softer{} or hard \parafac{} identifies the entry as important while the other does not. We plot the results in Figure \ref{app_fig:disagreement} as a function of the entry's true coefficient. We see that the entries that \softer{} identifies as important and hard \parafac{} does not happen uniformly over the entries' true coefficient. In contrast, when hard \parafac{} identifies an entry as important and \softer{} does not, it is more likely that the coefficient of this entry will be in reality small or zero.

When further investigating this feature of \parafac{}, we identified that the entries that it identifies as significant in disagreement to \softer{} are most often the ones that attribute to the coefficient tensor's block structure. This is evident in Figure \ref{app_fig:parafac_disagree} where we see that the entries with high identification by \parafac{} in contrast to \softer{} are the ones at the boundary of the truly non-zero entries.

\subsection{Simulation results with alternative coefficient tensors}
\label{app_subsec:sims_altern}

\begin{figure}[!t]
\begin{center}
\begin{minipage}{0.2\textwidth}
\includegraphics[width=\textwidth]{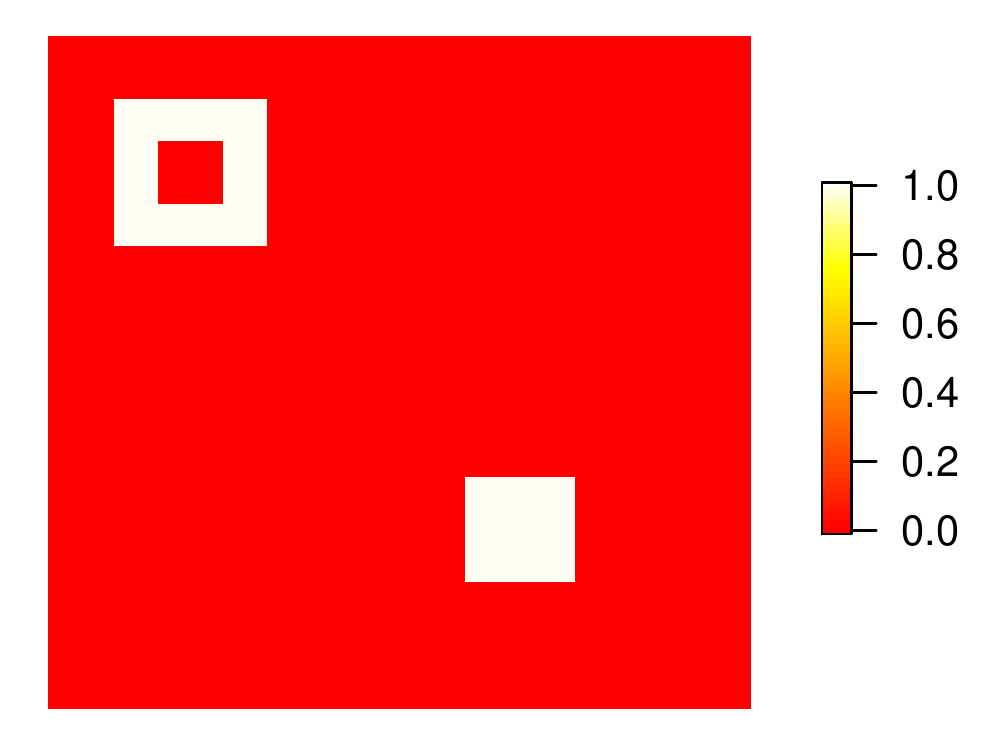} \\
\includegraphics[width=\textwidth]{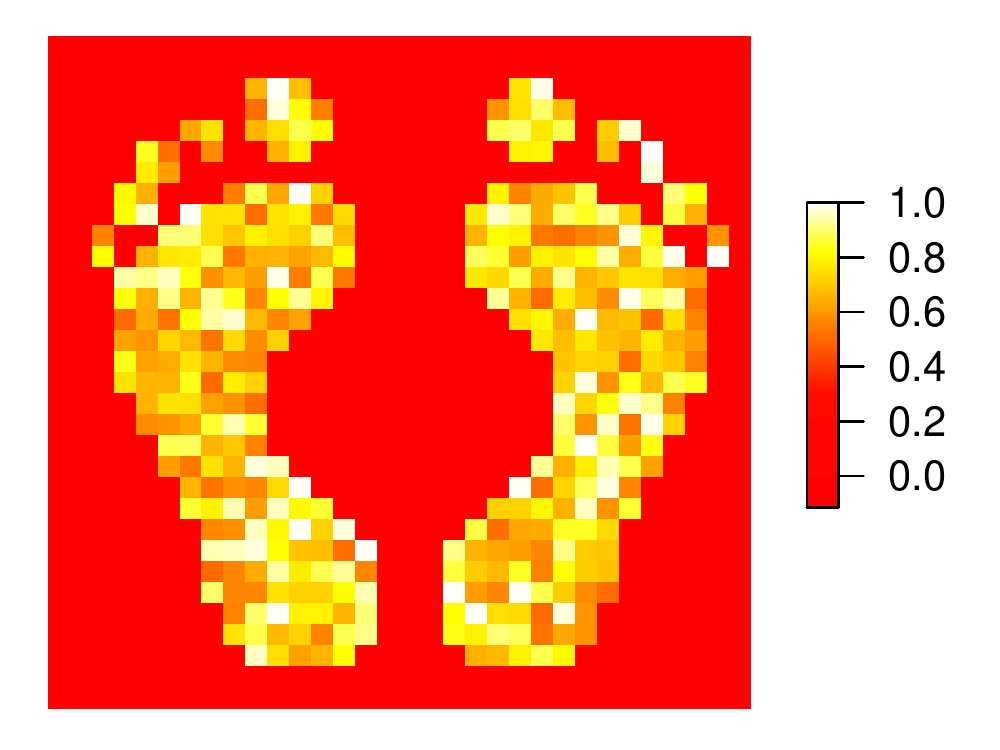} \\
\includegraphics[width=\textwidth]{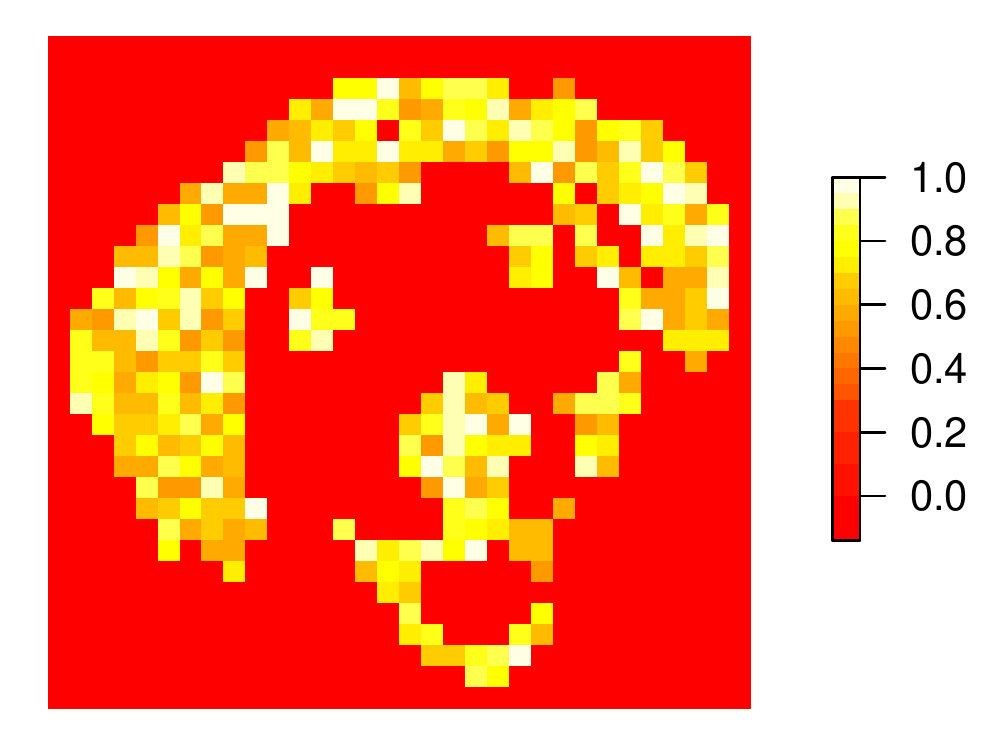}
\end{minipage}
\begin{minipage}{0.2\textwidth}
\includegraphics[width=\textwidth]{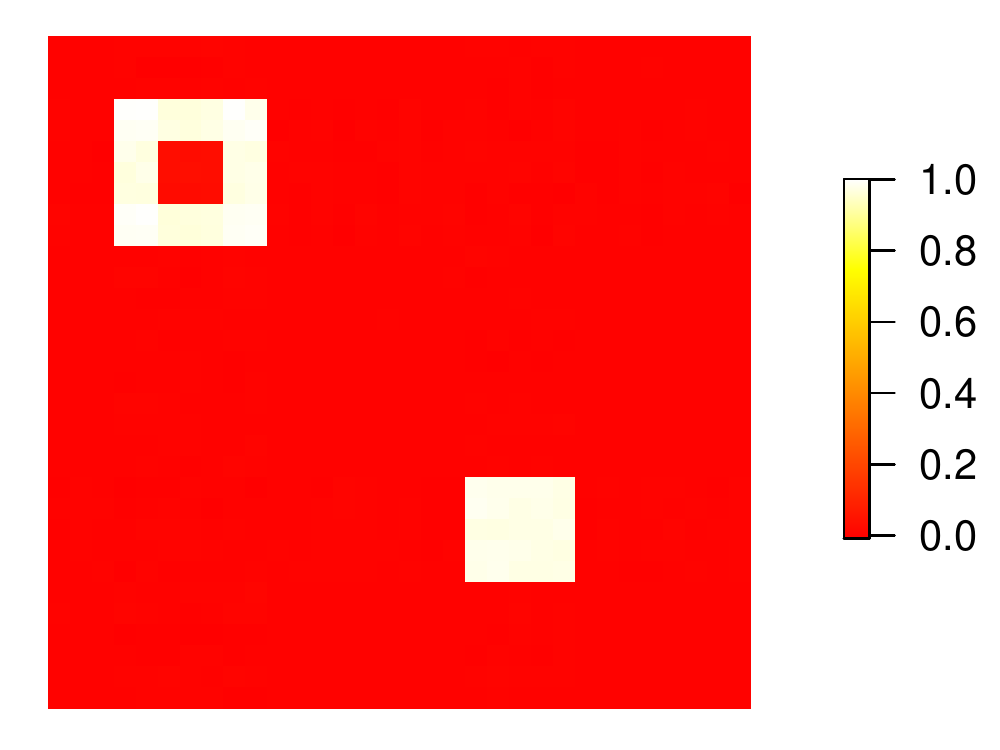} \\
\includegraphics[width=\textwidth]{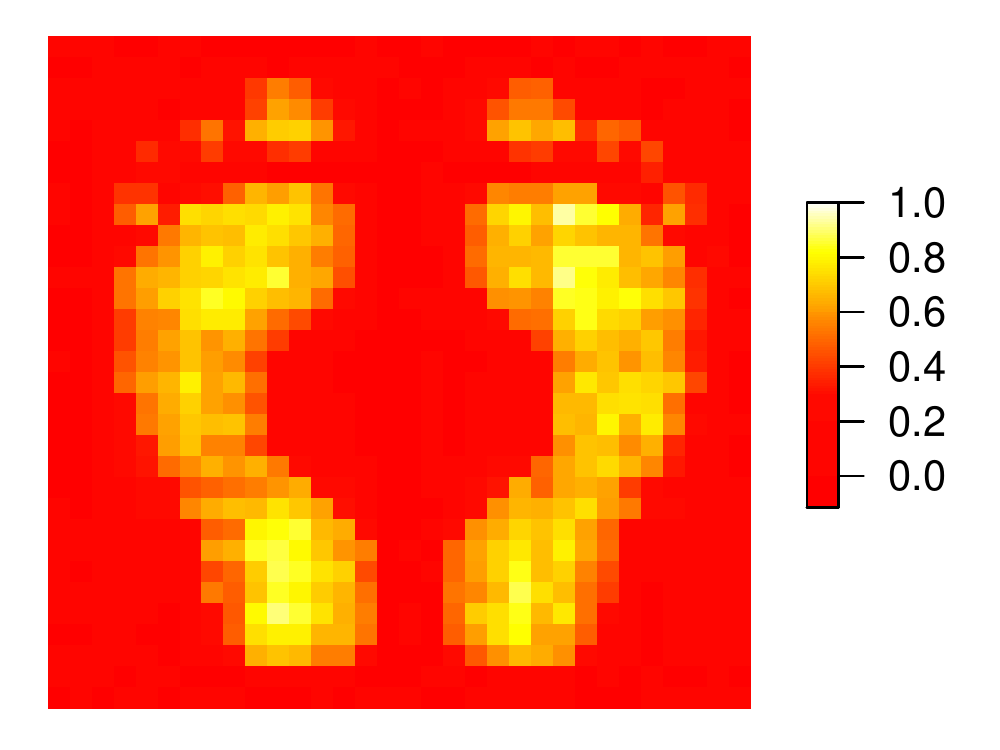} \\
\includegraphics[width=\textwidth]{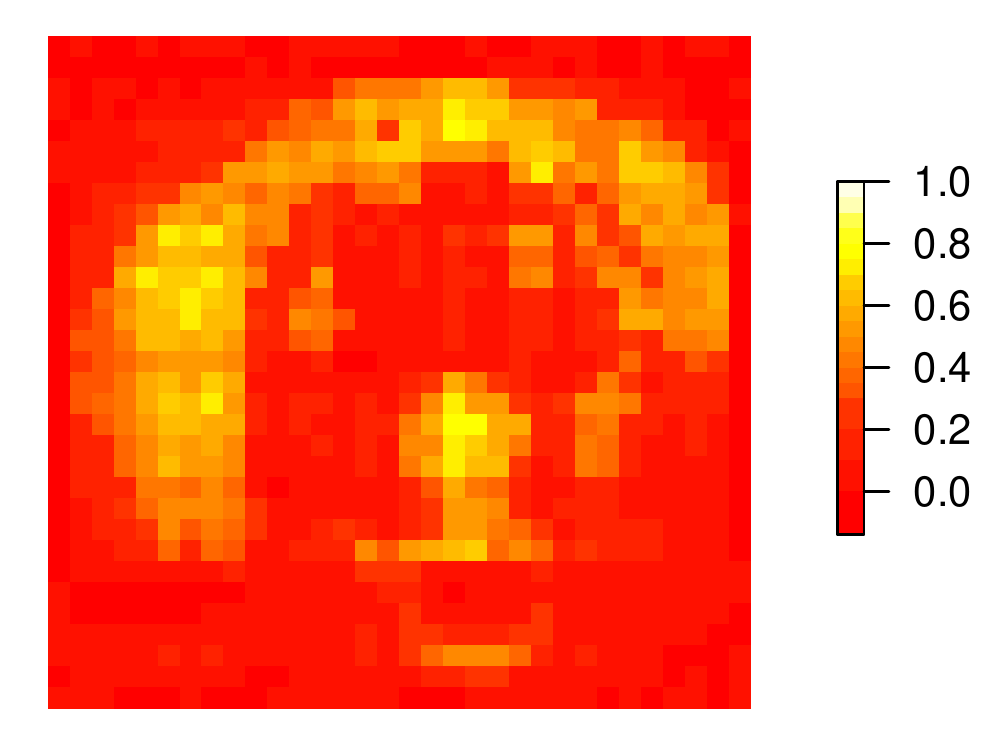}
\end{minipage}
\begin{minipage}{0.2\textwidth}
\includegraphics[width=\textwidth]{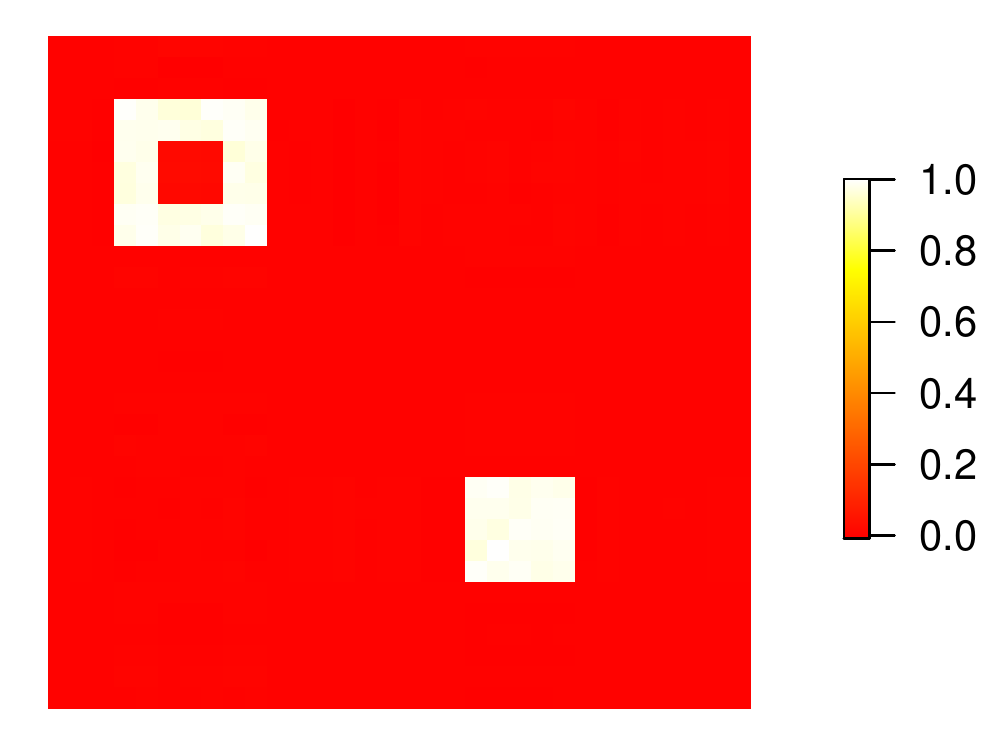} \\
\includegraphics[width=\textwidth]{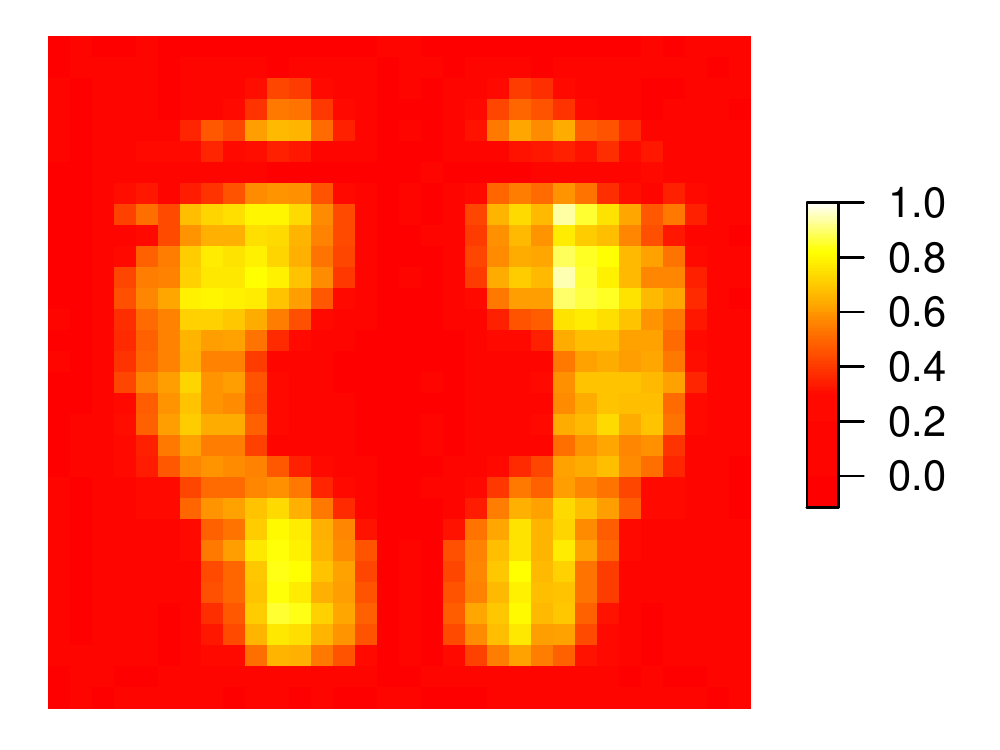} \\
\includegraphics[width=\textwidth]{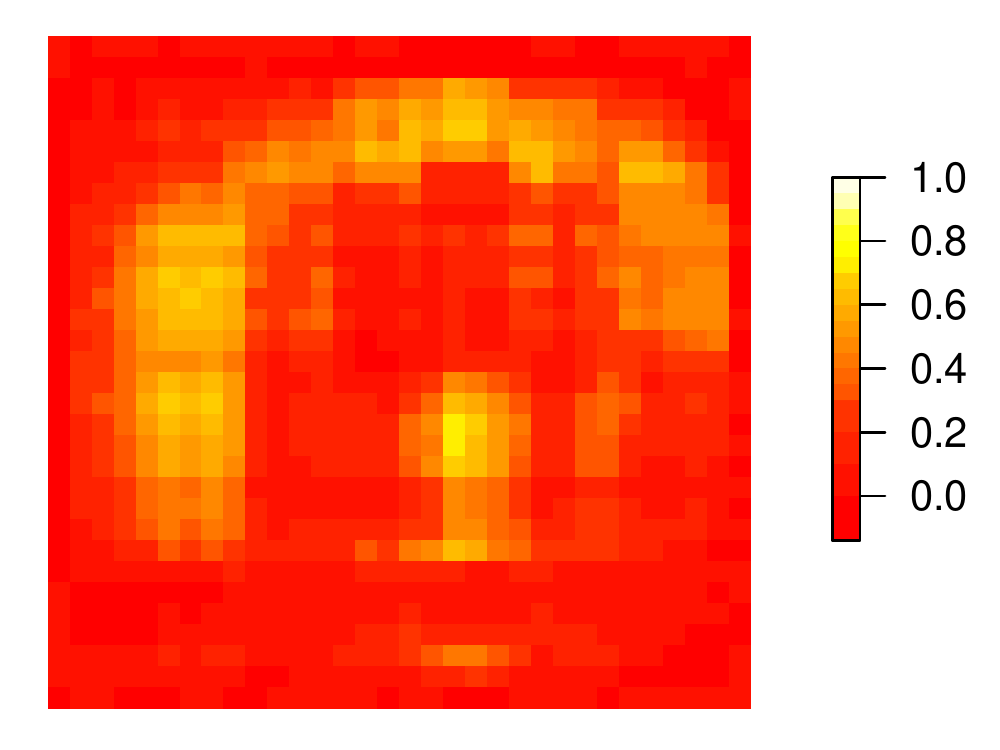}
\end{minipage}
\begin{minipage}{0.2\textwidth}
\includegraphics[width=\textwidth]{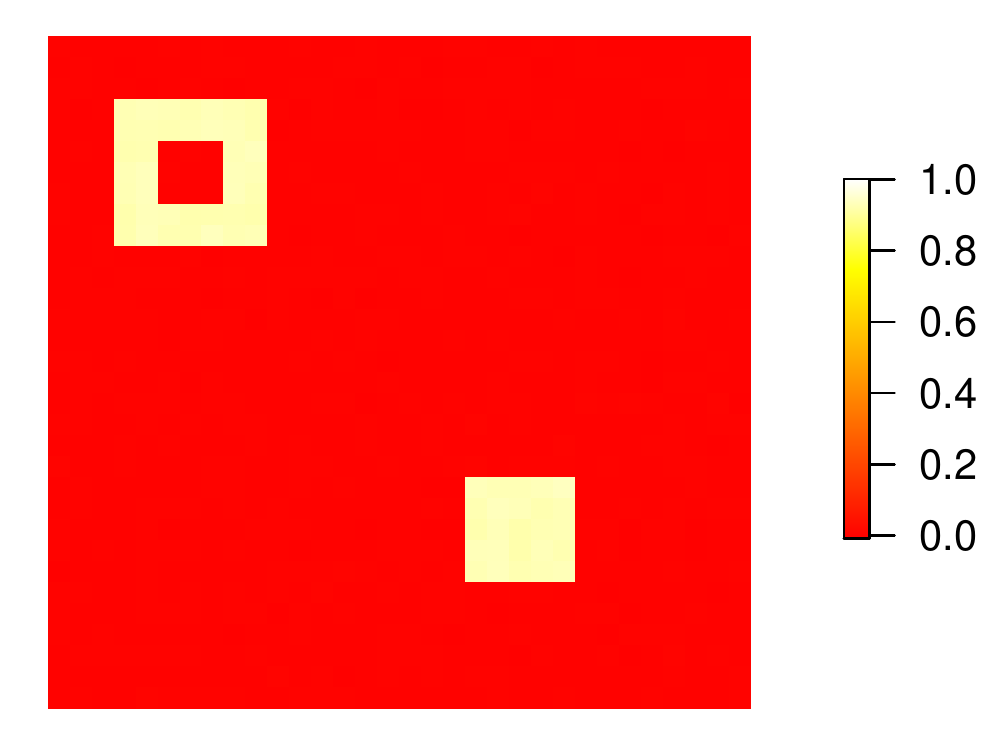} \\
\includegraphics[width=\textwidth]{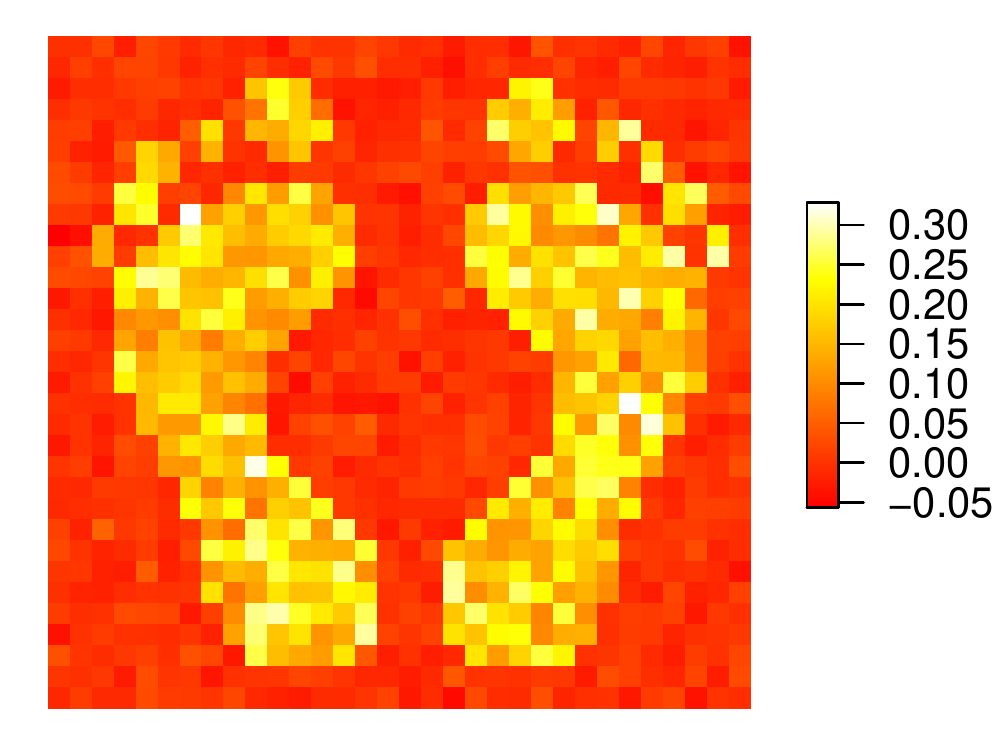} \\
\includegraphics[width=\textwidth]{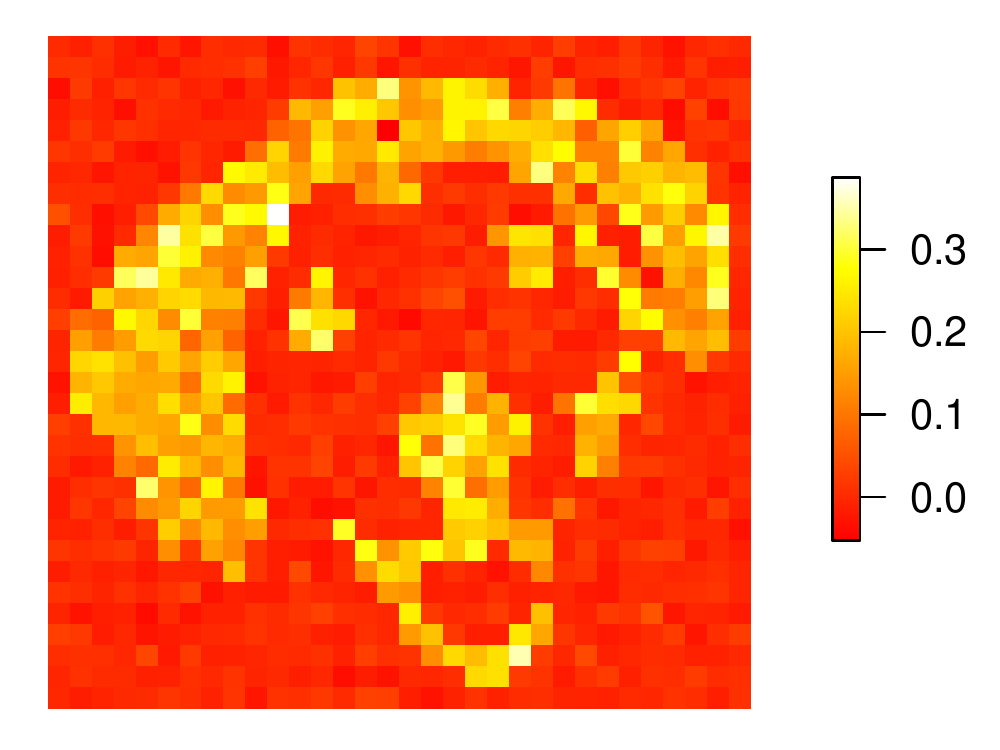}
\end{minipage}

\begin{minipage}{0.17\textwidth}
\subfloat[Truth]{\hspace{\textwidth}}
\end{minipage}
\begin{minipage}{0.22\textwidth}
	\subfloat[\softer{}]{\hspace{\textwidth}}
\end{minipage}
\begin{minipage}{0.2\textwidth}
	\subfloat[\parafac{}]{\hspace{\textwidth}}
\end{minipage}
\begin{minipage}{0.2\textwidth}
	\subfloat[Lasso]{\hspace{\textwidth}}
\end{minipage}
\end{center}
\caption{Simulation results for alternative coefficient matrices. True coefficient matrix and average across simulated data sets of the coefficient matrix posterior mean for \softer{} and hard \parafac{} and the penalized estimator for the Lasso. Note that the color scale is the same for the true, \softer{} and hard \parafac{} approaches, but different for the Lasso because the order of estimated coefficients for the Lasso is much smaller.}
\label{app_fig:sim_pictures}
\end{figure}

For a tensor predictor of dimension $32\times 32$ we also considered three alternative coefficient matrices. The constant squares represent a rank-3, sparse scenario. Both the hard \parafac{} and the Lasso are expected to perform well in this situation, and we are interested in investigating \softer{}'s performance when softening is not necessary.  The varying feet and dog scenarios are scenarios similarly to the dog and feet of the main text but for non-zero entries varying between 0.5 and 1.

\begin{table}[!t]
\centering
\begin{tabular}{lll lll}
& & & \softer{} & \parafac{} & Lasso \\ \hline

constant & Truly zero & bias & 0.0016 & 0.0014 & \textbf{0.0013}  \\
squares & & rMSE & 0.016 & \textbf{0.015} & 0.016  \\
& & coverage & 99.2\% & 99.7\% &  - \\[5pt]

& Truly non-zero & bias & 0.0211  & \textbf{0.017} & 0.073  \\
& & rMSE & \textbf{0.06}  & 0.076 & 0.093  \\
& & coverage & \textbf{90.9\%}  & 79.6\% & - \\[5pt]

& Prediction & MSE & \textbf{0.79}  & 1.25 & 1.32 \\
\hline

varying & Truly zero & bias & 0.06 & 0.076 & \textbf{0.014} \\
feet & & rMSE & \textbf{0.123} & 0.145 & 0.169  \\
& & coverage & 96.4\% & 90.7\% & --  \\[5pt]

& Truly non-zero & bias & \textbf{0.165} & 0.205 & 0.569  \\
& & rMSE & \textbf{0.244} & 0.279 & 0.661 \\
& & coverage & \textbf{87.7}\% & 73.8\% & --  \\[5pt]

& Prediction & MSE & \textbf{40.79} & 55.19 & 194 \\ \hline

varying & Truly zero & bias & 0.071 & 0.091 & \textbf{0.013} \\
dog & & rMSE & 0.159 & 0.176 & \textbf{0.152} \\
& & coverage & 98.3\% & 92.7 & -- \\[5pt]

& Truly non-zero & bias & \textbf{0.263} & 0.321 & 0.554 \\
& & rMSE & \textbf{0.358} & 0.398 & 0.644 \\
& & coverage & \textbf{81.2} & 63.2 & --  \\[5pt]

& Prediction & MSE & \textbf{67.6} & 85.3 & 159.2 \\ \hline
\end{tabular}
\caption{Average bias, root mean squared error, frequentist coverage of 95\% credible intervals among truly zero and truly non-zero coefficient entries, and predictive mean squared error for \softer{}, hard \parafac{} and Lasso for the simulation scenario with tensor predictor of dimensions $32 \times 32$ and sample size $n = 400$. Bold text is used for the approach performing best in each scenario and for each metric.}
\label{app_tab:sims_n400}
\end{table}

Figure \ref{app_fig:sim_pictures} shows the true and average estimated coefficient matrices in these additional simulations. Even though we plot the true, \softer, and hard \parafac{} matrices using a common scale, we plot the expected coefficients employing Lasso using a different color scale. That is because, we want to show that Lasso gets the correct structure, on average, but largely underestimates coefficients due to the assumption of sparsity. Further, \cref{app_tab:sims_n400} reports the average absolute bias, root mean squared error and 95\% coverage of the truly zero, and truly non-zero coefficients, and the prediction mean squared error. When the true underlying hard \parafac{} structure is correct, \softer{} is able to revert back to its hard version, as is evident by the simulation results for the constant squares coefficient matrix. Further, \softer{} performs better than the hard \parafac{} for the varying feet and varying dog scenarios. In all three scenarios, \softer{} has the best out-of-sample predictive ability.


\subsection{Simulation results for $32 \times 32$ tensor predictor and sample size $n = 200$}
\label{app_sec:sims_n200}

\begin{figure}[!t]
\centering
\begin{minipage}{0.21\textwidth}
\includegraphics[width=\textwidth]{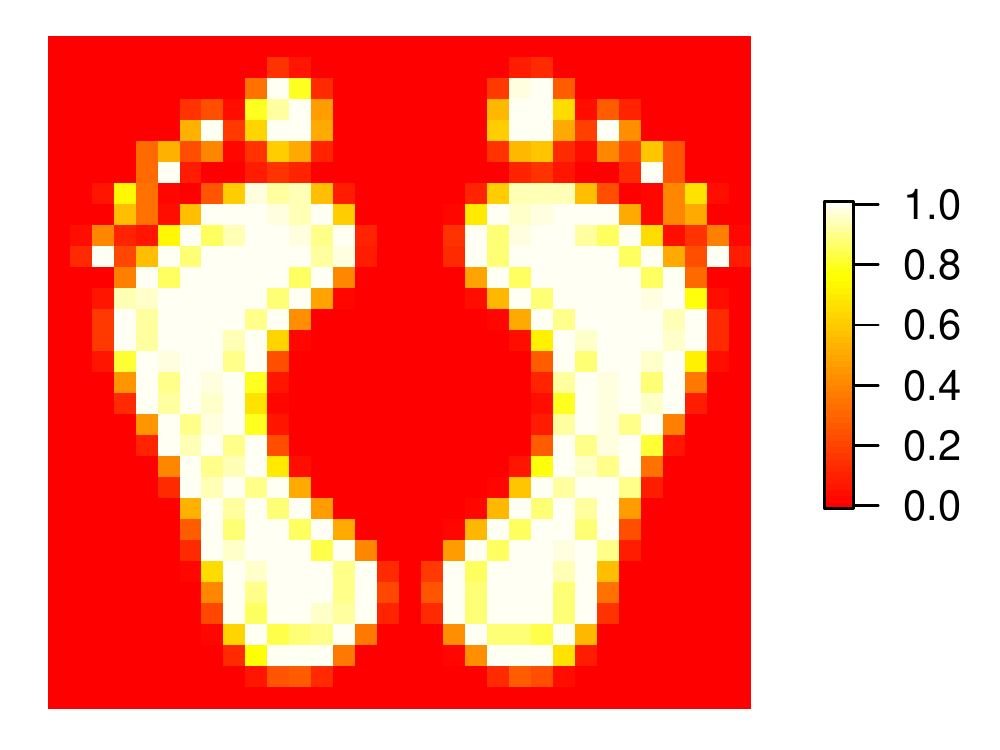} \\
\includegraphics[width=\textwidth]{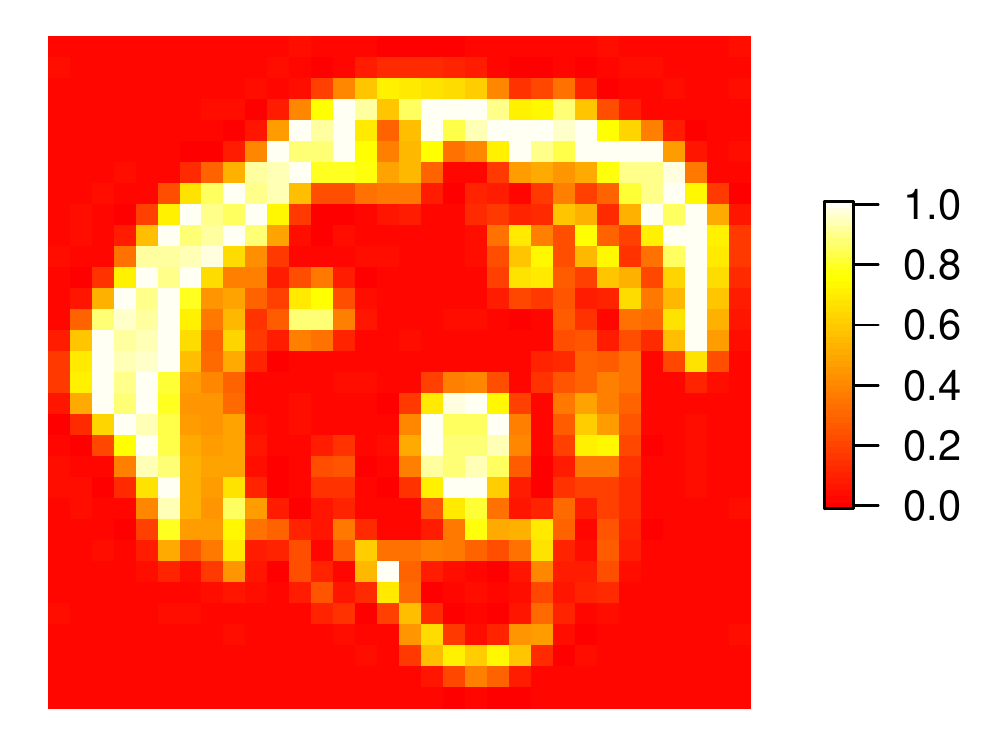} \\
\includegraphics[width=\textwidth]{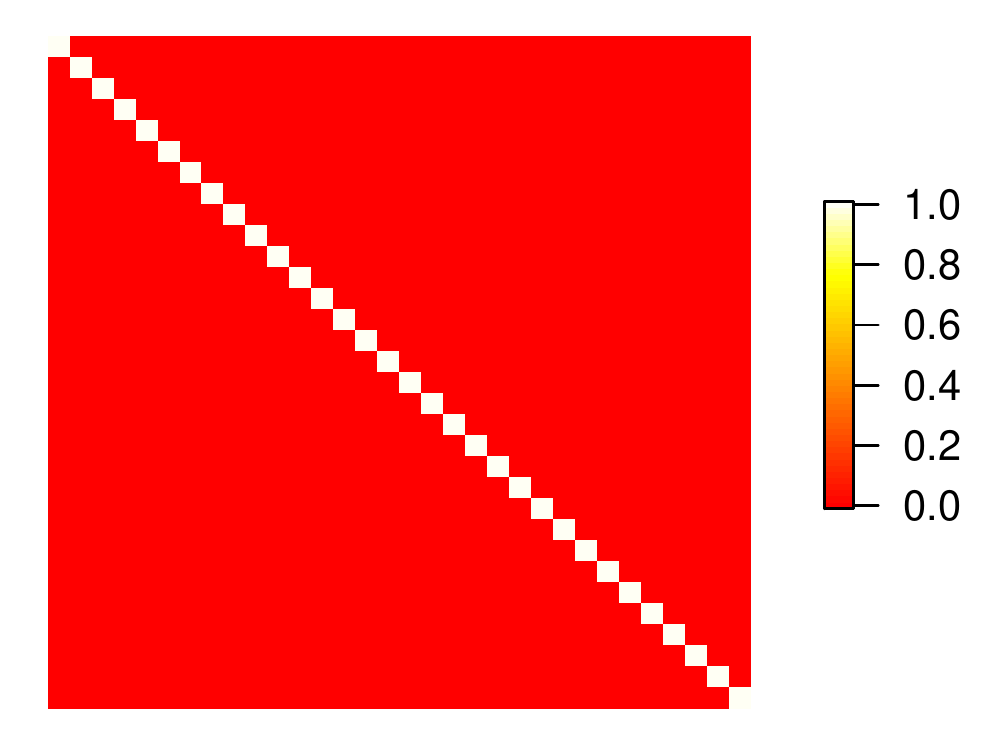}
\end{minipage}
\begin{minipage}{0.21\textwidth}
\includegraphics[width=\textwidth]{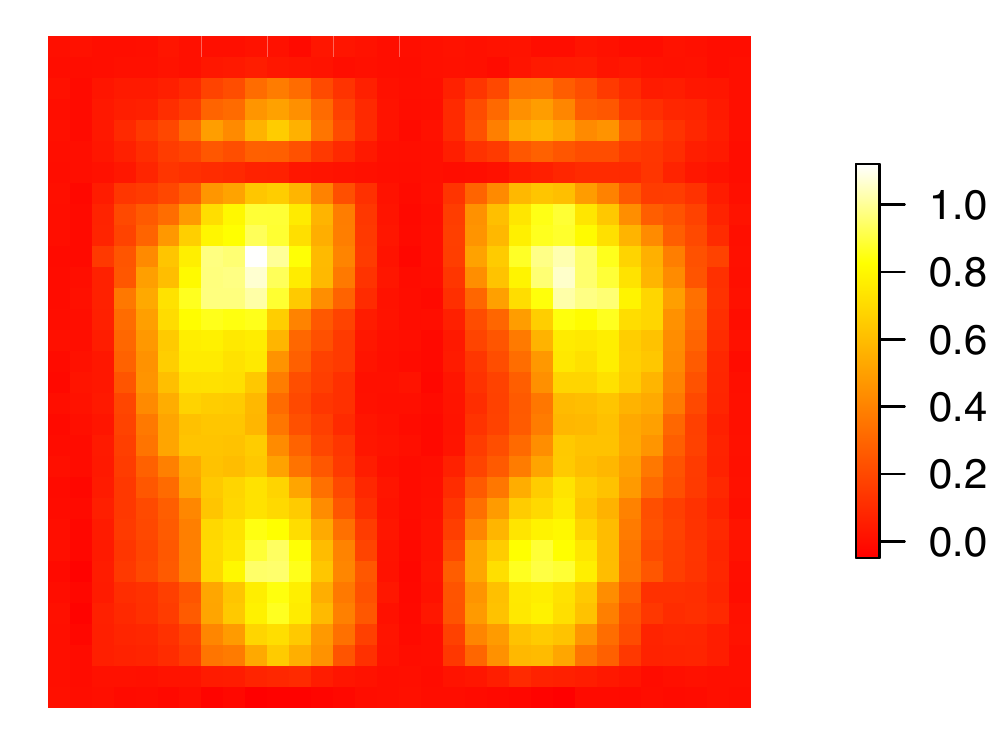} \\
\includegraphics[width=\textwidth]{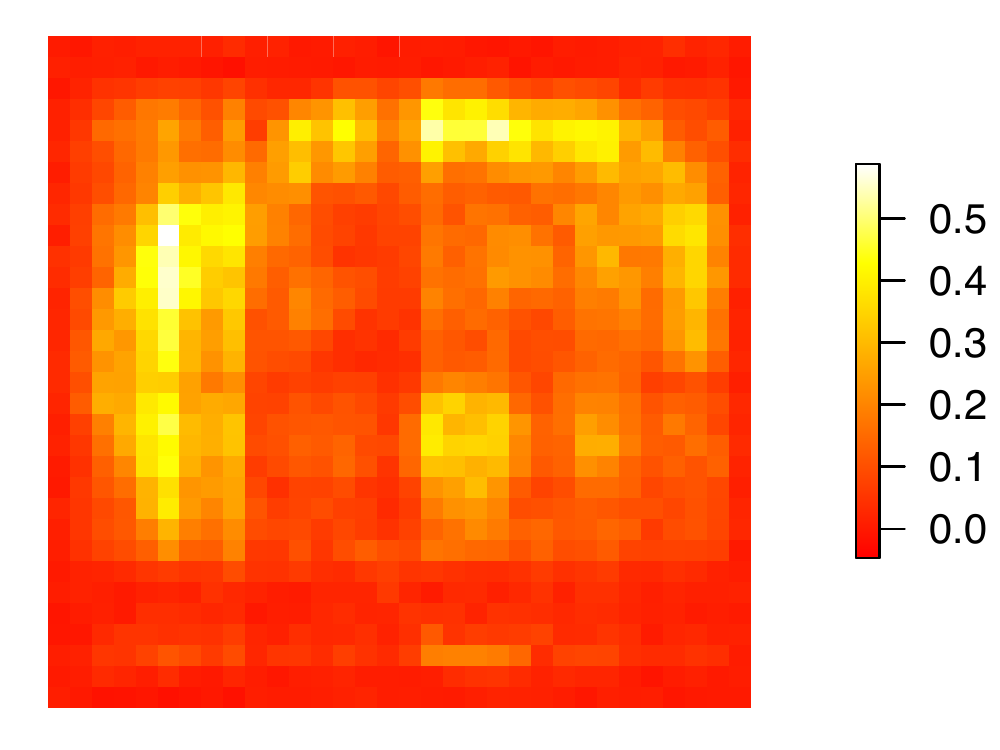} \\
\includegraphics[width=\textwidth]{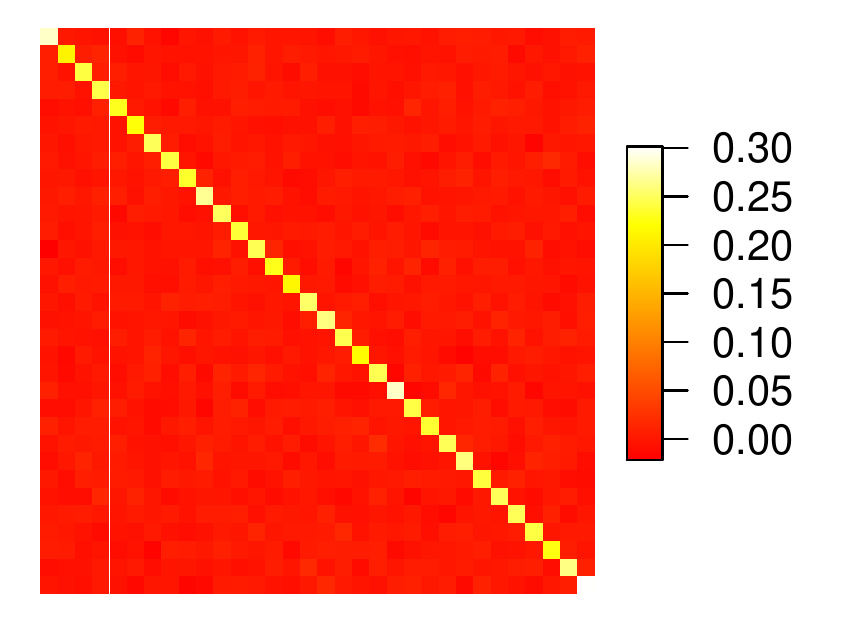}
\end{minipage}
\begin{minipage}{0.21\textwidth}
\includegraphics[width=\textwidth]{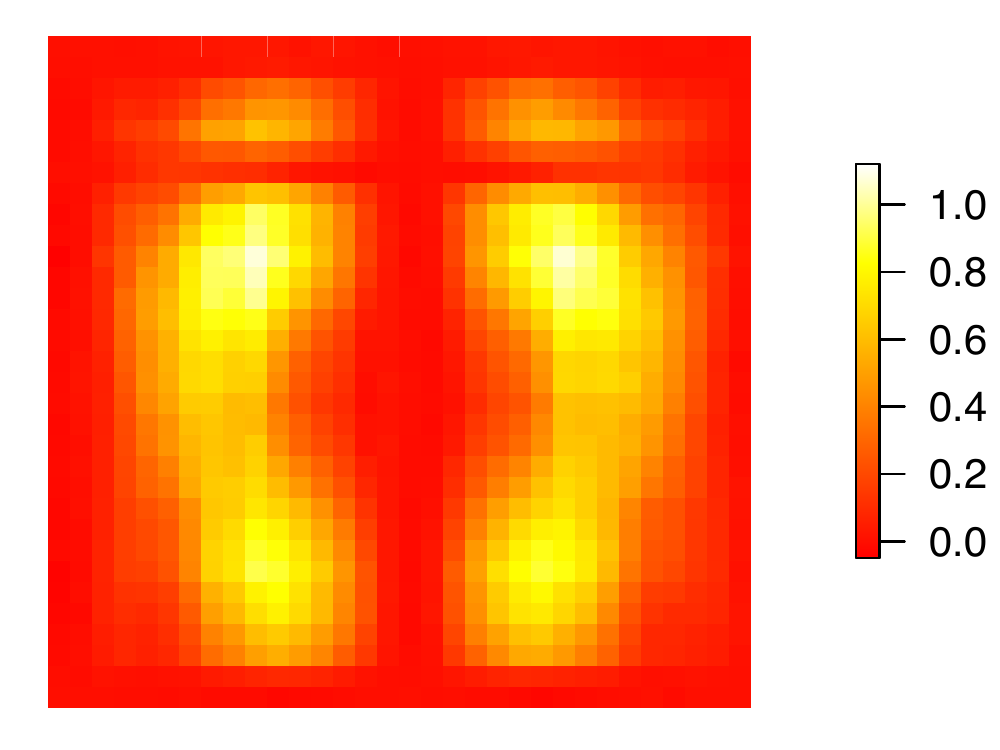} \\
\includegraphics[width=\textwidth]{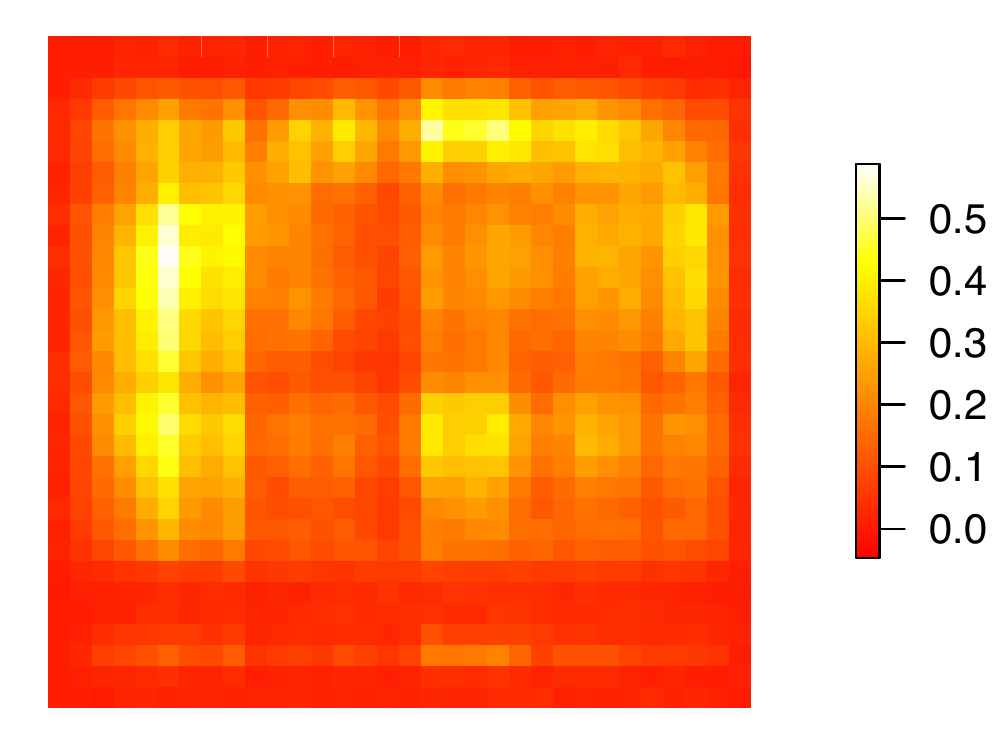} \\
\includegraphics[width=\textwidth]{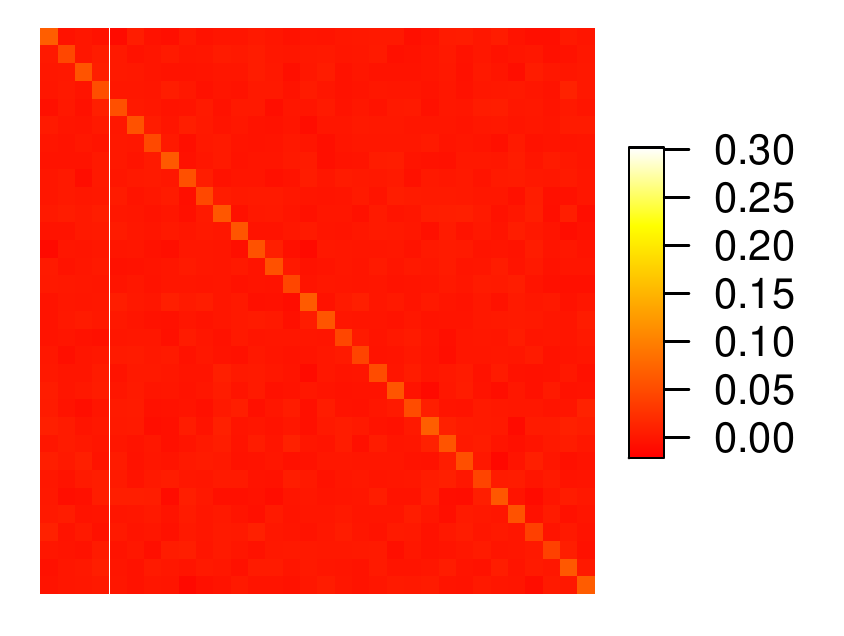}
\end{minipage}
\begin{minipage}{0.21\textwidth}
\includegraphics[width=\textwidth]{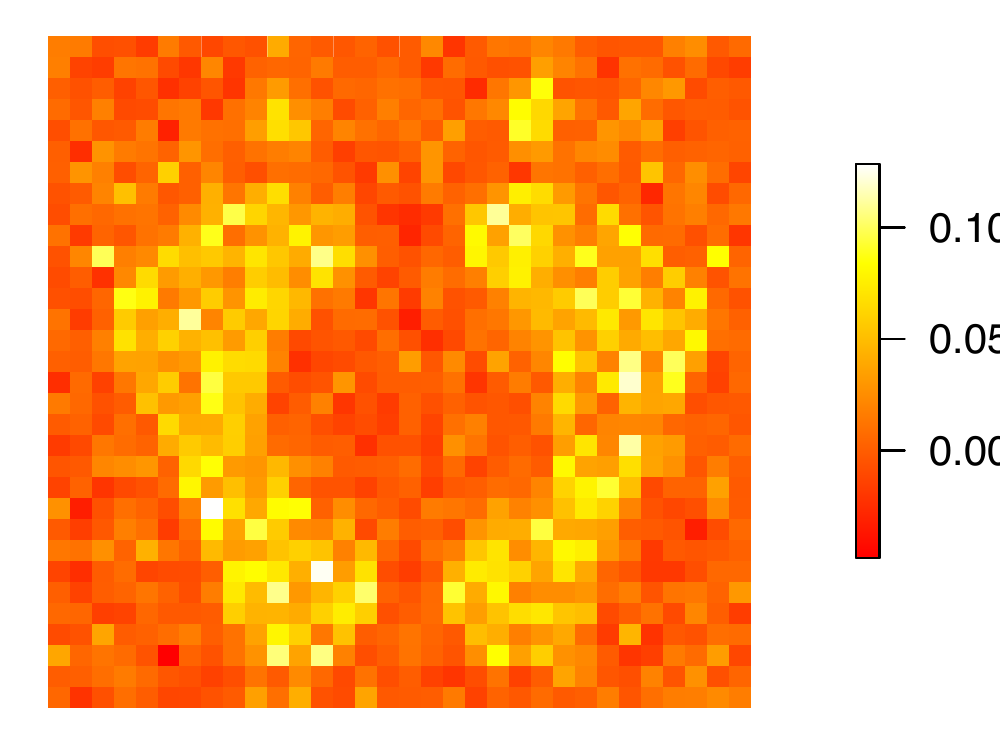} \\
\includegraphics[width=\textwidth]{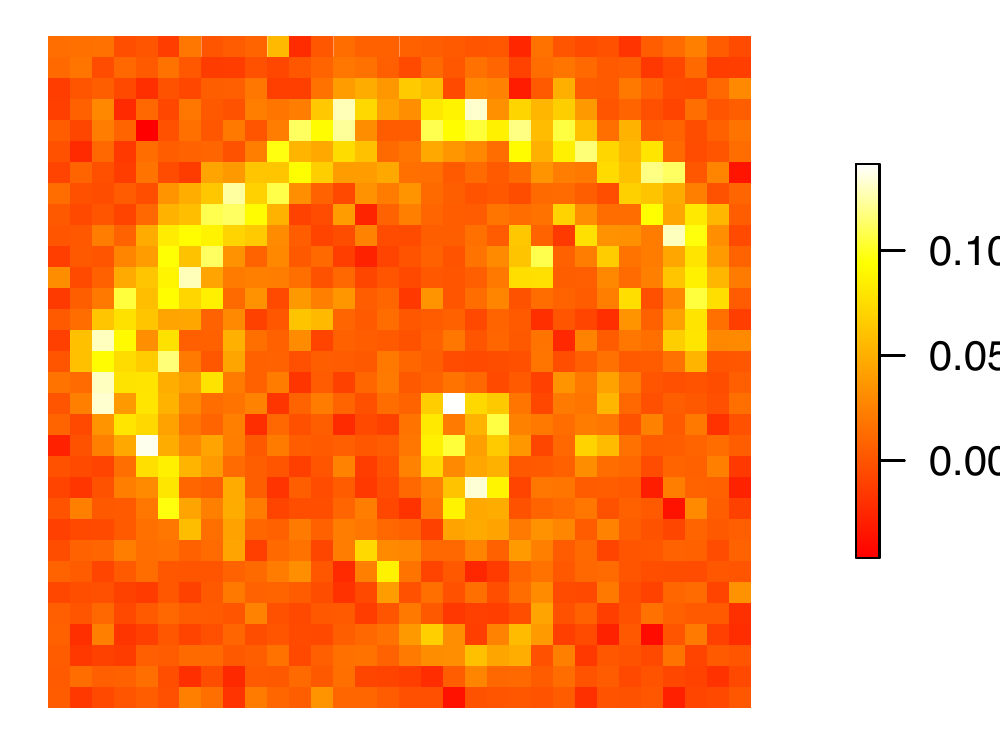} \\
\includegraphics[width=\textwidth]{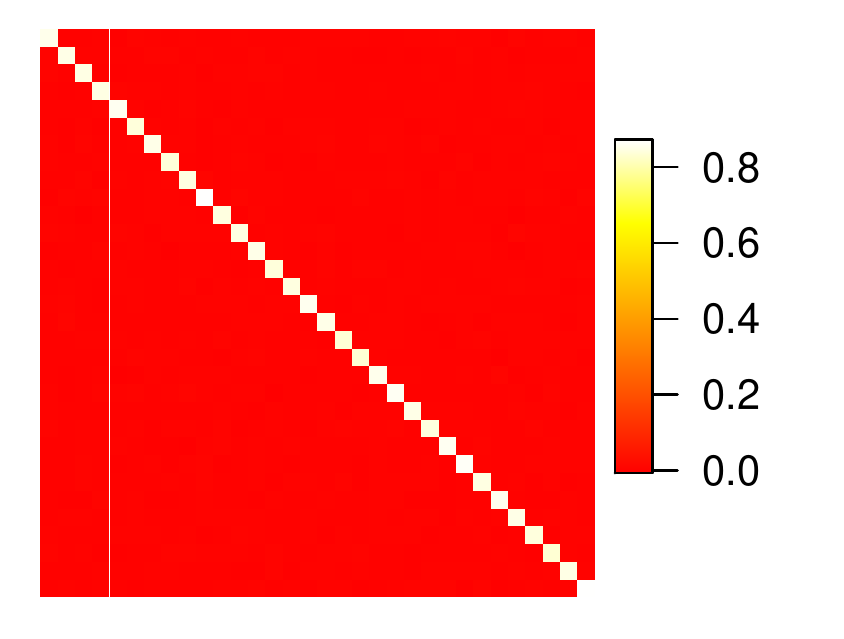}
\end{minipage}

\begin{minipage}{0.21\textwidth}
\subfloat[Truth]{\hspace{\textwidth}}
\end{minipage}
\begin{minipage}{0.23\textwidth}
	\subfloat[\softer{}]{\hspace{\textwidth}}
\end{minipage}
\begin{minipage}{0.17\textwidth}
	\subfloat[\parafac{}]{\hspace{\textwidth}}
\end{minipage}
\begin{minipage}{0.27\textwidth}
	\subfloat[Lasso]{\hspace{\textwidth}}
\end{minipage}

\caption{True coefficient matrix and average across simulated data sets of the coefficient matrix posterior mean for \softer{} and hard \parafac{}. Note that the true matrices might be shown at a different color scale than the estimated ones.}
\label{app_fig:sims_n200}

\end{figure}

Simulation results in this section represent a subset of the scenarios (dog, feet, diagonal) in \cref{subsec:sims_400} but for sample size $n = 200$. The general conclusions from \cref{subsec:sims_400} remain even when considering a smaller sample size. Figure \ref{app_fig:sims_n200} shows a plot similar to the one in Figure \ref{fig:sim_pictures} including the true coefficient matrices and average posterior mean or penalized estimate across data sets.
Again, the color scale of the results is the same for \softer{} and hard \parafac{}, but is different for the Lasso. Using different scales facilitates illustration of the underlying structure the methods estimate, even though different methods estimate different magnitude of coefficients. For example, Lasso estimates the feet structure, but non-zero coefficients are greatly underestimated around 0.1 (instead of 1). In contrast, in the truly sparse, diagonal scenario, Lasso estimates non-zero coefficients at about 0.8 whereas \softer{} estimates them to be close to 0.3, and hard \parafac{} near 0.06.

One of the main conclusions is that \softer{} deviates less from hard \parafac{} when the $(n, p)$ ratio is small, which is evident by the more rectangular structure in the mean coefficient matrix for the dog scenario, and the stronger shrinkage of the non-zero coefficients in the diagonal scenario. Further, \cref{app_tab:sims_n200} show the average absolute bias and root mean squared error for estimating the coefficient matrix entries, and the prediction mean squared error.

\begin{table}[!b]
\centering
\begin{tabular}{ll lll}
& & \softer{} & \parafac{} & Lasso \\[10pt]
feet & Truly zero & 0.06 (0.14) & 0.065 (0.15) & \textbf{0.01 (0.118)} \\ 
& Truly non-zero & \textbf{0.216} (0.332) & 0.229 \textbf{(0.328)} & 0.535 (0.585)  \\
& Prediction & 95.4 (76.6, 111.1) & \textbf{94 (78.8, 107.5)} & 312 (299, 323) \\[10pt]
dog & Truly zero & 0.042 (0.155) & 0.087 (0.164) & \textbf{0.017 (0.155)} \\
& Truly non-zero & \textbf{0.171} (0.264) & 0.174 \textbf{(0.254)} & 0.248 (0.305) \\
& Prediction & 110.6 (99.6, 121.1) & \textbf{101.2 (94, 107.8)}  & 177.3 (170.9, 183.4) \\[10pt]
diagonal & Truly zero & 0.005 (0.054) & 0.003 (0.038) & \textbf{0.002 (0.02)} \\
& Truly non-zero &  0.753 (0.773) & 0.945 (0.948) & \textbf{0.149 (0.181)} \\
& Prediction & 22.76 (21.04, 24.29) & 31.08 (30.1, 31.78) & \textbf{2.00 (1.57 2.32)}
\end{tabular}
\caption{Mean bias and rMSE among truly zero and truly non-zero coefficient entries (presented as bias (rMSE)), and average and IQR of the predictive mean squared error (presented as average (IQR)) for tensor predictor of dimensions $32 \times 32$ and $n = 200$. Bold text is used for the approach minimizing these quantities in each scenario.}
\label{app_tab:sims_n200}
\end{table}


\subsection{Simulation results for alternative rank of hard \parafac{}}
\label{app_subsec:D_7}

\begin{table}[!t]
\centering \small
\small
\begin{tabular}{lll ll}
& & & \softer{} \hspace{10pt} & \parafac{} \\ \hline

squares & Truly zero & bias &  0.003 & 0.005  \\
& & rMSE & 0.034 & 0.049  \\
& & coverage & 99.5\% & 98.6\% \\[5pt]

& Truly non-zero & bias & 0.085 & 0.104  \\
& & rMSE & 0.111 & 0.146   \\
& & coverage & 79.6\% & 70.5\% \\[5pt]

& Prediction & MSE & 5.17 & 8.76 \\
\hline

feet & Truly zero & bias & 0.035 & 0.041  \\
& & rMSE & 0.092 & 0.104   \\
& & coverage & 97.3\% & 96.7\%  \\[5pt]

& Truly non-zero & bias & 0.112 & 0.128  \\
& & rMSE & 0.181 & 0.199   \\
& & coverage & 89.2\% & 85.2\%  \\[5pt]

& Prediction & MSE & 30.69 & 37.15 \\ \hline

dog & Truly zero & bias & 0.079 & 0.078  \\
& & rMSE & 0.138 & 0.138 \\
& & coverage & 92.8\% & 94.7\%  \\[5pt]

& Truly non-zero & bias & 0.084 & 0.095  \\
& & rMSE & 0.157 & 0.166 \\
& & coverage & 92.4\% & 90.3\% \\[5pt]

& Prediction & MSE & 32.49 & 36.68  \\ \hline

diagonal & Truly zero & bias & 0.002 & 0.005 \\
& & rMSE & 0.02 & 0.06  \\
& & coverage & 100\% & 99.9\% \\[5pt]

& Truly non-zero & bias & 0.113 & 0.852 \\
& & rMSE & 0.128 & 0.861   \\
& & coverage & 93.3\% & 4.1\%  \\[5pt]

& Prediction & MSE & 1.43 & 28.09 \\ \hline
\end{tabular}
\caption{Simulation results for \parafac{} rank $D = 7$.}
\label{tab:sims_n400_D7}

\end{table}

In order to investigate the reliance of hard \parafac{} and \softer{} on the rank of the \parafac{} approximation, we considered the simulation scenarios of \cref{subsec:sims_400} for $D = 7$. Results are shown in \cref{tab:sims_n400_D7}. (Simulations comparing the performance of \softer{} and the hard \parafac{} for varying $D$ under an alternative scenario are included in \cref{app_subsec:sims_rank}).

Comparing \cref{tab:sims_n400_D7} to the results in \cref{subsec:sims_400} it is evident that the performance of \softer{} is remarkably unaltered when $D = 3$ or $D = 7$. Perhaps the only difference is in the dog simulations where bias and mean squared error for the truly zero coefficients is slightly increased when $D = 7$. This indicates that \softer{} is robust to the specification of the rank of the underlying hard \parafac{}.

In contrast, the hard \parafac{} approach shows some improvements in performance when $D=7$ compared to $D = 3$. This is evident when examining the bias and rMSE for the coefficient entries in the feet, dog, and diagonal scenarios. Specifically, the hard \parafac{} shows up to 15\% decreases in absolute bias and up to 10\% in rMSE. When examining the mean estimated coefficient matrices (not shown) we see the improvements in estimation are in picking up the toes (in the feet scenario) and the eyes (in the dog scenario). However, the improvements in performance are quite minor. We suspect that the reason is that the decrease in singular values of the true coefficient matrices is slow after the first three, indicating that adding a few additional components does not drive much of the approximation. Relatedly, it is likely that the Dirichlet prior on $\bm \zeta$ in \cite{Guhaniyogi2017bayesian}, along with a prior on Dirichlet parameter $\alpha$, effectively reduces the approximation rank to values smaller than 7.

Despite the improvements in performance for the hard \parafac{} when the rank is increased, \softer{} with either rank ($D = 3$ or 7) outperforms the hard \parafac{}. We suspect that the reason is that \softer{} allows for unstructured deviations from the underlying \parafac{} with $D = 3$, compared to structured increases in rank like the ones in the hard \parafac{}.

\subsection{Comparing \softer{} and the hard \parafac{} when varying the underlying rank $D$ and the true rank of the coefficient tensor}
\label{app_subsec:sims_rank}

We focus on the simulations in \cref{subsec:sims_rank} where we vary the true rank of the coefficient tensor and evaluate the relative performance of \softer{} and the hard \parafac{}. Here, we evaluate how \softer{} and the hard \parafac{} perform when also varying the rank $D$ used by each algorithm. Remember that the predictor is of dimension $20 \times 20$ and the sample size is set to 200.

\begin{figure}[!t]
\centering
\includegraphics[width=\textwidth]{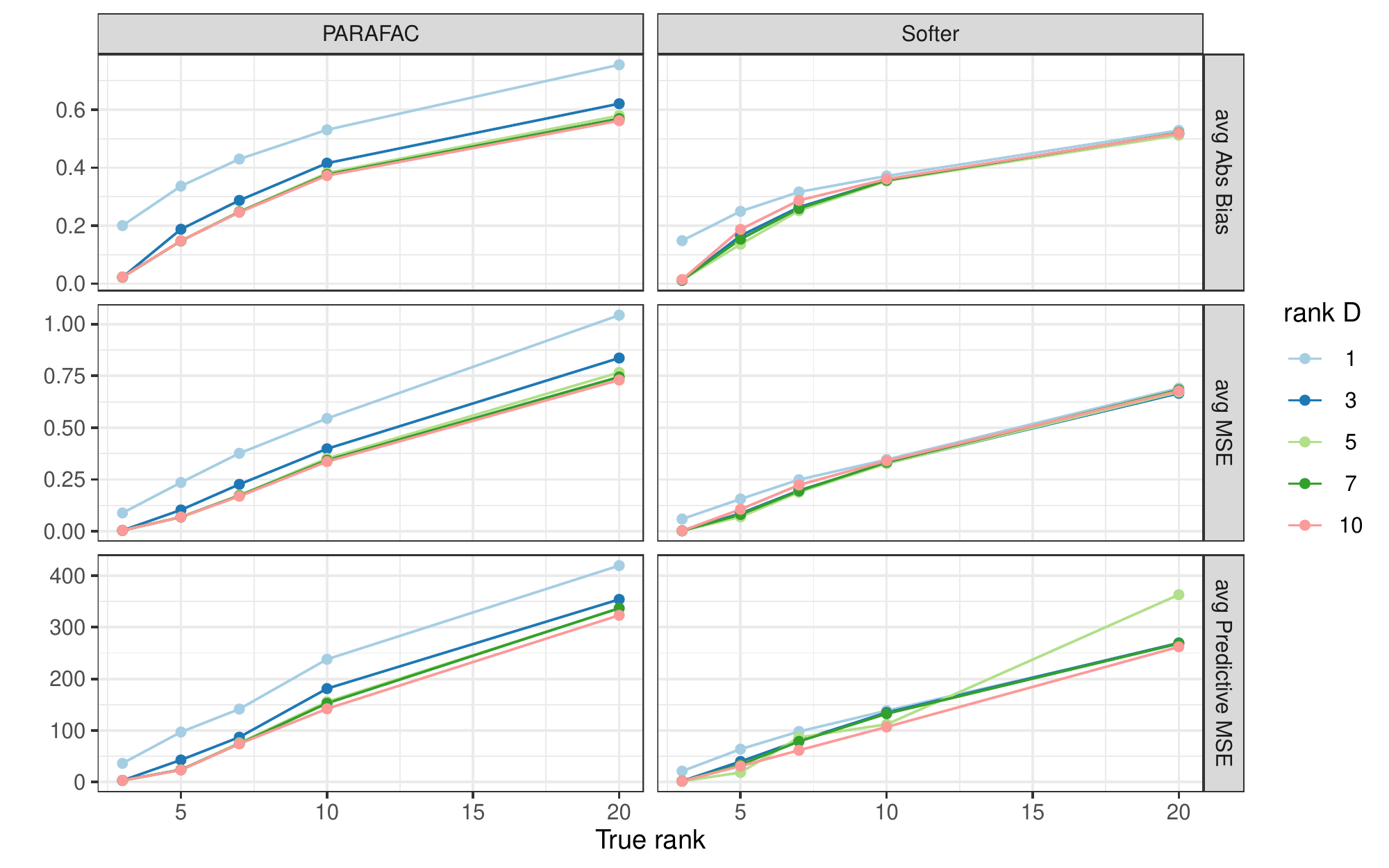}
\caption{\softer{} and hard \parafac{} Performance for Varying $D$. 
Bias, MSE and predictive MSE (rows) for varying algorithmic rank $D$ (illustrated by different colors) and varying rank of the true coefficient matrix (horizontal axis) for the hard \parafac{} (left) and \softer{} (right).
}
\label{app_fig:sims_varyingD}
\end{figure}

\cref{app_fig:sims_varyingD} shows the method performance for varying algorithmic rank $D$ and varying rank of the true coefficient matrix. Of course, as the true rank of the coefficient tensor increases, the performance of both methods deteriorates, irrespective of the algorithmic rank used. That is expected since the problem becomes harder with more parameters required to estimate the coefficient matrix, and sample size always equal to 200. As illustrated in \cref{subsec:sims_rank}, the performance of \softer{} is generally better than that of the hard \parafac{}, especially when the true rank is larger than the value of $D$ used.

Next, we compare how the same method performs for different values of $D$. The performance of the hard \parafac{} with higher values of $D$ is consistently better than the same method's performance for lower values of $D$. In contrast, \softer{} performance is strikingly similar across most values of $D$, illustrated by essentially indistinguishable lines, especially for larger values of the true coefficient tensor. These results again illustrate that the results from \softer{} are robust to the choice of $D$.

\clearpage

\section{Additional information on our brain connectomics study}
\label{app_sec:additional_application}

\subsection{Outcome information for our brain connectomics study}

\cref{app_tab:app_outcomes} shows the full list of outcomes we considered in our analysis. Both binary and continuous outcomes are considered. Information includes the category of trait. C/B represents continuous and binary outcomes.

{\footnotesize
\begin{longtable}{ll  L{5.3cm} ccc}
Category & Name & Description & \#obs & Type & Mean (SD)  \\ \hline
Cognition & ReadEng AgeAdj & Age adjusted reading score & 1,065 & C & 107.1 (14.8) \\
& PicVocab AgeAdj & Age adjusted vocabulary comprehension & 1,065 & C & 109.4 (15.2) \\
& VSPLOT TC & Spatial Orientation (Variable Short Penn Line Orientation Test) & 1,062 & C & 15 (4.44) \\[20pt]
Motor & Endurance AgeAdj & Age Adjusted Endurance & 1,063 & C & 108.1 (13.9)  \\
& Strength AgeAdj & Age Adjusted Strength & 1,064 & C & 103.6 (20.1) \\[20pt]
%
Substance Intake & Alc 12 Drinks Per Day & Drinks per day & 1,010 & C & 2.3 (1.57) \\
& Alc Hvy Frq 5plus & Frequency of $5+$ drinks & 1,011 & C & 3 (1.44) \\
& TB DSM Difficulty Quitting & Tobacco difficulty quitting & 280 & B & 74.6\% \\
& Mj Ab Dep & Marijuana Dependence & 1,064 & B & 9.3\%  \\[20pt]
%
Psychiatric and & ASR Intn T & Achenbach Adult Self-Report (Internalizing Symptoms) & 1,062 & C & 48.5 (10.7) \\
Life function & ASR Oth Raw & Achenbach Adult Self-Report (Other problems) & 1,062 & C & 9.1 (4.6) \\
& Depressive Ep & Has the participant experienced a diagnosed DSMIV major depressive episode over his/her lifetime & 1,035 & B & 9.2\% \\[20pt]
%
%
Emotion  & AngHostil Unadj & NIH Toolbox Anger and Affect Survey (Attitudes of Hostility) & 1,064 & C &  50.5 (8.58) \\[20pt]
%
Personality & NEOFAC Agreeableness & ``Big Five'' trait: Agreeableness & 1,063 & C & 32 (4.93) \\[20pt]
%
Health & BMI & Body Mass Index  & 1,064 & C & 26.4 (5.1) \\ \hline
\caption{List of outcomes with description and descriptive statistics.}
\label{app_tab:app_outcomes}
\end{longtable}}

\subsection{Additional Results}
\label{app_sec:application_additional_results}

\begin{figure}[!t]
\centering
\includegraphics[width = \textwidth]{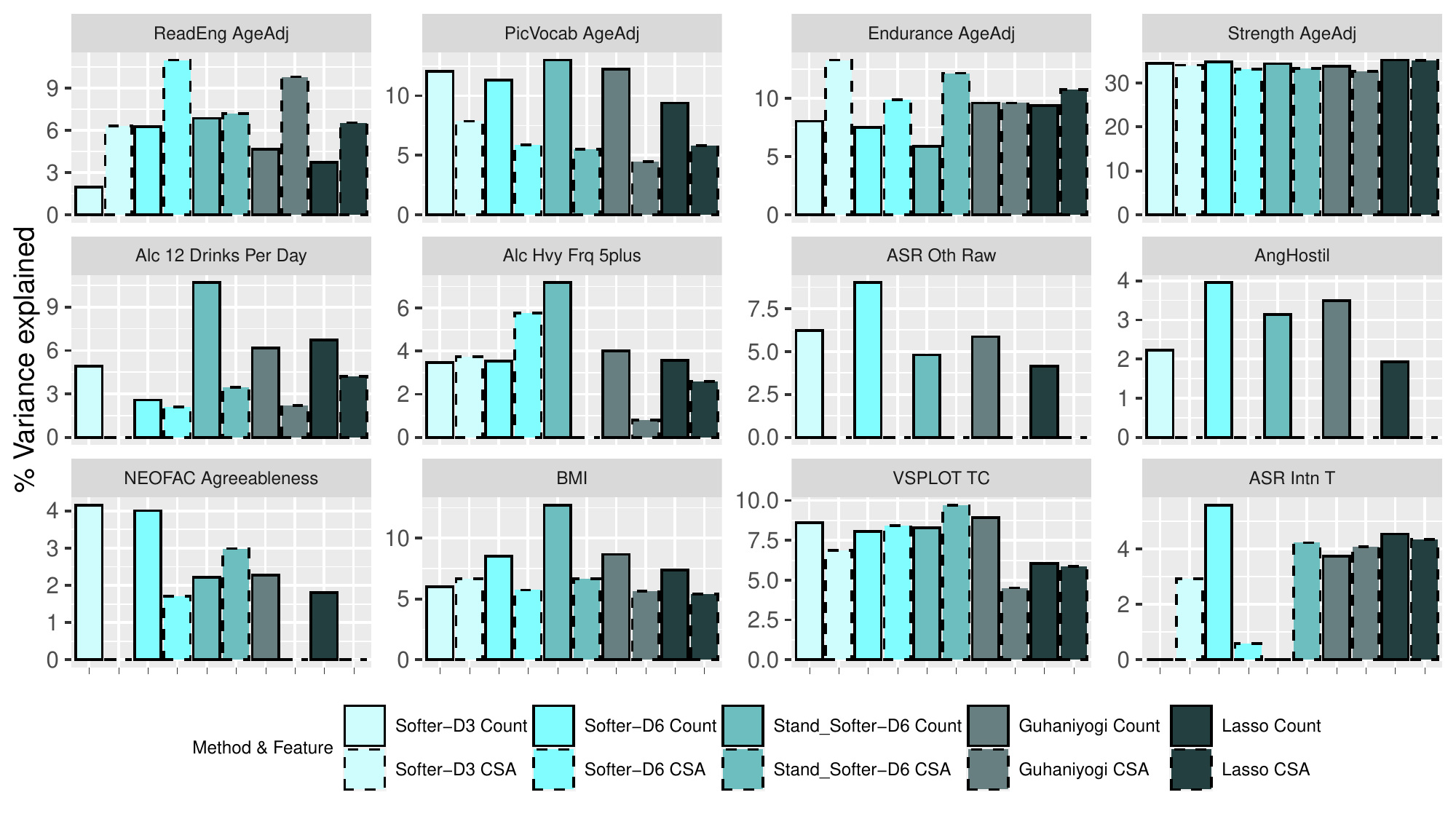}\\
\includegraphics[width = 0.75\textwidth]{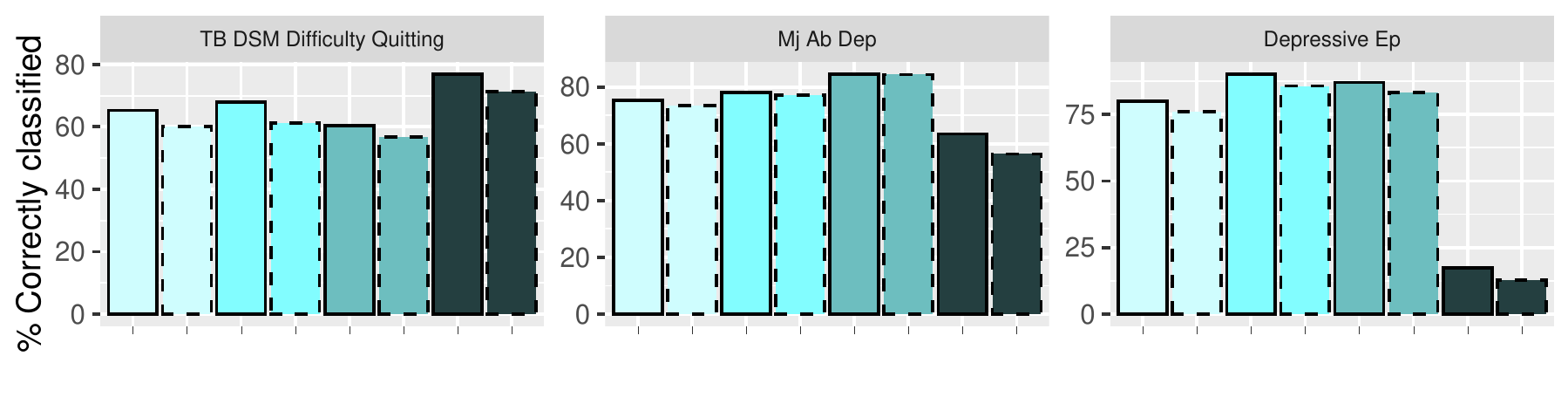}
\caption{Additional study results showing predicted variable explained and percentage of correctly classified held-out data based on a larger set of approaches.}
\label{app_fig:results_all}
\end{figure}

\cref{app_fig:results_all} shows predictive power of (1) symmetric \softer{} with $D = 3$ and (2) $D=6$, (3) standard \softer{} ignoring symmetry with $D = 6$, (4) hard \parafac{} and (5) Lasso. Results from (2), (4), and (5) are also shown in \cref{fig:app_predict}. Comparing results from (1) and (2) we see that increasing  the rank from $D = 3$ to  $D = 6$ improved predictions for a subset of outcomes. Ignoring the symmetry of the tensor predictor performed sometimes better and sometimes worse than accounting for symmetry directly into \softer{}, showing that the two approaches perform comparably for prediction (comparing (2) and (3)).

The two versions of \softer{} with $D = 3$ and $D = 6$ returned related results for the identified important entries. \cref{app_tab:app_connections} shows the connections identified based on \softer{} for $D = 3$.
When comparing the identified entries between symmetric \softer{} with rank 3 and 6, we see that the important connection for strength in \cref{tab:app_connections} is also identified by \softer{} with $D=3$, using either count or CSA as the tensor predictor. 
However, Softer with $D = 3$ identified a larger number of important connections for predicting strength than \softer{} with $D = 6$, though all of the connections involved the Precuneus region. 
The 3 connections identified by \softer{} with $D = 3$ using CSA were among the 5 connections identified by the same method using count as the predictor, an indication of stability in the identified brain connections across different features of brain connections.  These results indicate the interplay between increasing the number of parameters to be estimated when increasing the rank, with the increased flexibility that increasing the rank provides to estimation.
No important connections were identified for having had a depressive episode when using \softer{} with $D = 3$, and the identified connection for predicting VSPLOT includes the Banks of Superior Temporal Sulcus as one of the two regions, though in the different hemisphere and in combination with a different region for $D = 3$ and $D = 6$. It is also promising that the 2 out of 3 connections identified by \softer{} with $D = 3$ for CSA tensor predictor and endurance as the outcome correspond to the same pair of regions, though in the opposite hemisphere.

\begin{table}[!t]
\centering \small
\begin{tabular}{llllc}
Outcome & Feature & ROI 1 & ROI 2 &  \\ \hline

Strength & Count & (rh) Precuneus  & (rh) Superior Parietal & \textasteriskcentered{}\textasteriskcentered{} \\
& & & (lh) Superior Frontal & \textasteriskcentered{} \\
& & & (rh) Caudal Middle Frontal & \textasteriskcentered{} \\
& & & (rh) Isthmus of cingulate gyrus & \\
& & & (rh) Pericalcarine &  \\ \hline
& CSA & (rh) Precuneus  & (rh) Superior Parietal & \textasteriskcentered{}\textasteriskcentered{} \\
& & & (lh) Superior Frontal & \textasteriskcentered{} \\
& & & (rh) Caudal Middle Frontal & \textasteriskcentered{} \\
\hline
  
VSPLOT & CSA & (rh) Banks of Superior Temporal Sulcus & (lh) Superior Frontal & \\ \hline 

Endurance & CSA & (lh) Cuneus & (lh) Pericalcarine & \textasteriskcentered{} \\
& & (lh) Cuneus & (rh) Pericalcarine & \textasteriskcentered{} \\
& & (lh) Insula & (rh) Inferior Parietal & \\
\hline
\end{tabular}
\begin{flushleft}
\small
\textasteriskcentered{}\textasteriskcentered{}: connections that are related to those identified based on \softer{} with $D = 6$.\\
\textasteriskcentered{}: connections that are also identified based on alternative feature, outcome, or hemisphere.
\end{flushleft}
\caption{Identified connections based on \softer{} with $D = 3$.}
\label{app_tab:app_connections}
\end{table}


\subsection{Reproducibility of identified connections across subsamples}
\label{app_subsec:identified_reproducibility}

We fit \softer{} with $D = 6$ on 15 randomly chosen subsamples of our data which included 90\% of the observations. Since the sample size is smaller in the subsamples, we expect our power to detect important connections to also decrease. We investigate in how many of the subsamples we identify the connections marked as ``important'' in the full data set (with the same method), shown in \cref{tab:app_connections}.

\paragraph{Strength.} Using the count of streamlines as the tensor predictor, \softer{} with $D = 6$ identified one important connection between the ROIs Precuneus and Superior Parietal, both in the right hemisphere. The same connection was also identified in 10 out of 15 subsamples, while \softer{} did not identify any important connection in 4 of the subsamples. These results indicate that the identified connection for strength based on the count of streamlines stated in \cref{tab:app_connections} is reproducible across subsamples.

\paragraph{Depressive Episode.} Out of 15 subsamples, \softer{} with $D = 6$ identified important connections in 6 subsamples when using count of streamlines, and in 7 subsamples when using CSA as the tensor predictor. The connection between the Parahippocampal in the left hemisphere and the Lateral Orbitofrontal in the right hemisphere, was identified in all 13 instances with identified connections, based on either predictor (and in agreement with the results in \cref{tab:app_connections}). What's more, {\it any} of the connections identified across {\it any} subsample and for {\it either} predictor involved at least one of these two regions, implying that the importance of these two regions in predicting whether an individual has had a depressive episode is reproducible.

\paragraph{VSPLOT.} Using CSA as the tensor predictor, \softer{} with $D = 6$ identified no connections in any of the subsamples. This might imply that the connection in \cref{tab:app_connections} is not reproducible, or that the reduced sample size in the subsamples does not allow us to identify it.

\section{Symmetric and semi-symmetric soft tensor regression}
\label{app_sec:symmetric_softer}

\subsection{\softer{} for symmetric 2-mode predictor}

We start again from model \cref{eq:tensor_regression2} with $ Y_i = \mu + \C_i^T\bm \delta + \langle \tensor_i, \B \rangle_F + \epsilon_i. $ However, now $X_i$ is an $R \times R$ symmetric matrix with ignorable diagonal elements. This means that we can think of $\B$ as a real symmetric matrix with ignorable diagonal elements.

\subsubsection{Eigenvalue decomposition of real symmetric matrices}

We can still approximate $\B$ in the same way as in \parafac{} (SVD) by writing 
$\B = \sum_{d = 1}^D \gamma_1\tod \otimes \gamma_2\tod$
for some $D$ large enough and $\gamma_1\tod, \gamma_2\tod \in \mathbb{R}^R$.
However, this would not enforce that $\B$ is symmetric, since $\B_{j_1j_2} = \sum_{d = 1}^D \gamma_{1,j_1}\tod \gamma_{2,j_2}\tod \neq \sum_{d = 1}^D \gamma_{1j_2}\tod \gamma_{2j_1}\tod = \B_{j_2j_1}$. This implies that the entries of $\B$ would only be identifiable up to $\B_{j_1j_2} + \B_{j_2j_1}$.

Since $\B$ is a real symmetric matrix, it is diagonalizable and it has an eigenvalue decomposition. Therefore, we can think of approximating $\B$ using
\begin{equation}
\B = \sum_{d = 1}^D \xi\tod \gamma\tod \otimes \gamma\tod,
\label{app_eq:eigen_decomp}
\end{equation}
for sufficiently large $D$, where $\gamma\tod \in \mathbb{R}^R$ and $\xi\tod \in \mathbb{R}$. 
Note that the vectors $\gamma\tod$ here resemble the ones in the \parafac{} decomposition, but they are the same across the two tensor modes (matrix rows and columns). 

The main difference between using the eigenvalue-based approximation in \cref{app_eq:eigen_decomp}, compared to the \parafac{}-based approximation is the inclusion of the parameters $\xi\tod$. Here, $\xi\tod$ are necessary in order to have a eigenvalue decomposition employing vectors with real entries. In fact, excluding $\xi\tod$ from \cref{app_eq:eigen_decomp} can only be used to approximate positive definite symmetric matrices. To see this, take vector $\bm v \in \mathbb{R}^R$. Then,
$$
\bm v^T \Big(\sum_{d=1}^D \gamma\tod \otimes \gamma\tod \Big) \bm v =
\bm v^T \Big(\sum_{d=1}^D \gamma\tod \gamma\tod{}^T \Big) \bm v = 
\sum_{d=1}^D \bm v^T \gamma\tod (\bm v^T \gamma\tod)^T =
\sum_{d=1}^D (\bm v^T \gamma\tod)^2 \geq 0
$$
\subsubsection{Soft eigenvalue-based tensor regression}

In the case of tensor predictors without symmetries, \softer{} was built based on the \parafac{} (multimodal equivalent to SVD) approximation of the coefficient tensor. Instead, for symmetric matrices, \softer{} is based on the eigenvalue decomposition while still allowing for deviations in row (and column)-specific contributions. However, these deviations also have to account for the symmetric nature of the tensor predictor.

Write $\B = \sum_{d = 1}^D \xi\tod \B_1\tod \hadamard \B_2\tod$ similarly to \cref{eq:softB}, and assume prior distributions on all parameters as in \cref{subsec:bayesian}. Note that the parameters $\gamma\tod$ are not forced to be of norm 1 (as in classic eigenvalue decomposition), and they can have any magnitude. This allows us to restrict the parameters $\xi\tod$ to be in $\{-1, 1\}$, and base shrinkage of unnecessary ranks on shrinkage of the vectors $\gamma\tod$. Therefore, we assume a Bernoulli$(0.5)$ distribution over $\{-1, 1\}$ for parameters $\xi\tod$.

However, even though the symmetry of the underlying decomposition is enforced based on $\gamma\tod$, we need to ensure that it is also enforced when writing $\B = \sum_{d = 1}^D \xi\tod \B_1\tod \hadamard \B_2\tod$ using $\B_1\tod, \B_2\tod$.
Note that the \softer{} framework assumes that entries $\beta_{1,j_1j_2}\tod$ are centered around $\gamma_{j_1}\tod$, and similarly entries $\beta_{2, j_1j_2}\tod$ are centered around $\gamma_{j_2}\tod$. However, doing so does not necessarily lead to symmetric matrices $\B$ since
$$\B_{j_1j_2} = \sum_{d = 1}^D \beta_{1,j_1j_2}\tod \beta_{2, j_1j_2}\tod \neq \sum_{d = 1}^D \beta_{1,j_2j_1}\tod \beta_{2, j_2j_1}\tod = \B_{j_2j_1}.$$

We enforce symmetry of $\B$ by focusing only on the lower-triangular part. \softer{} for symmetric matrix predictor specifies row $i$'s contributions to entries $\B_{ij}$, $\beta_{1,ij}\tod$, as centered around $\gamma_i\tod$ only for $i > j$. Then, for $j > i$ we set $\beta_{1,ij}\tod = \beta_{1,ji}\tod$. 
Similarly, column $i$'s contributions to entries $\B_{ji}$, $\beta_{2, ji}\tod$ are centered around $\gamma_i\tod$ only for $i < j$, and for $j > i$ we set $\beta_{2, ji}\tod = \beta_{2, ij}\tod$. An equivalent way to enforce symmetry on $\B$ is to allow all entries in $\B_1\tod$ to have the same form as in \softer{}, and force $\B_2\tod = (\B_1\tod)^T$.
\subsubsection{Note on implementation using RStan}

Note that RStan cannot directly handle discrete parameters as $\xi\tod$. The most common approach to discrete parameters is to specify the likelihood integrating these parameters out. However, this approach is not easily applicable in our setting since $\xi\tod$ are entangled in the likelihood through their role in the coefficient matrix $\B$. For that reason, we take an alternative approach, and assume that $\xi\tod$ are continuous and specify a mixture of normals distribution on each of them: $\xi\tod \sim 0.5 N(-1, 0.001) + 0.5 N(1, 0.001)$. Since the parameters $\xi\tod$ are not directly of interest, and shrinkage of the contributions of component $d$ in a rank$-D$ decomposition is achieved through the prior on $\gamma\tod$, we expect that this approach will closely resemble results from a specification that defines $\xi\tod$ to be binary taking values in $\{-1, 1\}$ from a Bernoulli$(0.5)$ distribution.

\subsection{\softer{} for semi-symmetric 3-mode tensor}

In brain connectomics, and specifically in our study of features of brain connections and their relationship to traits, tensor predictors are often of dimensions $R \times R \times p$ and are semi-symmetric. Semi-symmetry means that the predictor $\tensor$ is symmetric along its first two modes and $\tensor_{j_1j_2j_3} = \tensor_{j_2j_1j_3}$. An example of such tensor includes $R$ brain regions along the first two modes and $p$ features of brain connection characteristics along its third mode. When these features are symmetric (feature of connection from region $i$ to region $j$ is the same as the feature of connection from region $j$ to region $i$), the tensor predictor is semi-symmetric. In such cases, the standard \softer{} approach could be applied, but entries of $\B$ would be identifiable only up to $\B_{j_1j_2j_3} + \B_{j_2j_1j_3}$. In order to account for the semi-symmetry in $\tensor$ we can enforce the same type of semi-symmetry in $\B$ by adopting a \parafac{}-eigenvalue decomposition hybrid.

Specifically, assume that $\B$ is a 3-mode semi-symmetric coefficient tensor corresponding to the semi-symmetric predictor $\tensor$. Then, for sufficiently large $D$, $\gamma\tod \in \mathbb{R}^R$ and $\rho\tod \in \mathbb{R}^p$, we can write 
\begin{equation}
\B = \sum_{d = 1}^D \gamma\tod \otimes \gamma\tod \otimes \rho \tod.
\label{app_eq:semi-parafac}
\end{equation}
This leads to a natural approximation for $\B$ for some value $D$ potentially smaller than the true one. \softer{} for semi-symmetric tensor predictor builds on \cref{app_eq:semi-parafac} while allowing for deviations in the row-specific contributions along the three modes.

We achieve that by specifying $\B = \sum_{d = 1}^D \B_1\tod \hadamard \B_2\tod \hadamard \B_3\tod$ for $\B_k\tod$ are tensors of dimensions $R \times R \times p$. The structure and specification of $\B_k\tod$ are as in \cref{subsec:bayesian} with small changes to account for the semi-symmetric structure in $\tensor$ and ensure that the estimated coefficient tensor is also semi-symmetric. Note that the $(j_1, j_2, j_3)$ entry of $\B$ is equal to
$$
\B_{j_1j_2j_3} = \sum_{d = 1}^D \beta_{1,j_1j_2j_3}\tod \beta_{2,j_1j_2j_3}\tod \beta_{3,j_1j_2j_3}\tod,
$$
and we want $\B_{j_1j_2j_3} = \B_{j_2j_1j_3}$. Borrowing from the symmetric case, we allow all row-specific contributions along mode 1, $\beta_{1, j_1j_2j_3}\tod$, to vary around the corresponding entry in the decomposition \cref{app_eq:semi-parafac}, $\gamma_{j_1}\tod$, and set $\B_{2,..j_3} = \B_{1, ..j_3}^T$. Further, we allow entries $\beta_{3,j_1j_2j_3}\tod$ to vary around $\rho_{j_3}\tod$ for $j_1 < j_2$, and set $\beta_{3,j_1j_2j_3}\tod = \beta_{3,j_2j_1j_3}\tod$ when $j_1 > j_2$. Doing so, ensures that $\B_{j_1j_2j_3} = \B_{j_2j_1j_3}$.

\newpage
\bibliographystyle{plainnat}
\bibliography{Softer,Clustering-Networks}

\end{document}